\documentclass[a4paper,10pt]{article}

\usepackage[utf8]{inputenc}

\usepackage{authblk}

\usepackage{ifthen,ifpdf}

\ifpdf
  \usepackage{hyperref}
  \hypersetup{colorlinks=false,allcolors=blue}
  \usepackage{hypcap}
  \pdfpagewidth=\paperwidth
  \pdfpageheight=\paperheight
\fi

\usepackage[utf8]{inputenc}
\usepackage[T1]{fontenc}

\usepackage{amsmath, amsthm, amssymb}
\usepackage{mathtools}
\usepackage{enumerate}
\usepackage{afterpage}
\usepackage{tabularx}
\usepackage{dsfont}

\usepackage{graphicx}

\usepackage{algorithm}
\usepackage{algpseudocode}
\usepackage{caption}
\usepackage[labelformat=simple]{subcaption}

\usepackage{color}
\usepackage{units}
\usepackage{array, multirow}
\usepackage{booktabs}

\newcolumntype{M}[1]{>{\vspace{3pt}\raggedleft\arraybackslash}m{#1}}

\usepackage{tikz}
\usepackage{pgfplots}
\usepackage{pgfplotstable}
\usepgfplotslibrary{units}
\pgfplotsset{compat=1.9}
\pgfplotstableset{col sep=comma}
\usepackage{bm}

\usepackage{cite}
\usepackage{anysize} 
\marginsize{2.5cm}{2.5cm}{2cm}{2cm}
\usepackage{stackengine}
\usepackage{wrapfig}
\usepackage{enumitem}
\usepackage[normalem]{ulem}
\theoremstyle{plain}

\theoremstyle{remark}

\numberwithin{equation}{section}

\newcommand{\norm}[1]{\|#1\|} 

\newcommand{\T}[1]{{#1}^{\sf  T}}

\newcommand{\fx}{\sty{ x}}
\newcommand{\fxi}{\sty{ \xi}}
\newcommand{\fq}{\sty{ q}}
\newcommand{\fa}{\sty{ a}}
\newcommand{\fv}{\sty{ v}}
\newcommand{\fsigma}{\mbox{\boldmath $\sigma$}}

\newcommand{\fe}{\sty{ e}}
\newcommand{\ffC}{\styy{ C}}

\newcommand{\ffP}{\styy{ P}}

\newcommand{\feps}{\mbox{\boldmath $\varepsilon $}}

\newcommand{\Sym}[1]{\textrm{Sym}(#1)}

\newcommand{\bfeps}{\bar{\feps}}

\newcommand{\sty}[1]{\mbox{\boldmath $#1$}}
\newcommand{\fu}{\sty{ u}}
\newcommand{\styy}[1]{{\mathbb{#1}}}
\newcommand{\fsymgrad}{{\sty{\nabla}^s}}

\DeclareMathOperator{\eps}{\varepsilon}

\newcommand{\ffZ}{{\styy{ Z}}}
\newcommand{\fzero}{\sty{ 0}}

\newcommand{\R}{{\styy{ R}}}

\newcommand{\uul}[1]{\underline{\underline{#1}}}
\newcommand{\ul}[1]{\underline{#1}}

\makeatletter
\newcommand{\citecomment}[2][]{\citen{#2}#1\citevar}
\newcommand{\citeone}[1]{\citecomment{#1}}
\newcommand{\citewo}[2][]{\citecomment[,~#1]{#2}}
\newcommand{\citevar}{\@ifnextchar\bgroup{;~\citeone}{\@ifnextchar[{;~\citewo}{]}}}
\newcommand{\citefirst}{\@ifnextchar\bgroup{\citeone}{\@ifnextchar[{\citewo}{]}}}
\newcommand{\cites}{[\citefirst}
\makeatother

\title{A stable and accurate X-FFT solver for\\linear elastic homogenization problems in 3D} 

\author[1]{Flavia Gehrig}
\author[1,2,3,*]{Matti Schneider}

\affil[1]{University of Duisburg-Essen, Institute of Engineering Mathematics}
\affil[2]{Fraunhofer Institute for Industrial Mathematics ITWM, Kaiserslautern}

\affil[3]{Center for Nanointegration Duisburg-Essen (CENIDE)}
\affil[*]{correspondence to: \texttt{matti.schneider@uni-due.de}}

\date{\today}

\begin{document}

\maketitle 
\begin{abstract}
\noindent Although FFT-based methods are renowned for their numerical efficiency and stability, traditional discretizations fail to capture material interfaces that are not aligned with the grid, resulting in suboptimal accuracy. To address this issue, the work at hand introduces a novel FFT-based solver that achieves interface-conforming accuracy for three-dimensional mechanical problems. More precisely, we integrate the extended finite element (X-FEM) discretization into the FFT-based framework, leveraging its ability to resolve discontinuities via additional shape functions. We employ the modified abs(olute) enrichment and develop a preconditioner based on the concept of strongly stable GFEM, which mitigates the conditioning issues observed in traditional X-FEM implementations. Our computational studies demonstrate that the developed X-FFT solver achieves interface-conforming accuracy, numerical efficiency, and stability when solving three-dimensional linear elastic homogenization problems with smooth material interfaces.
\end{abstract}

\newpage

\section{Introduction}\label{sec:intro}
\subsection{State of the art}
To understand the behavior of complex materials, developing efficient and accurate simulation techniques turns out to be essential. Over the last decade, methods based on the fast Fourier transform (FFT)~\cite{moulinec1994fast,moulinec1998numerical} became increasingly popular due to their efficiency, which rests on: 
\begin{itemize}[nosep]
\item their low memory footprint due to their matrix-free formulation,
\item their speed due to the use of state-of-the-art FFT implementations~\cite{frigo2005design,dalcin2019fast} and dedicated iterative solution schemes,
\item their use of regular grids, eliminating the need of expensive mesh generation and enabling to operate on $\mu$-CT data~\cite{chen2019analysis} directly,
\item their stability, which is rooted in their preconditioning strategy that leads to a mesh-independent upper bound on the condition number,
\item their parallelizability.
\end{itemize}
For an overview of FFT-based methods, we refer to the review articles~\cite{lebensohn2020spectral,lucarini2021fft,schneider2021review,gierden2022review}.

In the 1990s, Moulinec-Suquet~\cite{moulinec1994fast,moulinec1998numerical} introduced FFT-based methods by reformulating the cell problem as the Lippmann-Schwinger equation for elasticity~\cite{kroner1972statistical,zeller1973elastic} and solving it using an FFT-based iterative method known as the basic scheme.
Around 2014, the original method was decomposed into its constituent parts - discretization and solution scheme~\cite{brisard2012combining,willot2014fourier,vondvrejc2014fft,schneider2015convergence}. Moreover, the Fourier-based discretization of Mouline-Suquet was interpreted as a trigonometric collocation discretization~\cite{nguyen2011efficient} and a nonconforming Fourier-Galerkin discretization~\cite{vondvrejc2014fft}, while the basic scheme was identified as a variant of the gradient descent method~\cite{kabel2014efficient} and the Richardson iteration~\cite{mishra2016comparative}. These insights enabled incorporating alternative discretizations and solvers into FFT-based methods. By adapting solution schemes from large scale optimization~\cite{zeman2010accelerating,schneider2019barzilai,wicht2020quasi} the speed of FFT-based methods could be significantly improved. Moreover, discretizations based on voxel-wise constant fields~\cite{brisard2010fft,brisard2012combining,zecevic2021approximation}, Fourier-Galerkin methods~\cite{bonnet2007effective,monchiet2015combining,vondvrejc2016improved}, finite volumes~\cite{dorn2019lippmann,eloh2019development,ernesti2021fast}, finite differences~\cite{willot2015fourier,vidyasagar2017predicting,finel2025tetrahedron} and finite elements (FEs)~\cite{schneider2017fft,leuschner2018fourier,ladecky_optimal_2023} were introduced in the context of FFT-based computational micromechanics. 

For some discretizations, a superconvergence phenomenon was observed in FFT-based methods, where the effective stress converges at a quadratically higher rate than the local stress field~\cite{schneider2023superconvergence}. For linear elasticity problems that are discretized by trilinear hexahedral FEs, Ye-Chung~\cite{ye2023convergence} first proved the superconvergence of FFT-based methods and derived explicit convergence rates. Subsequently, Schneider~\cite{schneider2023effectiveness} proved that the convergence rates of the traditional Fourier-based discretization and the voxel finite element method (FEM) coincide. Put differently, for voxel finite element discretizations and the traditional Fourier-based discretization, the effective stress converges with the rate $\mathcal{O}\left(h\right)$ and the local stress/strain fields converge with the rate $\mathcal{O}\left(\sqrt{h}\right)$ to the exact solution, where the mesh parameter $h$ stands for the voxel edge length. These convergence rates are suboptimal compared to the quadratically higher rates of interface-conforming FEM with linear tetrahedral or bi-/trilinear hexahedral FEs~\cite{brenner_mathematical_2008,hughes_finite_1987}. Essentially, the suboptimal convergence rates of the voxel finite element discretizations and the traditional Fourier-based discretization result because of the inaccurate approximation capabilities of voxels for general interfaces~\cite{ramiere2008convergence}.

To achieve higher accuracy, composite voxels~\cite{gelebart2015filtering,kabel2015use} were introduced, which apply an averaging rule to the voxels that include an interface. Composite voxels with Voigt averaging~\cite{gelebart2015filtering,kabel2015use,jabs2025microstructure} were found to lead to the most accurate results for composites with voids, whereas composite voxels with Reuss averaging~\cite{gelebart2015filtering,kabel2015use,sterr2023homogenizing,sterr2025machine} are best suited for rigid inclusions. For finite material contrasts, composite voxels, which are furnished with the effective properties of an equivalent laminate~\cite{kabel2015use,mareau2017different,lendvai2024assumed} were found to yield the most accurate results. There are different ways to obtain the interface and subvolume fraction data required for the laminate mixing rule: Either, microstructure images with a higher resolution may be used~\cite{keshav2023fft,lendvai2025accurate} or a level-set description may be exploited~\cite{lendvai2024assumed}. Although improving the accuracy in linear elastic homogenization problems, the solution fields of the composite voxels still converge with the same rate as those of voxel FEM~\cite[Fig. 7B]{lendvai2024assumed}.

A different strategy to improve the accuracy of FFT-based methods was introduced by Zecevic et al.~\cite{zecevic2022new,zecevic2025achieving}. Their local mesh morphing approach involves mapping from an interface-conforming grid to a reference configuration, applying the FFT-based method in the reference configuration, and mapping back to the interface-conforming grid. For simple microstructures, the interface-conforming grid and the mapping operator can be computed explicitly~\cite{zecevic2022new}. More complex microstructures can be treated using alternative methods, such as balancing a system of springs connecting the nodes~\cite{zecevic2025achieving} or employing grid adaptation based on optimal transport~\cite{bellis2024numerical}. Nevertheless, generating the interface-conforming grid and mapping operator significantly increases the computational cost, and conditioning problems are expected for severely deformed grids.

Recently, a novel FFT-based framework~\cite{gehrig2025x} was introduced for two-dimensional thermal homogenization problems, which achieves interface-conforming accuracy while maintaining numerical efficiency and well-conditioning. This so-called X-FFT solver~\cite{gehrig2025x} integrates an extended finite element discretization with modified abs(olute) enrichment and a dedicated preconditioning strategy into FFT-based methods.
The extended finite element method (X-FEM) resolves discontinuities in a regular grid setting by locally enriching the polynomial approximation space of the classical FE shape functions with additional shape functions that incorporate the discontinuity. Based on X-FEM, \emph{weak} discontinuities, i.e., discontinuities in the gradient, as well as \emph{strong} discontinuities, i.e., discontinuities in the primary solution field, may be modeled. Initially, X-FEM was developed to resolve crack paths without any re-meshing during crack evolution~\cite{moes1999finite,belytschko1999elastic}. However, its application was extended to fluid mechanics~\cite{chessa2003extended,chessa2003enriched}, magnetic and coupled magneto-mechanical boundary value problems~\cite{spieler2013xfem,kastner2013higher}, poromechanics~\cite{prevost2016faults} and contact mechanics~\cite{liu2022interface}. In solid mechanics, X-FEM was used to model grain boundaries in polycrystals~\cite{simone2006generalized} (strong discontinuities) and material interfaces~\cite{sukumar_modeling_2001,moes_computational_2003} (weak discontinuities). 
The X-FEM is closely related to the generalized finite element method (GFEM)~\cite{strouboulis2000design,strouboulis2000generalized}. Both methods use standard FE shape functions to realize the partition of unity (PU)~\cite{babuvska1994special}. Initially, the GFEM employed global enrichments, whereas the X-FEM performs local enrichments on a subset of nodes. However, the characteristics of GFEM evolved over time, shifting towards local enrichments. Due to this development, the terms X-FEM and GFEM are often used interchangeably~\cite{fries_extendedgeneralized_2010,babuska_strongly_2017}. For a detailed history of the X-FEM and GFEM, we refer to section 1.1 of the review article by Fries and Belytschko~\cite{fries_extendedgeneralized_2010}. For simplicity, we will use the term X-FEM to refer to both methods for the remainder of this manuscript.

To model smooth material interfaces in X-FEM, enrichment functions that rely on the level set method~\cite{osher_fronts_1988} are suitable, as the level set method offers an implicit description of a smooth interface. As a result, applications to micromechanics require microstructure generation methods that provide a level-set representation of the primitives of the microstructure. Fortunately, there are respective methods available for spherical~\cite{williams2003random,torquato2010robust} and cylindrical inclusions~\cite{schneider2017sequential,schneider2022algorithm}, irregular shaped particles~\cite{garcia2009clustered,sonon2012unified,schneider2018modelling}, cellular materials~\cite{sonon2015advanced} and woven composites~\cite{sonon2013level}. A recently introduced approach encodes the level set information by a neural network in so-called Implicit Neural Representations (INRs) ~\cite{karki2025mechanics}. 

A suitable enrichment for modeling smooth material interfaces in X-FEM is the modified abs enrichment~\cite{moes_computational_2003}, which leverages the level set method and vanishes outside elements that comprise a material interface, thereby preventing parasitic terms in neighboring elements. The modified abs enrichment was shown to enable interface-conforming accuracy in X-FEM~\cite{moes_computational_2003,lian2013image}, among other enrichments~\cite{chessa2003extended,fries2008corrected}. Nevertheless, some formulations of the X-FEM suffer from ill-conditioned systems~\cite{fries_extendedgeneralized_2010,babuska_strongly_2017}.

To tackle the ill-conditioning, preconditioners based on Cholesky decompositions~\cite{bechet2005improved,menk2011robust}, domain decomposition strategies~\cite{menk2011robust,berger2012inexact,waisman2013adaptive} and multigrid methods~\cite{kergrene_stable_2016,lehrenfeld2017optimal} were developed. For two-dimensional thermal problems, Babu\v{s}ka et al.~\cite{babuska_stable_2012,babuska_strongly_2017} proved that, with a suitable enrichment and preconditioning strategy involving a diagonal preconditioner for the extended degrees of freedom, the condition number of the preconditioned X-FEM system matrix scales at the same rate as the underlying FEM system matrix with the mesh spacing $h$, namely, at the rate $\mathcal{O}(h^2)$. Moreover, they proved that for the so-called strongly stable X-FEM two mesh-independent bounds exist: one on the condition number of the preconditioned X-FEM system matrix and one on the condition number of the enriched block matrix~\cite{babuska_strongly_2017}. The group showed that the X-FEM with the modified absolute enrichment and a suitable preconditioning strategy leads to a strongly stable X-FEM~\cite{kergrene_stable_2016, babuska_strongly_2017} for two-dimensional thermal problems. The strongly stable X-FEM was extended to time-dependent problems, i.e., moving interfaces~\cite{zhu2020bdf}. However, the authors of this manuscript are not aware of any extension of the strongly-stable X-FEM to three-dimensional thermal problems or to linear elasticity for material interface problems. To improve the numerical efficiency of X-FEM, dedicated iterative solvers were developed. On the one hand, multigrid solvers~\cite{waisman2013adaptive,gong2024multi} and multigrid preconditioned conjugate gradient solvers~\cite{kergrene_stable_2016,lehrenfeld2017optimal} were tailored to X-FEM. On the other hand, FFT-based solvers were made available for X-FEM~\cite{gehrig2025x} in two-dimensional thermal problems based on the concept of strongly stable X-FEM~\cite{babuska_strongly_2017}.

\subsection{Contributions}
This work presents a novel X-FFT solver that solves three-dimensional mechanical problems with interface-conforming accuracy, overcoming a major limitation of classical FFT-based methods. 
Building on our recent work on X-FFT solvers for two-dimensional heat conductivity~\cite{gehrig2025x}, we extend the approach to three-dimensional mechanical problems. However, the recently introduced X-FFT solver heavily relies on the proof for the existence of a mesh-independent upper bound on the condition number in a strongly stable X-FEM, provided by Babu{\v{s}}ka et al.~\cite{babuska_strongly_2017}, which is limited to two-dimensional heat conductivity problems. 
To date, we are not aware of such a proof for three-dimensional mechanical homogenization problems. Moreover, mechanical problems often involve complex material properties and operators without minor symmetries, which potentially worsens the conditioning and increases the implementation effort compared to heat conductivity problems.
This work presents the modification of the recently introduced X-FFT solver~\cite{gehrig2025x} to address three-dimensional mechanical homogenization problems, discusses its implementation, and provides a comprehensive numerical analysis. Our primary research question is: 
\begin{center}
 \fbox{%
  \parbox{.9\textwidth}{\emph{Does the displacement-based X-FFT solver extend to FFT-based computational homogenization methods in 3D mechanics with the expected accuracy and efficiency?}}
 }	
\end{center}
Efficiency relies on the existence of a mesh-independent upper bound on the condition number of the preconditioned system matrix, ensuring that the system remains stable even when the mesh spacing or position changes. Our studies indicate that such a mesh-independent bound holds for 3D mechanical homogenization with smooth interfaces. Moreover, we observe interface-conforming accuracy in the local and effective error convergence rates. Currently, the X-FFT solver applies to linear elastic, non-porous, multiphase materials with periodic boundary conditions and a single interface per element. 

\newpage

\section{The X-FFT solver}\label{sec:discretizationSchemes}

\subsection{Discretization by extended finite elements}

We assume a three-dimensional cuboid cell
\begin{equation}
    Y = [0,\ell_1] \times [0,\ell_2] \times [0,\ell_3]
\end{equation}
with periodic boundary conditions, where a stress operator
\begin{equation}
\fsigma:Y\times\Sym{3}\rightarrow\Sym{3}
\end{equation}
is defined that maps for every point $\fx \in Y$ a strain tensor $\feps\in\Sym{3}$ to a stress tensor $\fsigma\left(\fx, \feps\right)$.
We aim to solve the weak form of the (quasi-) static equilibrium without microscopic body forces, so that the equation
\begin{equation}\label{eq:weakFormConti}
   \int_Y  \fsymgrad  \fv  : \fsigma\left(\fx, \feps(\fx)\right) dV = 0
\end{equation}
holds for all test fields $\fv$ in the first-order Sobolev space $H_{\#}^{1}(Y;\R^3)$ of periodic fields with vanishing mean. The strain field 
\begin{equation}\label{eq:definitionStrain}
    \feps(\fx) = \bfeps+\fsymgrad \fu(\fx)
\end{equation}
consists of the average strain $\bfeps\in\Sym{3}$ and the symmetric gradient of the displacement fluctuation $\fu \in H_{\#}^{1}(Y;\R^3)$.
In the context of linear elasticity, the stress operator
\begin{equation}
    \fsigma(\fx,\feps) = \ffC(\fx) : \feps(\fx)
\end{equation}
is defined by a linear mapping via the local stiffness tensor $\ffC(\fx)\in\mathrm{Lin}\left(\Sym{3}\right)$, which has minor and major symmetries. Moreover, we assume that the inequalities
\begin{equation}\label{eq: boundsStiffness}
    C_-\norm{\feps}^2\leq \feps:\ffC(\fx):\feps \leq C_+\norm{\feps}^2 \quad\text{for all } \feps\in\Sym{3}
\end{equation}
hold at all points $\fx \in Y$ for some positive constants $C_-,C_+\in\R$. The bounds encode that the local stiffness tensor field is uniformly positive definite on the strains and uniformly bounded.

\begin{figure}[htb]
    \centering
    \begin{subfigure}[c]{0.4\textwidth}
        \includegraphics[width=\textwidth]{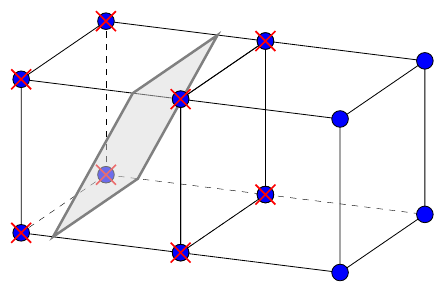}
    \end{subfigure}    
    \begin{subfigure}[c]{0.2\textwidth}
        \includegraphics[width=\textwidth]{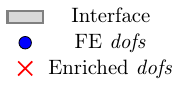}
    \end{subfigure}    
    \caption{Visualization of the location of FE \emph{dofs} and enriched \emph{dofs}.}
    \label{fig:dofVisualization}
\end{figure}

\begin{figure}
    \centering
    \begin{subfigure}[b]{0.27\textwidth}
      \includegraphics[width=\textwidth]{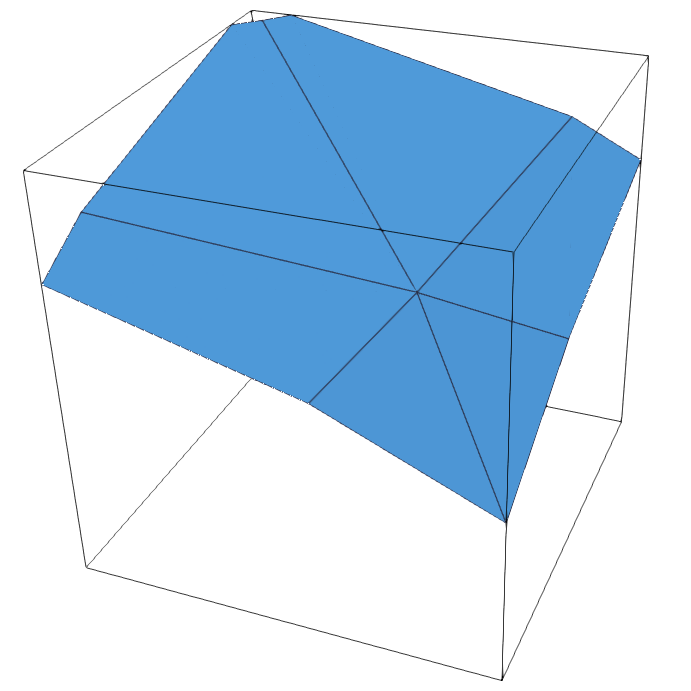}
      \subcaption{Interface}
    \label{fig:Interface}
    \end{subfigure}
     \begin{subfigure}[b]{0.35\textwidth}
      \includegraphics[width=0.8\textwidth]{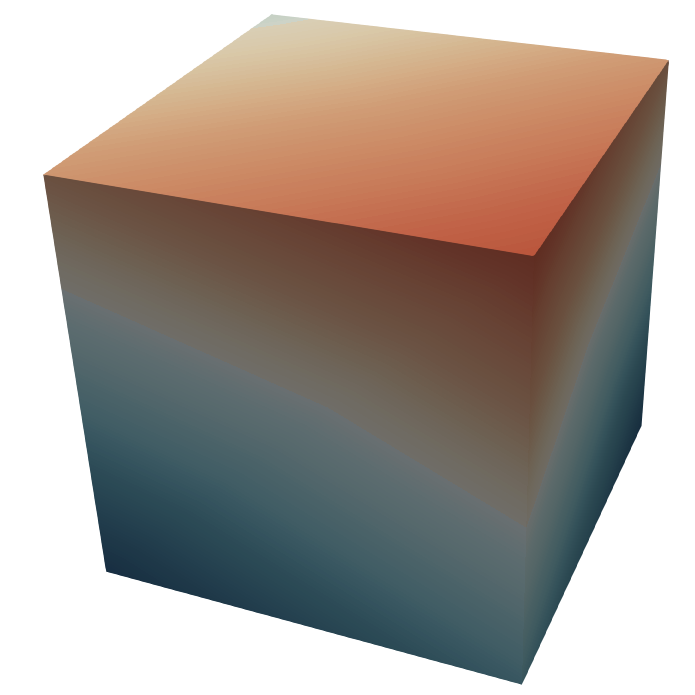}
      \includegraphics[width=0.14\textwidth]{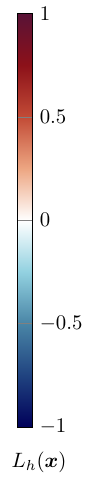}
      \subcaption{Level set representation}
    \label{fig:levelSet}
    \end{subfigure}
    \begin{subfigure}[b]{0.35\textwidth}
      \includegraphics[width=0.8\textwidth]{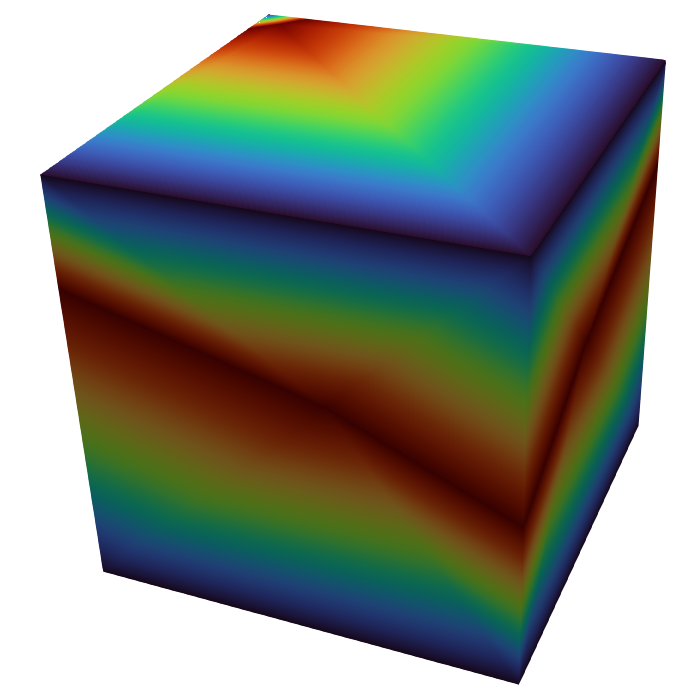}
      \includegraphics[width=0.12\textwidth]{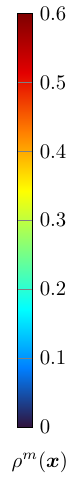}
      \subcaption{Modified abs enrichment}
    \label{fig:modified}
    \end{subfigure}
    \caption{Interface with level set representation and modified abs enrichment function of a voxel that is discretized by six $\ffP_1$ elements.}
\end{figure}

To discretize the weak form of the (quasi-) static equilibrium without microscopic body forces~\eqref{eq:weakFormConti} by extended finite elements on a regular grid, we define a suitable X-FEM approximation space $V_h$. With this space at hand, we seek a displacement fluctuation field $\fu_h\in V_h$ that satisfies the equation
\begin{equation} \label{eq:discreteFormConti}
   \int_Y  \fsymgrad  \fv_h  : \ffC:\left(\bfeps+\fsymgrad \fu_h\right) dV = 0
\end{equation}
for all test fields $\fv_h\in V_h$. 
To develop a suitable X-FEM approximation space, we adapt the X-FEM approximation space for two-dimensional thermal problems~\cite[eq.(2.16)]{gehrig2025x} to three-dimensional mechanical homogenization problems. The resulting X-FEM approximation space takes the form
\begin{equation}\label{eq:X-FEM_space}
	V_h = \left\{
		\sum_{i\in I} N_\mathrm{FE}^i \, \fu_\mathrm{FE}^i+\sum_{j\in J} N_\mathrm{X}^j \,  \fu_\mathrm{X}^j \, \middle| \,  i\in I, \quad j \in J, \quad \fu_\mathrm{FE}^i \in \R^3, \quad \fu_\mathrm{X}^j \in \R^3,  \quad\sum_{i \in I} \fu_\mathrm{FE}^i = \fzero
	\right\}
\end{equation}
with the nodes $i$, which are in the set of standard FE nodes $I$, and the enriched nodes $j$, which are in the set of enriched nodes $J$. The set of enriched nodes $J$ is a subset of the set of standard FE nodes $I$. Typically, the latter set has a much higher cardinality, because only the nodes of elements that contain an interface are enriched, as visualized in Fig.\ref{fig:dofVisualization}. The X-FEM approximation space for three-dimensional mechanical problems~\eqref{eq:X-FEM_space} consists of the standard nodal FE shape functions $N_\mathrm{FE}^i$, the standard nodal displacement fluctuations $\fu_\mathrm{FE}^i$, the enriched nodal FE shape functions $N_\mathrm{X}^j$, and the enriched nodal displacement fluctuations $\fu_\mathrm{X}^j$. The standard FE displacement fluctuations are forced to be mean free, so that~eq.\eqref{eq:discreteFormConti} admits a unique solution. The enriched shape functions take the form
\begin{equation}\label{eq:enrichedN}
    N_\mathrm{X}^j(\fx) = N_\mathrm{FE}^j(\fx) \, \rho(\fx), \quad \fx \in Y,
\end{equation}
where the scalar-valued function $\rho$ stands for the enrichment function. For matrix-inclusion composites, we seek to approximate the kink in the displacement fluctuation at the interface by the enrichment function $\rho$. To approximate these so-called \emph{weak discontinuities}, enrichment functions based on the level set description~\cite{osher_fronts_1988} of the interface may be used~\cite{sukumar_modeling_2001,moes_computational_2003}. A level set function $L:Y\rightarrow\R$ provides an implicit description of a composite comprising two phases $Y_\pm$ with interface $\mathcal{I}$ based on the definitions
\begin{align}
    \mathcal{I}=\{\fx \in Y\, | \, L(\fx) = 0\},\\
    Y_+=\{\fx \in Y\, | \, L(\fx) > 0\},\\
    Y_-=\{\fx \in Y\, | \, L(\fx) < 0\},
\end{align}
where the level set function is assumed to be smooth and at least Lipschitz continuous~\cite{osher_fronts_1988}.
For numerical reasons, it is convenient to normalize the level set function, such that its values encode the signed distance to the closest interface~\cite{lendvai2024assumed}.  
If the nodal level set values $L^i$ are available, i.e., via an analytical description of the interface, the continuous level set function may be approximated by linear interpolation via the FE shape functions in the form
\begin{equation}\label{eq:levelset}
	L_h(\fx) =  \sum_{i\in I} N_\mathrm{FE}^i(\fx) L^i.
\end{equation}
For an example of an interface with respective level set function, we refer to Fig.~\ref{fig:Interface} and Fig.~\ref{fig:levelSet}.
Sukumar et al.~\cite{sukumar_modeling_2001} pioneered using level set based enrichments for interface problems by introducing the \emph{abs enrichment}
\begin{equation}\label{eq:absEnrichment}
	\rho^\mathrm{a}(\fx) =  \left\vert\sum_{i\in I} N_\mathrm{FE}^i(\fx) L^i\right\vert.
\end{equation}
However, the~\emph{abs enrichment}~\eqref{eq:absEnrichment} leads to poor conditioning due to linear dependencies between the FE and the enriched shape functions.
As a remedy, Mo\"{e}s et al.~\cite{moes_computational_2003} introduced the~\emph{modified abs enrichment}, which takes the form
\begin{equation}\label{eq:modifiedabsEnrichment}
	\rho^\mathrm{m}(\fx) =  \sum_{i\in I} N_\mathrm{FE}^i(\fx) |L^i| - \left\vert\sum_{i\in I} N_\mathrm{FE}^i(\fx) L^i\right\vert, \quad \fx \in Y.
\end{equation} 
The~\emph{modified abs enrichment}~\eqref{eq:modifiedabsEnrichment} is constructed by subtracting the \emph{abs enrichment}~\eqref{eq:absEnrichment} from its linear interpolant. By construction, the~\emph{modified abs enrichment}~\eqref{eq:modifiedabsEnrichment} vanishes in all elements that do not contain an interface, so that linear dependencies between the FE and enriched shape functions are prevented. For a visualization of the~\emph{modified abs enrichment} of the interface shown in Fig.~\ref{fig:Interface}, we refer to Fig.~\ref{fig:modified}.

\subsection{Preconditioning with internal scaling}\label{sec:preconditioning}
In the previous paragraph, we discretized the weak form of the (quasi-) static equilibrium without microscopic body forces by extended finite elements on a regular grid for linear elasticity. To decrease the effort required to solve the resulting system~\eqref{eq:discreteFormConti}, we introduce a preconditioning strategy based on concepts that were recently developed for two-dimensional thermal problems~\cite{babuska_strongly_2017,gehrig2025x,gehrig2025improved}.

Traditionally, the condition number of X-FEM systems scales with the rate $\mathcal{O}(h^{-4})$ in the mesh spacing $h$~\cite{babuska_stable_2012}. Babu{\v{s}}ka et al.~\cite{babuska_stable_2012,babuska_strongly_2017} proved that for some conditions on the enrichment space and for two-dimensional thermal problems, the condition number of the X-FEM system reduces to the rate of an underlying voxel-FEM system, namely the rate $\mathcal{O}(h^{-2})$. Moreover, in their so-called strongly stable X-FEM~\cite{babuska_strongly_2017}, applying a diagonal preconditioner to the enriched degrees of freedom guarantees mesh-independent upper bounds on the condition numbers of the preconditioned X-FEM system and the preconditioned system of the  enriched degrees of freedom. Based on the inequalities that Babu{\v{s}}ka et al.~\cite{babuska_strongly_2017} use to derive these mesh-independent upper bounds, the authors of this manuscript developed an X-FFT preconditioner~\cite[Sec.3]{gehrig2025x} for two-dimensional thermal problems that leads to a mesh-independent upper bound on the condition number of the preconditioned X-FFT system.
In a subsequent publication, the authors~\cite{gehrig2025improved} introduced an internal scaling of the enriched shape functions to prevent numerical instabilities that could arise if the enriched shape functions and the standard FE shape functions have different order of magnitude.
To construct a preconditioner for the work at hand, we transfer the strategies that were previously developed for 2D thermal homogenization problems~\cite{gehrig2025x,gehrig2025improved} to three-dimensional linear elastic unit-cell problems. However, we note that a proof extending Babu{\v{s}}ka et al.~\cite[§5]{babuska_strongly_2017} to three-dimensional mechanics is currently not known to the authors. Rather, we report computational experiments, see section~\ref{sec:computationalInvestigations}.

Building on the previously developed internal scaling strategy~\cite{gehrig2025improved}, we introduce the scaled enriched shape function 
\begin{equation}\label{eq:scaledEnrichedFunction}
	\tilde{N}_\mathrm{X}^j =  \left(\uul{D^0}\right)^{-\frac{1}{2}}_{jj} N_\mathrm{X}^j
\end{equation}
and the scaled enriched displacement fluctuation 
\begin{equation}\label{eq:scaledEnrichedU}
	\tilde{\fu}_\mathrm{X}^j =  \left(\uul{D^0}\right)^{\frac{1}{2}}_{jj} \fu_\mathrm{X}^j
\end{equation}
with the scaling factors
\begin{equation}\label{eq:XFFT_diagonal}
	\left(\uul{D^0}\right)_{jj} = \int_Y \norm{\nabla^s N_\mathrm{X}^j}^2\, dV
\end{equation}
that are based on the formerly introduced diagonal preconditioner~\cite[eq.(3.10)]{gehrig2025x}. We use the tilde symbol to denote re-scaled variables in the manuscript at hand. The scaled enriched shape functions are designed to improve the numerical stability by preventing the enriched shape functions from having a different order of magnitude than the standard FE functions.
More precisely, for the standard FE functions at the node $k\in I$, by definition, the interpolation condition 
\begin{equation}
	N_\mathrm{FE}^k(x_p) = \delta_{kp}
\end{equation}
holds at the node $p\in I$. To obtain values in the same order of magnitude for the scaled enriched shape functions, we require the condition
\begin{equation}\label{eq:L2X}
	\norm{\nabla^s \tilde{N}_\mathrm{X}^j}_{L^2} = 1,
\end{equation}
to hold at all enriched nodes $j\in J$, which is fulfilled by the scaled enriched shape function~\eqref{eq:scaledEnrichedFunction} defined above.
We note that we rely on the symmetrized gradient $\nabla^s$ in the definition of the scaling factors~\eqref{eq:XFFT_diagonal} for three-dimensional linear elasticity, a choice that is motivated by the kinematic compatibility condition.

The adapted X-FFT approximation space~\eqref{eq:X-FEM_space} takes the form
\begin{equation}\label{eq:X-FEM_scaledspace}
	\tilde{V}_h = \left\{
		\sum_{i\in I} N_\mathrm{FE}^i \, \fu_\mathrm{FE}^i+\sum_{j\in J} \tilde{N}_\mathrm{X}^j \,  \tilde{\fu}_\mathrm{X}^j \, \middle| \,  i\in I, \quad j \in J, \quad \fu_\mathrm{FE}^i \in \R^3, \quad \tilde{\fu}_\mathrm{X}^j \in \R^3,  \quad\sum_{i \in I} \fu_\mathrm{FE}^i = \fzero
	\right\}.
\end{equation}
After selecting a basis of the adapted X-FEM approximation space~\eqref{eq:X-FEM_scaledspace}, we may rewrite the discretized weak form of the (quasi-) static equilibrium~\eqref{eq:discreteFormConti} as the linear system
\begin{equation}\label{eq:linearSystem}
	\uul{\tilde{A}}\,\ul{\tilde{u}} = \ul{\tilde{b}}
\end{equation}
with the scaled system matrix 
\begin{equation}
	\uul{\tilde{A}} = \begin{bmatrix}
		\uul{A_{11}} & \uul{\tilde{A}_{12}}\\
		\uul{\tilde{A}_{21}} & \uul{\tilde{A}_{22}}
	\end{bmatrix} \in \R^{3(n_\mathrm{FE} + n_\mathrm{X})\times 3(n_\mathrm{FE} + n_\mathrm{X})},
\end{equation}
the scaled solution vector
\begin{equation}
	\ul{\tilde{u}} = \begin{bmatrix}
		\ul{u_{1}}\\
		\ul{\tilde{u}_{2}}
	\end{bmatrix} \in \R^{3(n_\mathrm{FE} + n_\mathrm{X})}
\end{equation}
and the scaled right-hand side vector
\begin{equation}
	\ul{\tilde{b}} = \begin{bmatrix}
		\ul{b_{1}}\\
		\ul{\tilde{b}_{2}}
	\end{bmatrix} \in \R^{3(n_\mathrm{FE} + n_\mathrm{X})}.
\end{equation}
The number $n_\mathrm{FE}$ denotes the number of FE nodes and the number $n_\mathrm{X}$ accounts for the number of enriched nodes, i.e., the nodes of elements that contain an interface.

The upper left block matrix 
\begin{equation}\label{eq:block11}
	\left(\uul{A_{11}}\right)_{ij} = \int_Y \nabla^s N_\mathrm{FE}^i : \ffC : \nabla^s N_\mathrm{FE}^j \, dV, \quad i,j=1,2,\ldots,n_\mathrm{FE},
\end{equation}
and the upper vector $\ul{u_{1}}\in \R^{3n_\mathrm{FE}}$ comprise solely classical FE degrees of freedom and are independent of the internal scaling. The lower right block matrix 
\begin{equation}\label{eq:block22}
	\left(\uul{\tilde{A}_{22}}\right)_{ij} = \int_Y \nabla^s \tilde{N}_\mathrm{X}^i : \ffC :\nabla^s \tilde{N}_\mathrm{X}^j \, dV, \quad i,j=1,2,\ldots,n_\mathrm{X}
\end{equation}
and the lower vector $\ul{\tilde{u}_{2}}\in \R^{3n_\mathrm{X}}$ depend solely on the enriched degrees of freedom, whereas the off-diagonal matrices 
\begin{align}
	\left(\uul{\tilde{A}_{12}}\right)_{ij} &= \int_Y \nabla^s N_\mathrm{FE}^i : \ffC: \nabla^s \tilde{N}_\mathrm{X}^j \, dV, \quad i=1,2,\ldots,n_\mathrm{FE}, \quad j=1,2,\ldots,n_\mathrm{X}, \label{eq:scaled_block12}\\
	\left(\uul{\tilde{A}_{21}}\right)_{ij} &= \int_Y \nabla^s \tilde{N}_\mathrm{X}^i  :\ffC: \nabla^s N_\mathrm{FE}^j \, dV, \quad i=1,2,\ldots,n_\mathrm{X}, \quad j=1,2,\ldots,n_\mathrm{FE}, \label{eq:scaled_block21}
\end{align}
quantify the interaction between the classical FE and the enriched degrees of freedom.
To precondition the linear system~\eqref{eq:linearSystem}, we rely on the X-FFT preconditioner~\cite{gehrig2025improved} that was originally developed for two-dimensional thermal problems and takes the form
\begin{equation}\label{eq:scaledPreconditioner}
\uul{\tilde{P}} = \begin{bmatrix}
		\uul{A_{11}^0}\\&\uul{I}
	\end{bmatrix} \in \R^{3(n_\mathrm{FE}+n_\mathrm{X})\times 3(n_\mathrm{FE}+n_\mathrm{X})}
\end{equation}
for three-dimensional mechanical problems. The X-FFT solver~\eqref{eq:scaledPreconditioner} consists of the identity matrix $\uul{I}\in \R^{3n_\mathrm{X}\times 3n_\mathrm{X}}$ and the constant coefficient preconditioner 
\begin{equation}\label{eq:constantCoefficientPreconditioner}
	\left(\uul{A_{11}^0}\right)_{ij} = \int_Y \nabla^s N_\mathrm{FE}^i : \nabla^s N_\mathrm{FE}^j \, dV, \quad i,j=1,2,\ldots,n_\mathrm{FE},
\end{equation}
which uses the symmetrized gradient $ \nabla^s$ in consistency with the scaling factors~\eqref{eq:XFFT_diagonal}.
Applying the block preconditioner~\eqref{eq:scaledPreconditioner}, the linear system~\eqref{eq:linearSystem} takes the form
\begin{equation}\label{eq:preconditionedXFFT}
	\uul{\tilde{P}}^{-1}\uul{\tilde{A}}\,\ul{\tilde{u}} = \uul{\tilde{P}}^{-1}\ul{\tilde{b}}.
\end{equation}
For voxel-based FEs, the constant coefficient preconditioner~\eqref{eq:constantCoefficientPreconditioner} may be inverted efficiently via the fast Fourier transform. Details on the implementation of Green's operator are provided in the section below. 

\section{Implementation}\label{sec:implementation}
\begin{figure}[h]
    \centering 
    \includegraphics[width=0.16\textwidth]{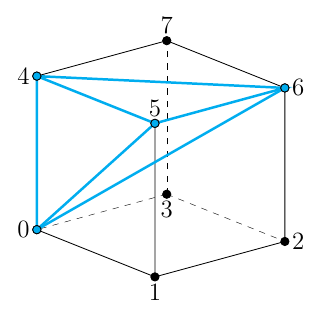}
    \includegraphics[width=0.16\textwidth]{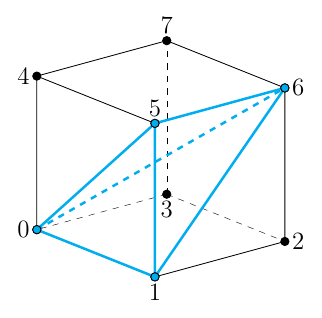}
    \includegraphics[width=0.16\textwidth]{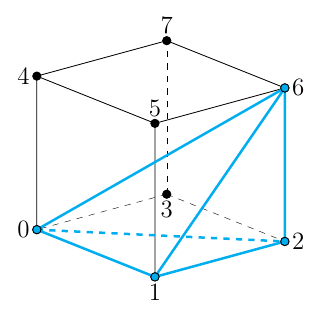}
    \includegraphics[width=0.16\textwidth]{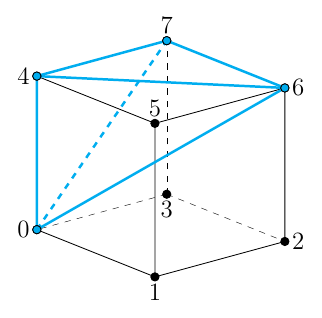}
    \includegraphics[width=0.16\textwidth]{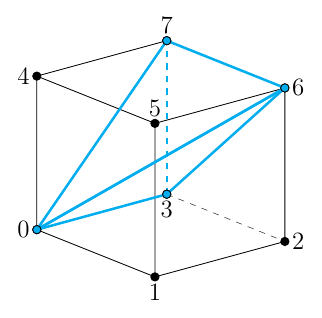}
    \includegraphics[width=0.16\textwidth]{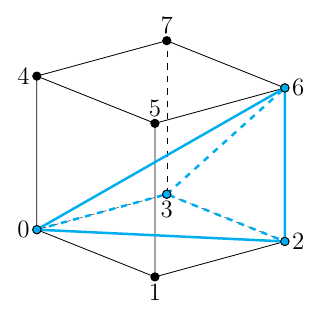}
\caption{Division of the voxel into six $\ffP_1$ elements~\cite{sadarjoen1998particle}.}
\label{fig:P16}
\end{figure}

Our X-FFT solver applies an iterative scheme to solve the linear system~\eqref{eq:linearSystem} utilizing the preconditioner~\eqref{eq:scaledPreconditioner}. For efficient implementation it is convenient to treat the global linear system 
\begin{equation}
    \uul{\tilde{A}}\,\ul{\tilde{u}}-\ul{\tilde{b}} = \ul{0}
\end{equation}
with a divide-and-conquer strategy on the element level, such that the form
\begin{equation}\label{eq:linearSystem_element}
    \sum_{e=1}^{n_e} \T{\uul{\Lambda_e}} \left(\uul{\tilde{A}_e}\,\ul{\tilde{u}_e}-\ul{\tilde{b}_e}\right) = \ul{0}
\end{equation}
results, where the matrix $\uul{\Lambda_e}\in\R^{3(n_\mathrm{FE}+n_\mathrm{X})\times4n_\mathrm{n}}$ associates the global degrees of freedom with the element nodal degrees of freedom.
We assume that each voxel consists of six $\ffP_1$ elements as depicted in Fig.~\ref{fig:P16}. 
Via the utilized X-FFT approximation space~\eqref{eq:X-FEM_scaledspace} nodes that are not located on elements containing an interface, i.e., nodes $k\in I\setminus J$, have solely three standard FE degrees of freedom per node ($n_\mathrm{n}=3$). In contrast, nodes that belong to enriched elements, i.e., nodes $k\in J$, have six degrees of freedom per node ($n_\mathrm{n}=6$).
The relations
\begin{align}
\ul{\tilde{u}_e} &= \uul{\Lambda_e}\ul{\tilde{u}},\\
\uul{\tilde{A}_e} &=  \int_{Y_e} \T{\uul{\nabla^s \tilde{N}_e}(\fx)} \uul{C}(\fx)\uul{\nabla^s \tilde{N}_e}(\fx) \,dV, \label{eq:Kapp}\\
\ul{\tilde{b}_e} &=  \int_{Y_e} \T{\uul{\nabla^s \tilde{N}_e}(\fx)} \uul{C}(\fx)\,\ul{\bar{\eps}} \, dV, \label{eq:b}
\end{align}
hold with the symmetrized gradient of the scaled nodal shape functions $\uul{\nabla^s \tilde{N}_e}(\fx)\in\R^{6\times4n_\mathrm{n}}$ and the local stiffness matrix $\uul{C}$. Note that our implementation relies on the Voigt-Mandel notation for all matrices. At a point $\fx \in Y$, the gradient of the nodal shape function may be computed using the scalar gradient of the nodal shape function via the relation
\begin{equation}\label{eq:shapeFunctionGradient}
    \uul{\nabla^s \tilde{N}_e}(\fx) = \uul{M}\,  \uul{\nabla \tilde{N}_e}^\mathrm{scalar}(\fx) \otimes \uul{I}_{3\times3}
\end{equation}
with the identity matrix $\uul{I}_{3\times3}$ and the dyadic product $\otimes$. The matrix $\uul{M}$ extracts the symmetrized (strain) part in Voigt-Mandel notation from a full deformation gradient.
For a reference tetrahedron with nodes $ABCD$, the scalar gradient of the nodal shape function takes the form
\begin{equation}
    \ul{\tilde{N}_e}^\mathrm{scalar}(\fx) = \begin{bmatrix}N^A_\mathrm{FE}(\fx) &N^B_\mathrm{FE}(\fx)&N^C_\mathrm{FE}(\fx)& N^D_\mathrm{FE}(\fx)& \tilde{N}^A_\mathrm{X}(\fx)& \tilde{N}^B_\mathrm{X}(\fx)& \tilde{N}^C_\mathrm{X}(\fx)&\tilde{N}^D_\mathrm{X}(\fx)
    \end{bmatrix}
\end{equation}
for enriched elements and reduces to the first four entries for non-enriched elements. 
Using the modified abs enrichment~\eqref{eq:modifiedabsEnrichment}, and the internal scaling factors~\eqref{eq:XFFT_diagonal}, the representation
\begin{equation}
\tilde{N}^j_\mathrm{X}(\fx)= \left(\uul{D^0}\right)^{-\frac{1}{2}}_{jj} \rho^\mathrm{m}(\fx) N^j_\mathrm{FE}(\fx)
\end{equation}
follows for the scaled enriched shape functions. For linear FE shape functions $N^j_\mathrm{FE}$, the modified abs enrichment $\rho^\mathrm{m}$~\eqref{eq:modifiedabsEnrichment} is piecewise linear, such that the enriched shape function $\tilde{N}^j_\mathrm{X}$ is quadratic in space. For enriched elements, the shape function gradient~\eqref{eq:shapeFunctionGradient} is thus linear in space.

\begin{figure}[tb]
    \centering 
    \includegraphics[width=0.16\textwidth]{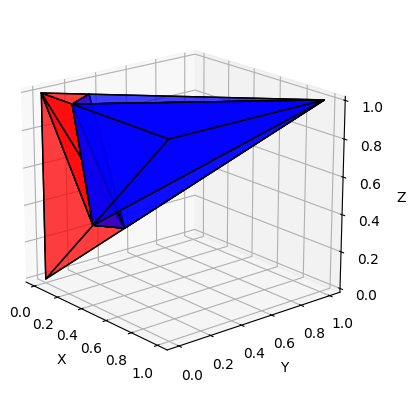}
    \includegraphics[width=0.16\textwidth]{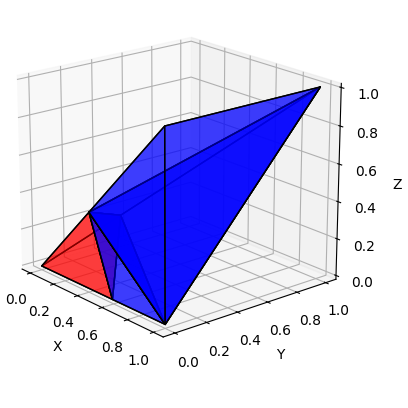}
    \includegraphics[width=0.16\textwidth]{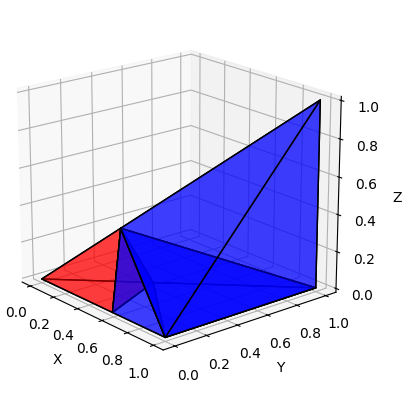}
    \includegraphics[width=0.16\textwidth]{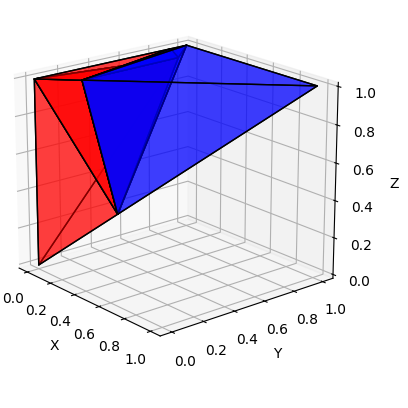}
    \includegraphics[width=0.16\textwidth]{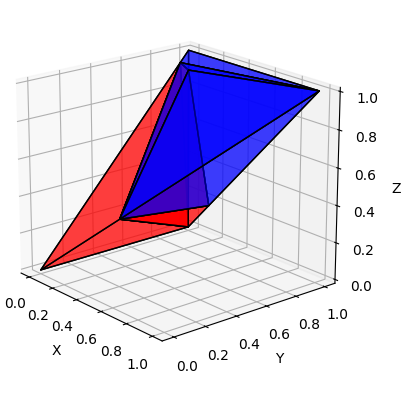}
    \includegraphics[width=0.16\textwidth]{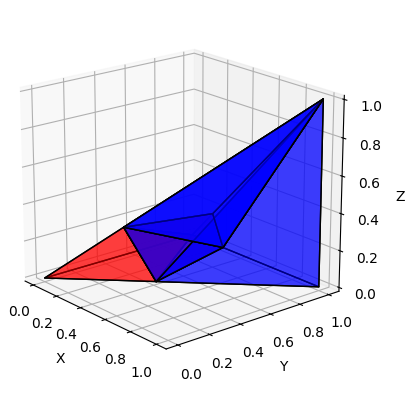}
\caption{Division of the $\ffP_1$ elements into subtetrahedra for integration.}
\label{fig:subdivision}
\end{figure}

We aim to compute the integrals of the element stiffness matrix~\eqref{eq:Kapp} and the element vector~\eqref{eq:b} via quadrature. The modified abs enrichment leads to a jump in the shape function gradient at the interface so that a discontinuous integrand results. However, when dividing the $\ffP_1$ into subtetrahedra~\cite{moes1999finite,sukumar_modeling_2001} that resolve the linearized interface, exact integration via quadrature is possible. Schweiger and Arridge~\cite[Fig. 2]{schweiger2017basis} provide an overview of the different subdivisions resulting from a tetrahedron cut by a single plane. For a specific example of such a subdivision, we refer to Fig.~\ref{fig:subdivision}, where the colors blue and red denote the two different phases. To reduce superfluous subdivisions, we assume a node to be intersected by the interface, if the absolute level set value at the node is below $10^{-6}$. For each subtetrahedron, we compute the quadrature points and weights according to Shunn and Ham~\cite[Appendix F]{shunn2012symmetric}. Each voxel consists of six $\ffP_1$ elements and the maximum number of subtetrahedra per $\ffP_1$ element is six~\cite[Fig. 2]{schweiger2017basis}. As the integrand of the element stiffness matrix~\eqref{eq:Kapp} is quadratic, four quadrature points per subtetrahedron are required~\cite[Appendix F]{shunn2012symmetric}. Therefore, a maximum of 144 quadrature points per voxel are needed. With the quadrature points $\fq_\alpha$ and the weights $w_\alpha$ at hand, the element stiffness matrix~\eqref{eq:Kapp} and the element vector~\eqref{eq:b} take the form
\begin{align}
\uul{\tilde{A}_e}^\mathrm{appr} &=  \sum_{\alpha=1}^{q_e} w_\alpha \T{\uul{\nabla \tilde{N}_e}(\fq_\alpha)} \uul{C}(\fq_\alpha)\uul{\nabla \tilde{N}_e}(\fq_\alpha), \label{eq:Kapp_q}\\
\ul{\tilde{b}_e}^\mathrm{appr} &=  \sum_{\alpha=1}^{q_e} w_\alpha \T{\uul{\nabla \tilde{N}_e}(\fq_\alpha)} \uul{C}(\fq_\alpha)\, \ul{\bar{\eps}}, \label{eq:b_q}
\end{align}
where the number $q_e$ denotes the number of required quadrature points per element.
The average stress per element may be computed by the equation
\begin{equation}
   \langle \uul{\sigma}(\ul{\tilde{u}_e})\rangle_{Y_e}  = \uul{\tilde{S}_e}^\mathrm{appr} \ul{\tilde{u}_e}
\end{equation}
with the matrix 
\begin{equation}\label{eq:Sapp_q}
\uul{\tilde{S}_e}^\mathrm{appr} = \sum_{i=\alpha}^{q_e} w_i \uul{C}(\fq_\alpha)\uul{\nabla \tilde{N}_e}(\fq_\alpha) 
\end{equation}
that results from the quadrature scheme used.
With these quantities at hand, we may compute the left-hand side of the linear system~\eqref{eq:linearSystem_element} at each iteration by the residual vector
\begin{equation}\label{eq:resVector}
    \ul{r}\left(\ul{\tilde{u}}\right) =  \sum_{e=1}^{n_e} \T{\uul{\Lambda_e}}  \left(\uul{\tilde{A}_e}^\mathrm{appr}\,\ul{\tilde{u}_e}-\ul{\tilde{b}_e}^\mathrm{appr}\right). 
\end{equation}
Following Schneider et al.~\cite{schneider2022voxel}, our implementation relies on the stopping criterion 
\begin{equation}
    \mathrm{res}_k \stackrel{!}{\leq} \mathrm{tol} \, \norm{\langle \uul{\sigma}(\ul{\tilde{u}}^k)\rangle_Y}
     \label{eq:stoppingCriterion}
\end{equation}
with the residual
\begin{equation}
    \text{res}_k = \sqrt{\T{\ul{r}(\ul{\tilde{u}}^k)} \, \left(\uul{\tilde{P}}\right)^{-1} \, \ul{r}(\ul{\tilde{u}}^k)},
    \label{eq:convergence}
   \end{equation}
the desired tolerance $\mathrm{tol}$, and the average stress
\begin{equation}
    \langle \uul{\sigma}(\ul{\tilde{u}})\rangle_Y  =  \sum_{e=1}^{n_e} \langle \uul{\sigma}(\ul{\tilde{u}_e})\rangle_{Y_e}. 
\end{equation}
The residual~\eqref{eq:convergence} is computed using the scaled X-FFT preconditioner~\eqref{eq:scaledPreconditioner} which is a constant block diagonal matrix that consists of the constant coefficient preconditioner~\eqref{eq:constantCoefficientPreconditioner} and the identity matrix. In practice, we use well-established strategies from the literature~\cite{schneider2017fft,leuschner2018fourier,ladecky_optimal_2023} to precompute and cache Green's operator, which is the inverse of the constant coefficient preconditioner~\eqref{eq:constantCoefficientPreconditioner}. 
More precisely, we follow Leuschner and Fritzen~\cite[eqs. (47),(48)]{leuschner2018fourier} and Schneider~\cite[eq. (57)]{schneider2022voxel}, and represent the action of Green's operator for each Fourier frequency $\fxi\in\ffZ^3$ in Fourier space by applying the matrix $\uul{\widehat{A_{11}^0}}^{-1}(\fxi)$, which is the inverse of a $3\times3$ Hermitian matrix at non-zero frequency. This representation is valid because a constant coefficient operator on a regular periodic grid corresponds to matrix-valued Fourier multipliers in Fourier space~\cite{rudin1962fourier}.
However, in contrast to the referred articles~\cite{leuschner2018fourier,schneider2022voxel}, in the manuscript at hand, there is a strict distinction between voxels and elements: Each voxel comprises six $\ffP_1$ elements. Up to this point of the manuscript, the X-FFT solver was developed solely on element level. As the description via voxels is natural for the FFT, we augment our description by an additional layer of abstraction. More precisely, we introduce in addition to the global element counter $e\in\{1,\dots,n_e\}$, a voxel counter $v\in\{1,\dots,n_e/6\}$, and a local element counter $\ell\in\{1,\dots,6\}$, such that the relation
\begin{equation}
    e = 6v + \ell
\end{equation}
holds. 
Based on the description via voxels, we assemble the matrix $\uul{A_{11,\mathrm{v}}^0}\in\R^{24\times24}$ for a reference voxel using the equation
\begin{equation}\label{eq:A0Voxel}
\uul{A_{11,\mathrm{v}}^0} = \frac{1}{6} \sum_{\ell=1}^{6} \uul{\Lambda^\mathrm{v}_\ell}\, \T{\uul{\fsymgrad N_{\mathrm{FE},\ell}}}(\fq_\ell)\, \uul{\fsymgrad N_{\mathrm{FE},\ell}}(\fq_\ell),
\end{equation}
where each matrix $\uul{\Lambda^\mathrm{v}_\ell}\in\R^{24\times 12}$ associates the \emph{dofs} of the voxel with the \emph{dofs} of one of the six $\ffP_1$ elements within that voxel. Since the symmetrized gradient of the FE shape functions is constant, only one quadrature point per $\ffP_1$ element is required to compute the homogeneous FE stiffness matrix of a voxel via quadrature, as shown in eq.~\eqref{eq:A0Voxel}.
We compute the matrix representing the action of Green's operator by the formula
\begin{equation}\label{eq:GreenHermitian}
    \uul{\widehat{A_{11}^0}}^{-1}(\fxi) = \begin{cases}
    \uul{0}, \quad \text{if} \quad \fxi=\fzero,\\
    \left(\left(\uul{Z}(\fxi)\right)^{H} \uul{A_{11,\mathrm{v}}^0}\,\uul{Z}(\fxi)\right)^{-1},\quad \text{otherwise},
    \end{cases}
\end{equation}
where $H$ denotes Hermitian conjugation, consisting of complex conjugation and transposition.
The matrix $\uul{Z}\in\R^{24\times3}$ encodes the relative position of the FE nodes within a voxel via the relation
\begin{equation}
    \left(\uul{Z}\right)(\fxi) = \begin{bmatrix} 1 \\ 1 \\ 1 \end{bmatrix}\otimes\ul{c}(\fxi)\,\begin{bmatrix} 1 & 1 & 1 \end{bmatrix},
\end{equation}
with the dyadic product $\otimes$, where the vector $\ul{c}\in\R^{8}$ takes the form
\begin{equation}
   \left(\ul{c}\right)_{j}(\fxi) =\exp\left(2\pi i h\, \sum_{k=1}^{3} Y_{jk} \xi_k\right)\quad\text{for}\quad j=1,\ldots,8,
\end{equation}
and the matrix
\begin{equation}
   \uul{Y}= \T{\begin{bmatrix}
0 & 1 & 1 & 0 & 0 & 1 & 1 & 0\\
0 & 0 & 1 & 1 & 0 & 0 & 1 & 1\\
0 & 0 & 0 & 0 & 1 & 1 & 1 & 1
\end{bmatrix}}
\end{equation}
encodes the node numbering of the FE nodes in the voxel.
For computational speed up, we precompute and cache the (pseudo-) inverse of the Hermitian matrix in eq.~\eqref{eq:GreenHermitian}, which requires storing nine scalar values per frequency, as discussed in section 3.1 of Schneider~\cite{schneider2022voxel}.

\begin{algorithm}
\caption{Displacement-based linear CG scheme for the X-FFT system}\label{alg:solutionSchemeXFEM}
\begin{algorithmic}[1]
\State $\ul{\tilde{u}} \gets \ul{0}$
\State $\ul{f} \gets \ul{r}(\ul{\tilde{u}})$ \label{line:resVector} \Comment{Compute residual vector via eq.~\eqref{eq:resVector}}
\State {$\begin{bmatrix}\ul{d}\\\mathrm{res}\end{bmatrix} \gets \begin{bmatrix}\uul{\tilde{P}}^{-1}\, \ul{f}\\
 \sqrt{\T{\ul{f}}\uul{\tilde{P}}^{-1}\, \ul{f}}\end{bmatrix}$}\label{line:precondition} \Comment{Green's operator is applied in Fourier space using eq.~\eqref{eq:GreenHermitian}}
\State $\mathrm{res}_\mathrm{old} \gets \mathrm{res}$
\State $k \gets 1$
\While{$k<\mathrm{maxit}$ and $\mathrm{res}>\mathrm{tol}$}
\State $k \gets k+1$
\State $\ul{z} \gets \ul{r}(\ul{d})$ \label{line:z}
\State $\alpha \gets \mathrm{res}_\mathrm{old}^2/\langle \ul{z},\ul{d}\rangle_{L_2}$\label{line:alpha}
\State {$\begin{bmatrix}\ul{\tilde{u}}\\\ul{f} \end{bmatrix}\gets \begin{bmatrix}\ul{\tilde{u}} + \alpha \ul{d}\\ \ul{f} - \alpha \ul{z}\end{bmatrix}$}
\State {$\begin{bmatrix}\ul{z}\\\mathrm{res}\end{bmatrix} \gets \begin{bmatrix}\uul{\tilde{P}}^{-1}\, \ul{f}\\
\sqrt{\T{\ul{f}}\uul{\tilde{P}}^{-1}\, \ul{f}}\end{bmatrix}$}\label{line:precondition2}
\State $\beta \gets \mathrm{res}^2/\mathrm{res}_\mathrm{old}^2$
\State $\ul{d}\gets \ul{z} + \beta \ul{d}$
\State $\mathrm{res}_\mathrm{old} \gets \mathrm{res}$
\EndWhile
\end{algorithmic}
\end{algorithm}
We may solve the X-FFT system~\eqref{eq:linearSystem_element} using the basic scheme~\cite{moulinec1994fast,moulinec1998numerical}, the Barzilai-Borwein scheme~\cite{schneider2019barzilai}, the linear conjugate gradient scheme~\cite{zeman2010accelerating} or the nonlinear conjugate gradient scheme~\cite{schneider2020dynamical}, among others~\cite[Tab.4]{schneider2021review}.
The basic scheme~\cite{moulinec1994fast,moulinec1998numerical}, renowned for its robustness and reliability, is interpretable as a gradient descent method~\cite{kabel2014efficient} or a Richardson iteration~\cite{mishra2016comparative}. With the residual~\eqref{eq:convergence}, the basic scheme may be implemented on two displacement fields.
Nevertheless, the basic scheme only converges slowly to the solution of the Lippmann–Schwinger equation, with an iteration count that increases at a rate of order $\mathcal{O}(\kappa)$ as the material contrast $\kappa$ increases.
The Barzilai-Borwein scheme~\cite{schneider2019barzilai} accelerates the basic scheme with an advanced step size selection strategy, without requiring any additional memory footprint. The linear conjugate gradient (linear CG) method~\cite{zeman2010accelerating} is particularly well-suited for solving linear systems. Compared to the basic scheme, it does not require any internal parameters and improves the rate of the iteration count until convergence towards the rate $\mathcal{O}(\sqrt{\kappa})$~\cite{schneider2021review}. However, the implementation of linear CG requires four displacement fields, resulting in an increased memory footprint.
The nonlinear conjugate gradient (nonlinear CG) method~\cite{schneider2020dynamical} is a generalization of linear CG towards unconstrained nonlinear optimization problems and operates on three displacement fields. While it offers a general-purpose solver for FFT-based computational micromechanics, its iteration count until convergence may be up to twice as high as that of linear CG in the special case of linear elasticity~\cite{schneider2020dynamical}.

The nonlinear CG~\cite{schneider2020dynamical} and the Barzilai-Borwein~\cite{schneider2019barzilai} scheme rely on the step size of the basic scheme. As the basic scheme is a Richardson iteration~\cite{mishra2016comparative}, the optimal step size may be computed based on the maximum and minimum eigenvalue of the system matrix.
To compute the optimal step size for the X-FFT system at hand, the maximum and minimum eigenvalue of the preconditioned X-FFT system matrix would be required. However, computing these eigenvalues by assembling the preconditioned X-FFT system matrix would destroy the efficiency of the solver.
As a band-aid solution, we choose the step size
\begin{equation}\label{eq:stepsize}
    s_0 = \frac{2}{C_- + C_+},
\end{equation}
where the parameter $C_-$ ($C_+$) denotes the smallest (largest) stiffness eigenvalue of all individual materials comprising the matrix-inclusion problem. We acknowledge that the step size~\eqref{eq:stepsize} is not optimal. However, since our X-FFT system~\eqref{eq:linearSystem_element} is currently limited to linear elasticity, the most suitable choice for solving the system is the linear CG scheme, which does not rely on the step size.
Algorithm~\eqref{alg:solutionSchemeXFEM} outlines a displacement-based linear CG scheme that is adapted to our X-FFT solver. The other mentioned solution schemes may be used analogously.
We highlight that the computation of the residual vector, in the lines~\ref{line:resVector},~\ref{line:z}, and the action of the Green's operator, in the lines~\ref{line:precondition},~\ref{line:precondition2}, may be evaluated in parallel.
In addition, we may precompute and cache the voxel stiffness matrix and voxel average stress matrix, which are computed based on the element matrices~\eqref{eq:Kapp_q} and~\eqref{eq:Sapp_q} using the extraction matrices $\uul{\Lambda^\mathrm{v}_\ell}$. With this caching strategy at hand, the quadrature is computed only once per simulation instead of once per iteration. All in all, the caching strategy reduces the computational time associated with the high number of quadrature points required for subdivision integration at the cost of an increased memory footprint. We refer to the computational investigations in section~\ref{sec:hashin} for an analysis of the costs and benefits of the caching strategy.

For the implementation of the X-FFT solver, we use Python with Cython extensions and the pyfftw wrapper to FFTW3~\cite{frigo2005design}. Moreover, we leverage OpenMP parallelization to take advantage of multi-core processors.

\section{Computational investigations}\label{sec:computationalInvestigations}

\subsection{Setup}\label{sec:setup}

 \begin{table}[htb]
        \centering
        \begin{tabular}{l|ll}
        Shortcut &Discretization of homogeneous voxels & Discretization of interface voxels\\
        \hline
        P1& voxel-segmented Galerkin-FEM                       & - \\
        &with $\ffP_1$ elements&\\
         \hline
        Q1R&rotated staggered grid discretization~\cite{willot2015fourier}           & - \\
        &&\\
        \hline
        CoVo&rotated staggered grid discretization~\cite{willot2015fourier}          & composite voxel \\
        & &with laminate mixing rule~\cite{merkert_efficient_2015,keshav2023fft,lendvai2024assumed}\\
        \hline
        X-FEM &voxel-segmented Galerkin-FEM                  & X-FEM~\eqref{eq:X-FEM_scaledspace}\\
        &with $\ffP_1$ elements&\\
        \end{tabular}
        \caption{Discretizations.}
        \label{tab:Discretizations}
\end{table}

Our computational investigations comprise matrix inclusion problems with smooth and non-smooth interfaces. To validate the results of the X-FFT solver, we compare them to the results of other FE discretizations in FFT-based homogenization. The respective discretizations are listed in Tab.~\ref{tab:Discretizations}. We note that the rotated staggered grid discretization~\cite{willot2015fourier} may be interpreted as a voxel-segmented trilinear hexahedral Galerkin-FEM discretization with reduced integration~\cite{schneider2017fft}. For the composite voxels in the CoVo discretization, we use the level set-based strategy of Lendvai and Schneider~\cite{lendvai2024assumed} to compute the interface normals and volume fractions. 
In the CoVo discretization, the interface in each voxel is approximated by a single plane. Therefore, we correct the level set values using regression~\cite[Sec. 3.2]{lendvai2024assumed} to obtain the best-fitting linear level-set function.

We quantify the error of a scalar $a \in\R_{>0}$ w.r.t a reference value $a_\mathrm{ref}\in\R$ by the relative error
\begin{equation}
    e(a, a_\mathrm{ref}) = \left|\frac{a - a_\mathrm{ref}}{a_\mathrm{ref}}\right|.
    \label{eq:relError}
\end{equation}
The $L^2$-norm of a FE discretized vector field $\fa$ is evaluated by quadrature and takes the form
\begin{equation}
    \norm{\fa}_{L^2} = \sqrt{\sum_{\alpha=1}^{n_\mathrm{int}} w_\alpha\,\fa(\fq_\alpha) \cdot \fa(\fq_\alpha)}
    \label{eq:L2NormGauss},
\end{equation}
where the parameters $w_\alpha$ and $\fq_\alpha$ denote the respective quadrature weights and quadrature points. 

For a conforming Galerkin method, we note that based on eq.~\eqref{eq: boundsStiffness} the error between the simulated and exact effective elastic energy, may be bounded from above by a value which is proportional to the quadratic error of the simulated local strain field $\feps_h$ to the exact solution $\feps_*$ via the relation
\begin{equation}\label{eq:upperBoundOnEffStiffnessError}
     \bar{\feps}:\left(\ffC^\mathrm{eff}_*-\ffC^\mathrm{eff}_h\right):\bar{\feps}\leq C_+\norm{\feps_* - \feps_h}_{L^2}^2
\end{equation}
with the simulated effective stiffness $\ffC^\mathrm{eff}_h$ and the exact effective stiffness $\ffC^\mathrm{eff}_*$.
Moreover, the error in the local strain fields is bounded from above by a value proportional to the square root of the error in the effective elastic energy via the relation
\begin{equation}\label{eq:upperBoundOnLocalStrainError}
    \norm{\feps_* - \feps_h}_{L^2} \leq \sqrt{\frac{\ \bar{\feps}:\left(\ffC^\mathrm{eff}_*-\ffC^\mathrm{eff}_h\right):\bar{\feps}}{C_-}}.
\end{equation}
The positive constants $C_-,C_+\in \R$ appearing in the inequalities above refer to the bounds on the stiffness~\eqref{eq: boundsStiffness}. The derivation of the inequalities is discussed in Appendix~\ref{apx:Bounds}, based on previous work of Schneider~\cite{schneider2022superaccurate}. 
We observe that, based on the inequalities~\eqref{eq:upperBoundOnEffStiffnessError} and~\eqref{eq:upperBoundOnLocalStrainError}, the error in the effective elastic energy is proportional to the squared error of the local strain fields for conforming Galerkin methods. Consequently, the error in the effective stress is proportional to the squared error of the local stress fields for these methods. Among the discretizations listed in Tab.~\ref{tab:Discretizations}, the X-FEM discretization and the P1 discretization are conforming Galerkin methods, so that the inequalities~\eqref{eq:upperBoundOnEffStiffnessError} and~\eqref{eq:upperBoundOnLocalStrainError} hold respectively.

All studies below were computed on a workstation with 1.15 TB RAM and 96 cores (2.4GHz).

\subsection{Hashin's neutral inclusion}\label{sec:hashin}
\subsubsection*{Microstructure}
\begin{figure}[htb]
    \begin{subfigure}[b]{0.3\textwidth}
        \centering
  \includegraphics[width=\textwidth]{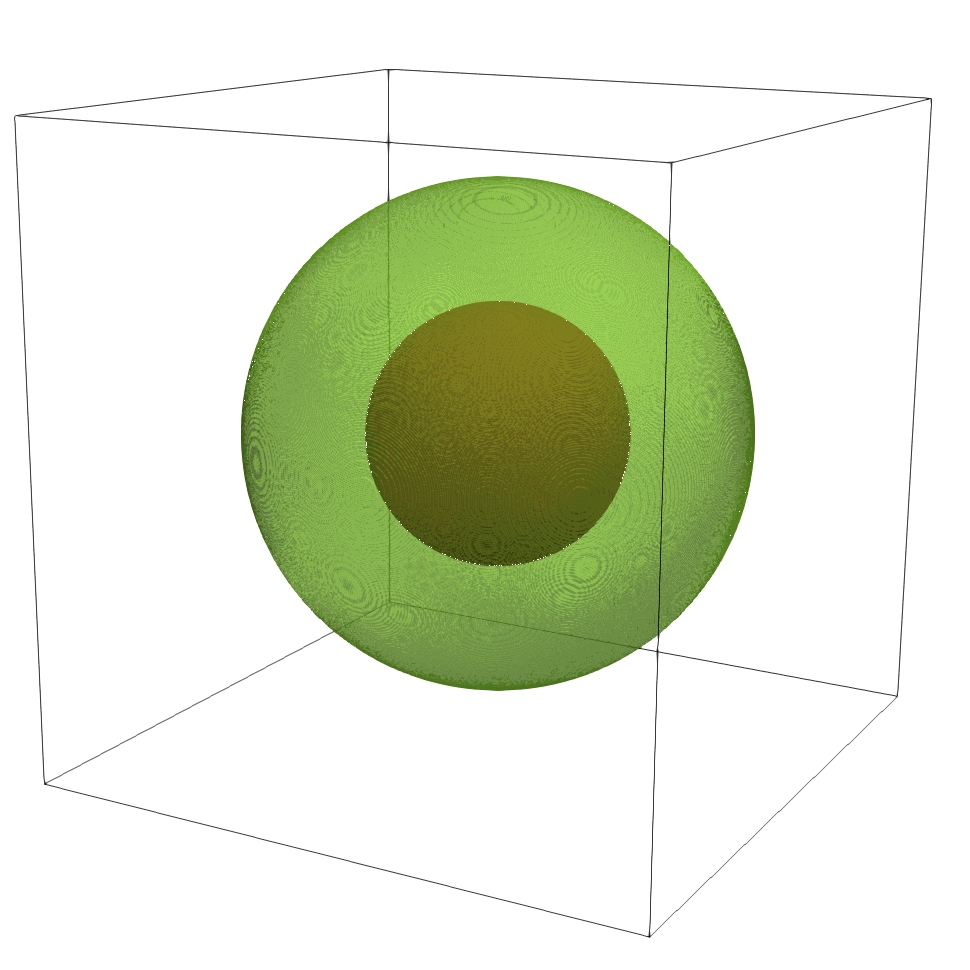}
        \subcaption{Geometry}\label{fig:geometry}
    \end{subfigure}
    \begin{subtable}[b]{0.55\textwidth}
        \centering
        \begin{tabular}{l|lll}
          & Matrix &Coating&Inclusion\\
        \hline
        Bulk modulus in MPa& 1.000000& 0.808024& 8.080240 \\
        Poisson's ratio&0.25&0.25&0.25\\
        Radius in $\mu$m&-&$2\pi$&$6/5e$
        \end{tabular}
        \subcaption{Parameters}\label{tab:material}
    \end{subtable}
    \caption{Geometry and parameters of Hashin's neutral inclusion problem ($e$ refers to Euler's number).}
\end{figure}

Our first study is dedicated to Hashin's neutral inclusion problem. Due to its analytical solution~\cite{hashin1960elastic} that is available for the setup described below, we may assess the quality of the X-FFT solutions. 
We consider a cubic cell with an edge length of $16\mu\text{m}$ that contains a coated sphere, which is embedded in a matrix phase, as shown in Fig.~\ref{fig:geometry}. 
All three phases are assumed to be isotropic, and a macroscopic uniform compression with loading $\bfeps=\sty{I}$ is prescribed. We resolve the cell by \mbox{$16,32,\dots,1024$} voxels per dimension, denoted by the voxel count $N$ per axis, which is the inverse of the mesh parameter $h$.
Using the convergence criterion~\eqref{eq:stoppingCriterion}, the computations in this study were solved up to the tolerance $\mathrm{tol}=10^{-7}$ using 4 threads per computation.
The utilized radii and material parameters are given in Tab.~\ref{tab:material} - rounded to six digits for clarity. To allow for an analytical solution, the material parameters are chosen such that the analytical solution of the effective bulk modulus coincides with the matrix bulk modulus. For the respective equation on the effective bulk modulus and the analytical solution of the local displacement field, we refer to Schneider et al.~\cite[Sec. 4.1.1]{schneider2016computational}.

\subsubsection*{Solution scheme}

\begin{figure}[htb]
  \centering
    \begin{subfigure}[c]{0.40\textwidth}
        \centering 
        \includegraphics[width=\textwidth]{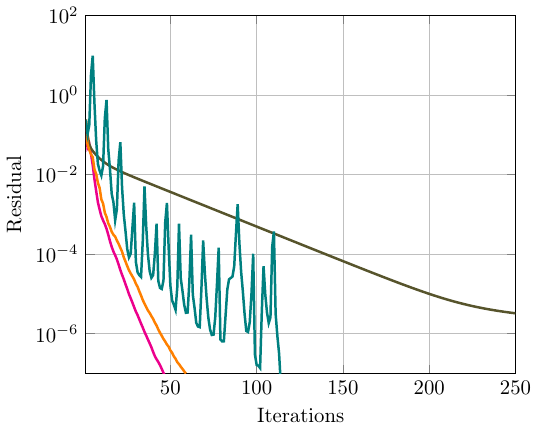}
    \end{subfigure}
    \begin{subfigure}[c]{0.18\textwidth}
        \fbox{\includegraphics[width=\textwidth]{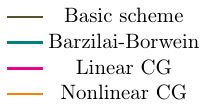}}\\
    \end{subfigure}
\caption{Solver convergence for X-FEM at voxel count \mbox{$N=128$} in Hashin's neutral inclusion.}
\label{fig:solverHashin}
\end{figure}

For X-FEM at voxel count \mbox{$N=128$}, we compare the convergence of linear CG~\cite{zeman2010accelerating} to other solvers, namely to the basic scheme~\cite{moulinec1994fast,moulinec1998numerical}, the Barzilai-Borwein scheme~\cite{schneider2019barzilai} and the nonlinear CG scheme~\cite{schneider2020dynamical}. The respective convergence of the residual~\eqref{eq:convergence} with the iteration count of the different solvers is shown in Fig.~\ref{fig:solverHashin}. 
We observe that the residual of the basic scheme converges monotonically but slowly with increasing iteration count, reaching the desired tolerance at iteration count 668. However, the convergence rate is anticipated to worsen with increasing material contrast. In contrast, the residual of the Barzilai-Borwein scheme does not converge monotonically with increasing iteration count, but reaches the desired tolerance $10^{-7}$ at iteration count 115. The nonlinear CG scheme requires 60 iterations to reach the desired tolerance, while linear CG achieves this in only 48 iterations. This decreased iteration count of linear CG in linear elasticity is well-documented in the literature~\cite{schneider2020dynamical}. Therefore, we choose to use linear CG for the remainder of this section.

\subsubsection*{Accuracy}
\begin{figure}[htb]
  \centering
	\fbox{\includegraphics[width=0.75\textwidth]{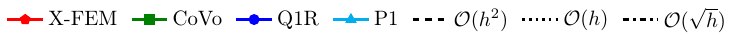}}\\
   \begin{subfigure}[b]{0.40\textwidth}
        \centering 
        \includegraphics[width=\textwidth]{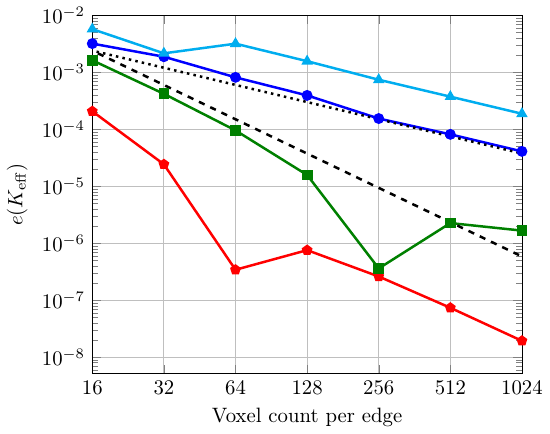}
        \subcaption{Effective bulk modulus error}
        \label{fig:effBulkError}
    \end{subfigure}
    \begin{subfigure}[b]{0.40\textwidth}
        \centering 
        \includegraphics[width=\textwidth]{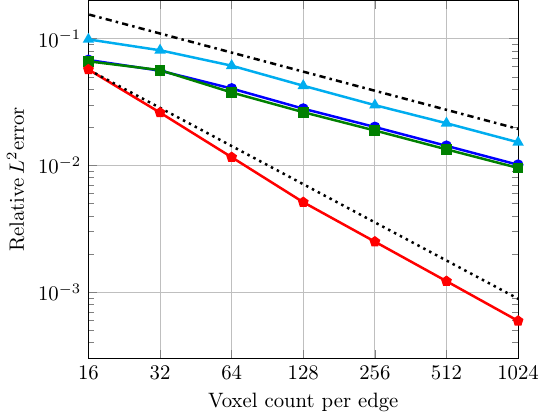}
        \subcaption{Local strain error}
        \label{fig:locStrainError}
    \end{subfigure}
    \begin{subfigure}[b]{0.40\textwidth}
        \centering 
        \includegraphics[width=\textwidth]{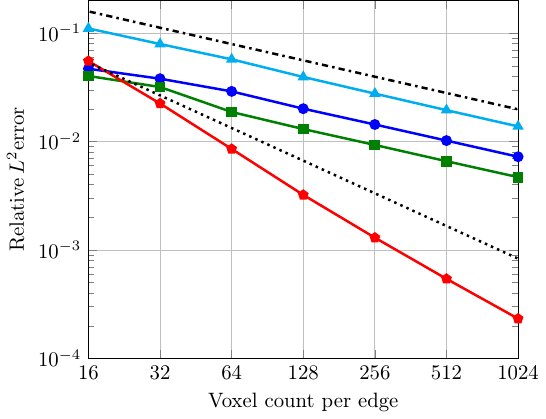}
        \subcaption{Local stress error}
        \label{fig:locStressError}
    \end{subfigure}
    \begin{subfigure}[b]{0.40\textwidth}
        \centering 
        \includegraphics[width=\textwidth]{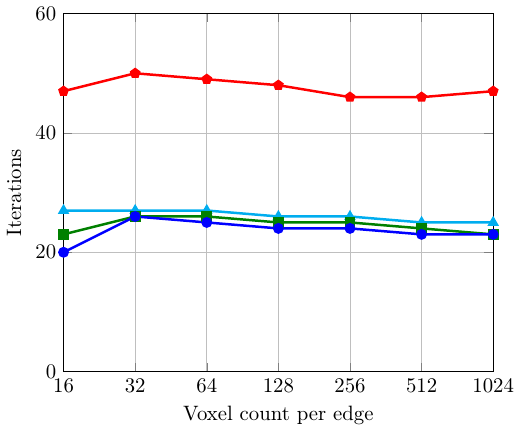}
        \subcaption{Iteration count}
        \label{fig:iterationHashin}
    \end{subfigure}
\caption{Accuracy and iteration count in Hashin's neutral inclusion.}
\end{figure}

With the analytical solution at hand, we compute the relative error in the effective bulk modulus using eq.~\eqref{eq:relError}. The resulting errors with respect to the mesh parameter $h$ are shown in Fig.~\ref{fig:effBulkError}. We observe that the relative error in the effective bulk modulus is lowest for X-FEM at all resolutions. Moreover, a decrease in the error at the rate $\mathcal{O}(h^2)$ is observed as the mesh parameter decreases, which is consistent with previous observations on the convergence of error in the energy norm for X-FEM~\cite[Fig. 6]{moes_computational_2003}. The CoVo discretization shows the second-lowest error in the effective bulk modulus. Surprisingly, for CoVo, the error in the bulk modulus of Hashin's neutral inclusion decreases superlinearly with increasing resolution. These findings agree with the observations made by Lendvai et al.~\cite{lendvai2025accurate} for Hashin's neutral inclusion, but differ from the convergence behavior observed for more complex microstructures~\cite{lendvai2024assumed,lendvai2025accurate}. The analytical solution of Hashin's neutral inclusion is independent of the Poisson's ratios, but depends on the phases bulk moduli. The bulk modulus is a measure of the infinitesimal pressure increase relative to the resulting decrease of the volume. One possible reason for the superlinear error convergence of the CoVo discretization in the bulk modulus of Hashin’s neutral inclusion is CoVo’s ability to accurately approximate the volume fraction if the level-set based normal computation~\cite{lendvai2024assumed} is used instead of the traditional subvoxel-based normal-estimation technique~\cite{merkert_efficient_2015}. For a thorough explanation of the issues with the traditional subvoxel-based normal-estimation technique, we refer to section 3.2 of Lendvai et al.~\cite{lendvai2025accurate}. At all resolutions, the Q1R discretization and the P1 discretization show the highest error in the effective bulk modulus at all resolutions, following the rate $\mathcal{O}(h)$ with decreasing mesh parameter.

To quantify the error in the local fields, the relative $L^2$-error in the local strain and stress fields to the exact solution is computed based on eq.~\eqref{eq:relError} and eq.~\eqref{eq:L2NormGauss}. In general, we observe similar trends for the local strain in Fig.~\ref{fig:locStrainError} and the local stress in Fig.~\ref{fig:locStressError}. In both cases, X-FEM shows the least error at most resolutions and a linear error convergence with decreasing mesh parameter is observed. The rate $\mathcal{O}(h)$ coincides with the rate that is achieved by \emph{interface-conforming} FEM with linear shape functions~\cites[Theorem 5.4.8]{brenner_mathematical_2008}[p.190]{hughes_finite_1987}. The P1, Q1R and Covo discretization, show only a square root convergence rate in the relative $L^2$-error in the local strain and stress fields with decreasing mesh parameter. The highest error is observed for the P1 discretization followed by the Q1R and the CoVo discretization. 
We observe that for the conforming Galerkin discretizations, namely, the X-FEM discretization and the P1 discretization, the error in the effective bulk modulus decreases with the squared rate of the $L^2$-error in the local strain and stress fields, as proved by the inequalities~\eqref{eq:upperBoundOnEffStiffnessError},~\eqref{eq:upperBoundOnLocalStrainError}. However, our observations on the error convergence rates of CoVo demonstrate that the inequalities~\eqref{eq:upperBoundOnEffStiffnessError},~\eqref{eq:upperBoundOnLocalStrainError} may not be fulfilled if no conforming Galerkin discretization is used.

\subsubsection*{Numerical efficiency}

    \begin{figure}[htb]
    \centering
       \fbox{\includegraphics[width=0.95\textwidth]{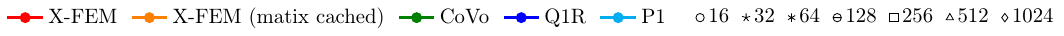}}
        \begin{subfigure}[b]{0.33\textwidth}
        \includegraphics[width=\textwidth]{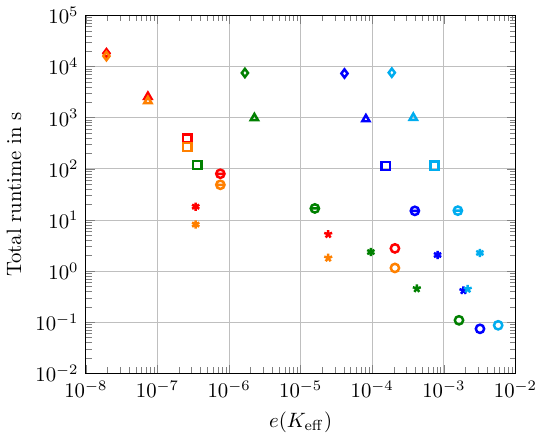}
        \caption{Total runtime}
        \label{fig: TimeHashin}
        \end{subfigure}
        \begin{subfigure}[b]{0.32\textwidth}
        \includegraphics[width=\textwidth]{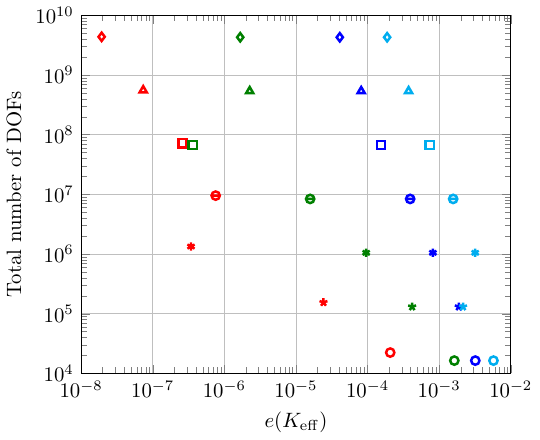}
        \caption{Degrees of freedom}
        \label{fig: DOFHashin}
        \end{subfigure}
        \begin{subfigure}[b]{0.32\textwidth}
        \includegraphics[width=\textwidth]{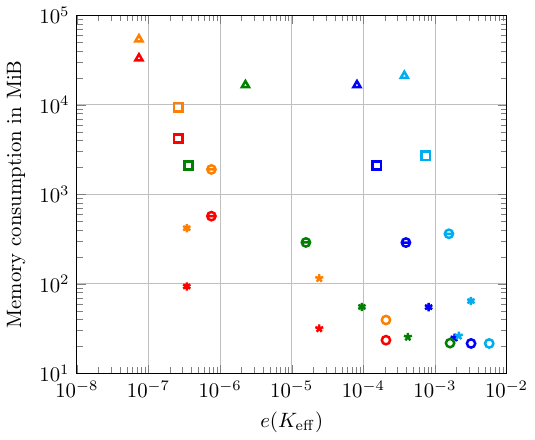}
        \caption{Measured memory consumption}
        \label{fig: MemoryHashin}
        \end{subfigure}
        \caption{Numerical efficiency for Hashin's neutral inclusion.}
    \end{figure}

For the iteration count until convergence of linear CG, we observe in Fig.~\ref{fig:iterationHashin} that the discretizations P1, Q1R and CoVo require around 20-27 iterations. In contrast, for X-FEM 46-50 iterations are required until convergence. Nevertheless, the iteration counts of X-FEM appear stable with respect to the mesh parameter. 

In Fig.~\ref{fig: TimeHashin}, we show the total runtime versus the error in the effective bulk modulus. We observe that X-FEM shows a higher runtime than the P1, Q1R and CoVo discretization for all resolutions. At low resolutions, the runtime of X-FEM is approximately ten times higher than for the other discretizations. At high resolutions, the runtime of X-FEM is approximately twice the runtime of the other discretizations. The higher runtime of X-FEM results partly from the higher iteration count. Moreover, for X-FEM, matrix-vector products of larger dimension have to be evaluated due to the additional degrees of freedom. Last but not least, the subdivision strategy used for numerical integration comes with computational overhead, which may be reduced at low resolutions by precomputing and caching the voxel stiffness matrix. Nevertheless, if we desire an error in the effective bulk modulus that is below 0.005\%, X-FEM offers the fastest solution at resolution \mbox{$N=64$} with a total runtime of $8$s when caching the voxel stiffness matrix and a total runtime of $18$s without caching.

To assess the efficiency of the discretizations, not only the iteration count and runtime, but also the memory footprint, which depends on the number of degrees of freedom, is relevant. For all discretizations considered, three degrees of freedom per node are required for each standard FE field. In the X-FEM discretization, three additional degrees of freedom (\emph{dofs}) are added to the approximation space for each enriched node. 
To evaluate whether the higher accuracy of X-FEM justifies its increased number of \emph{dofs}, we show in Fig.~\ref{fig: DOFHashin} the total number of \emph{dofs} per field versus the relative error in the effective bulk modulus. At low resolutions, the difference in the number of \emph{dofs} is pronounced, i.e., at voxel count \mbox{$N=16$}, X-FEM requires 36\% more \emph{dofs} than the other discretizations. However, this difference decreases quadratically with increasing voxel count so that at voxel count \mbox{$N=1024$} the difference is only 0.5\%. Above all, we observe that for a desired accuracy of 0.1\%, X-FEM offers the cheapest result in terms of \emph{dofs}: For X-FEM, a study with voxel count \mbox{$N=16$} is sufficient to reach an error in the effective bulk modulus that is below 0.1\%. To reach an accuracy below 0.1\%, for CoVo a voxel count \mbox{$N=32$} would be necessary, Q1R needs a resolution of \mbox{$N=64$} and for P1 a study with voxel count \mbox{$N=256$} would be required.

We measured the total memory consumption of our implementation with \emph{valgrind}~\cite{nethercote2007valgrind}. The results are shown in Fig.~\ref{fig: MemoryHashin}, for all resolutions except for 1024 voxels per edge, as \emph{valgrind} crashed during those measurements. We observe that the trends are similar to those observed for the \emph{dofs}. However, X-FEM shows a slightly higher increase in memory footprint than the other discretizations compared to the observations made for the \emph{dofs}. This increase is caused by additional local variables that are allocated for X-FEM. In case the voxel matrices are cached, the increase in memory consumption is significantly higher for X-FEM. Weighting the large increase in memory consumption against the small decrease in runtime at low resolutions, caching the voxel matrices appears not to be worth the cost and is thus neglected in the rest of the manuscript. 
Still, for a desired error in the effective stress that is below 0.1\%, X-FEM offers the cheapest result in terms of total memory - confirming the results for the \emph{dofs}, discussed above. Due to the close resemblance of the trends observed in Fig.~\ref{fig: DOFHashin} and Fig.~\ref{fig: MemoryHashin}, for the rest of the manuscript, we refrain from measuring the total memory consumption and limit our discussion to the number of \emph{dofs}.
Concerning the numerical efficiency in Hashin's neutral inclusion, we conclude that the X-FFT solver offers the best balance of accuracy and efficiency across all resolutions.

\subsubsection*{Contrast dependence}

\begin{figure}[htb]
  \centering
	\fbox{\includegraphics[width=0.4\textwidth]{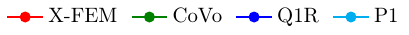}}\\
    \begin{subfigure}[b]{0.40\textwidth}
        \centering 
        \includegraphics[width=\textwidth]{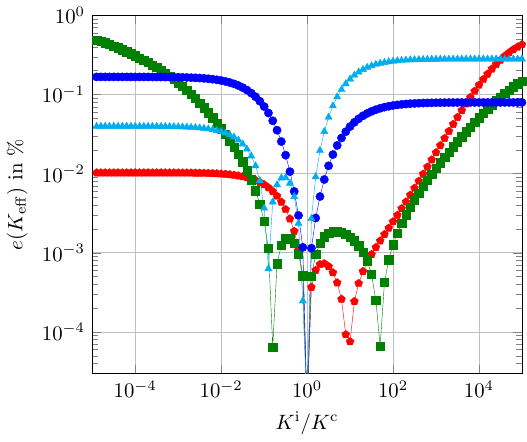}
        \subcaption{Accuracy}\label{fig: contrastAccuracy}
    \end{subfigure}   
    \begin{subfigure}[b]{0.40\textwidth}
        \centering 
        \includegraphics[width=\textwidth]{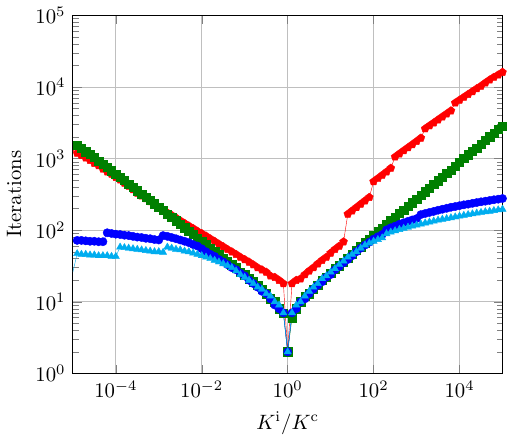}
        \subcaption{Iterations}\label{fig: contrastIterations}
    \end{subfigure}
\caption{Contrast study for Hashin's neutral inclusion at voxel count \mbox{$N=128$}.}
\end{figure}

To analyze the effect of the material contrast on the accuracy, we compute the error of the effective bulk modulus for different ratios of the bulk moduli of the inclusion and coating phase, denoted by the parameter \mbox{$K^\mathrm{i}/K^\mathrm{c}$}. If the parameter \mbox{$K^\mathrm{i}/K^\mathrm{c}$} is low, we model an almost void inclusion, whereas a high bulk moduli ratio denotes an almost rigid inclusion.
The connection of the parameter \mbox{$K^\mathrm{i}/K^\mathrm{c}$} with the material contrast $\kappa$ is given by the relation
\begin{equation}
    \kappa = \begin{cases}
        K^\mathrm{i}/K^\mathrm{c}, \quad \text{if } K^\mathrm{i}\geq K^\mathrm{c},\\
        K^\mathrm{c}/K^\mathrm{i},  \quad\text{if } K^\mathrm{c}> K^\mathrm{i}.
    \end{cases}
\end{equation}
In Fig.~\ref{fig: contrastAccuracy}, we show the error of the effective bulk modulus in Hashin's neutral inclusion at voxel count \mbox{$N=128$} for different parameters \mbox{$K^\mathrm{i}/K^\mathrm{c}$}. We observe that X-FEM shows the least error in the effective bulk modulus for almost void inclusions, i.e., at parameters \mbox{$K^\mathrm{i}/K^\mathrm{c}\in\left[10^{-5},0.04\right)$} and at moderately stiff inclusions, i.e., at parameters \mbox{$K^\mathrm{i}/K^\mathrm{c}\in\left(1;25\right]$}.
For moderately compliant inclusions, i.e., at parameters \mbox{$K^\mathrm{i}/K^\mathrm{c}\in\left[0.04;1\right)$} and for stiff inclusions, i.e., at parameters \mbox{$K^\mathrm{i}/K^\mathrm{c}\in\left(25,2\cdot10^4\right]$}, CoVo shows the least error in the effective bulk modulus. If the bulk moduli of the inclusion and the coating phase coincide, Hashin's neutral inclusion reduces to a homogeneous microstructure, such that all discretizations perform equally accurate. For almost rigid inclusions, i.e., at parameters \mbox{$K^\mathrm{i}/K^\mathrm{c}\in\left(2\cdot10^4,10^5\right]$}, Q1R shows the least error in the effective bulk modulus.

The material contrast is expected to have significant impact on the convergence rate of the residual~\eqref{eq:convergence}, as the condition number of the preconditioned systems is bounded by a contrast-dependent constant for all considered discretizations.
We assess the influence of the specific discretization on this behavior. To that end, we investigate the iteration count for Hashin's neutral inclusion at voxel count \mbox{$N=128$} for different ratios of the bulk moduli of the inclusion and coating phase. Our results, shown in Fig.~\ref{fig: contrastIterations}, indicate that the iteration count of Q1R is smallest for small material contrasts, while the iteration count of P1 is smallest for large material contrasts. For small material contrasts, the iteration count of CoVo is close to that of Q1R, whereas for large material contrasts it increases considerably. For nearly rigid inclusions with parameter $K^\mathrm{i}/K^\mathrm{c}=10^5$, CoVo requires 2859 iterations until reaching the desired tolerance of $10^{-7}$. For X-FEM, the increase in the iteration count is even higher for nearly rigid inclusions, so that, with the parameter $K^\mathrm{i}/K^\mathrm{c}=10^5$ the iteration count increases to 16081 iterations. For inclusions with material properties similar to those of voids, the iteration count of X-FEM is slightly below that of CoVo. At the parameter $K^\mathrm{i}/K^\mathrm{c}=10^{-5}$, X-FEM requires 1316 iterations and CoVo requires 1696 iterations until reaching the desired tolerance. For all discretizations considered, the iteration count appears to increase at the rate $\mathcal{O}(\sqrt{\kappa})$ as the material contrast increases, which coincides with the expected rate for linear CG~\cite{schneider2021review}.

\subsection{Smooth rock inclusions in a cement matrix}
\subsubsection*{Microstructure}
\begin{figure}[htb]
    \centering
        \begin{subfigure}[b]{0.48\textwidth}
        \centering
        \includegraphics[width=0.82\textwidth]{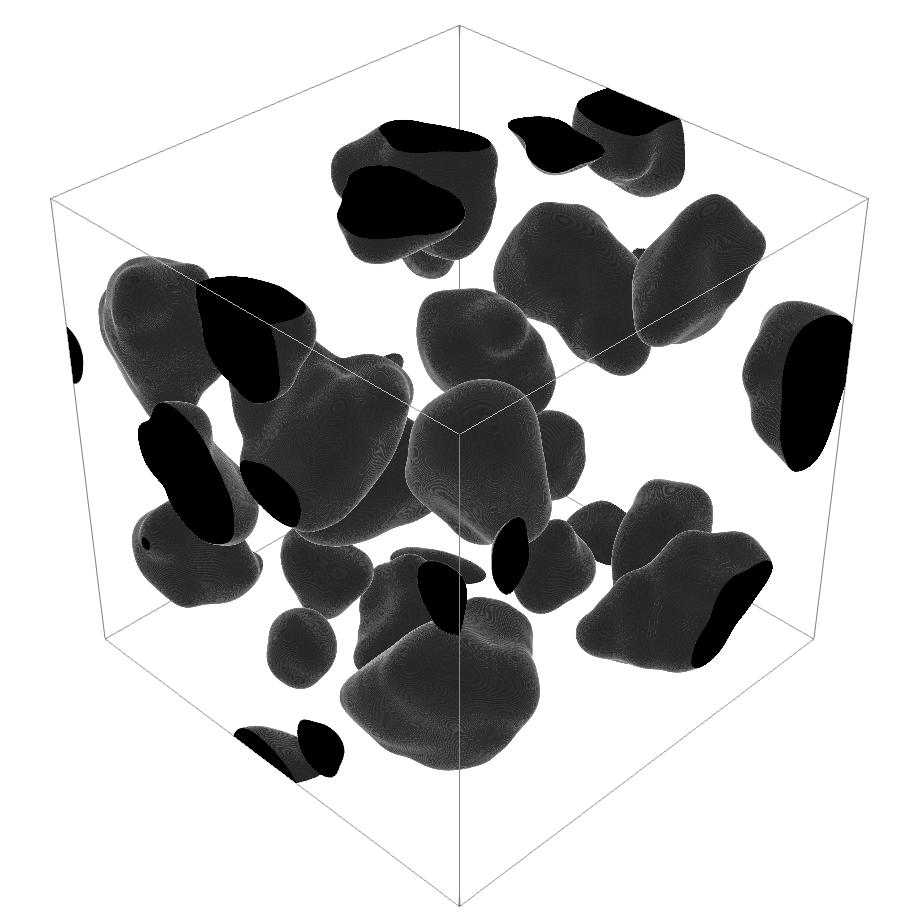} 
        \includegraphics[width=0.16\textwidth]{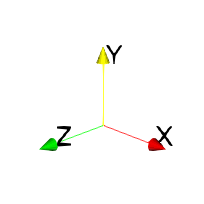}
         \subcaption{Geometry}
         \label{fig:microstructureGrain}    
        \end{subfigure}
         \begin{subfigure}[b]{0.48\textwidth}
        \centering 
          \includegraphics[width=0.82\textwidth]{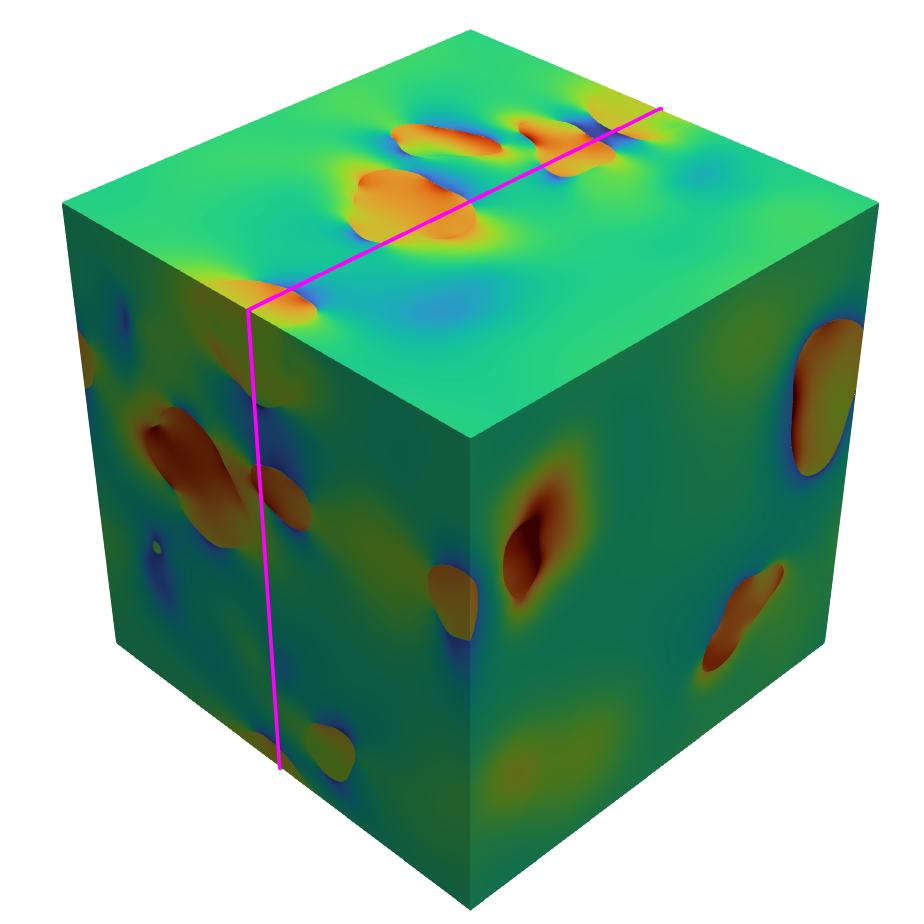}
          \includegraphics[width=0.16\textwidth]{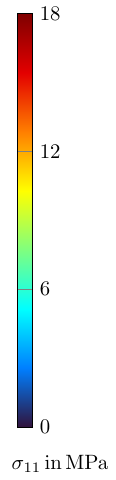}
         \subcaption{Reference}\label{fig:referenceGrain}
    \end{subfigure}
    \caption{Geometry and reference solution of the rock-cement microstructure.}
\end{figure}
To analyze whether the X-FFT solver maintains its accuracy and efficiency for more complex microstructures, we consider smooth rock inclusions in a cement matrix in this section. The respective microstructure was generated by forming each grain as a cluster of spheres and packing the grains using the Mechanical Contraction Method~\cite{williams2003random,schneider2018modelling}. In this context, we apply an exponential smoothing of the grains' surface~\cite{donval2025directional}. Due to the description as a cluster of spheres, the level set description of each particle is naturally available. The resulting microstructure is shown in Fig.~\ref{fig:microstructureGrain} with an edge length \mbox{$512\mu\text{m}$} and features a volume fraction of $15\%$ of the grains. We use the material parameters listed in Tab.~\ref{tab:materialGrains} for the cement matrix (gypsum) and the rocks (Portland cement) and prescribe unidirectional loading with the strain $\bar{\feps}_{11}=0.1\%$. The desired tolerance of the stopping criterion~\eqref{eq:stoppingCriterion} of the solver is set to $10^{-5}$ for this study. All computations of this study were run on 16 threads.

 \begin{table}[htb]
        \centering
        \begin{tabular}{l|ll}
          & Cemented matrix&Rocks\\
          & (Gypsum) &(Portland cement)\\
          \hline
        Young's modulus&6 GPa&35 GPa\\
        Poisson's ratio&0.3&0.22
        \end{tabular}
        \caption{Material parameters for the rock-cement microstructure~\cite[Tab.3]{lin2022systematic}.}
        \label{tab:materialGrains}
\end{table}

\subsubsection*{Solution scheme}

\begin{figure}[htb]
  \centering
    \begin{subfigure}[c]{0.40\textwidth}
        \centering 
        \includegraphics[width=\textwidth]{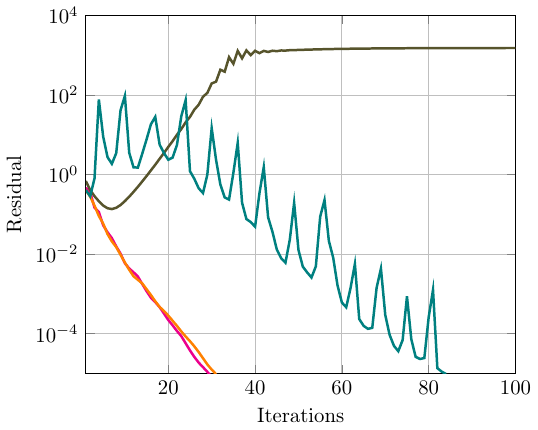}
    \end{subfigure}
    \begin{subfigure}[c]{0.18\textwidth}
        \fbox{\includegraphics[width=\textwidth]{legendSolvers.pdf}}\\
    \end{subfigure}
\caption{Solver convergence for X-FEM at voxel count \mbox{$N=128$} for the rock-cement microstructure.}
\label{fig:solverGrains}
\end{figure}

We compare the convergence of linear CG~\cite{zeman2010accelerating}, the basic scheme~\cite{moulinec1994fast,moulinec1998numerical}, the Barzilai-Borwein scheme~\cite{schneider2019barzilai}, and the nonlinear CG scheme~\cite{schneider2020dynamical}, for X-FEM at voxel count \mbox{$N=128$}. The convergence of the residual~\eqref{eq:convergence} with the iteration count of each solver is shown in Fig.~\ref{fig:solverGrains}. The residual of the basic scheme increases after reaching a minimum value and remains constant after approximately 50 iterations. This behavior indicates that the chosen step size~\eqref{eq:stepsize} is suboptimal, i.e., too large. The Barzilai-Borwein scheme shows its characteristic non-monotonically convergence behavior and reaches the desired tolerance $10^{-5}$ at iteration count 85. The nonlinear CG scheme requires 32 iterations to reach the desired tolerance, while linear CG achieves this in 31 iterations. Based on these results, we choose to use linear CG for this section.

\subsubsection*{Accuracy}
\begin{figure}[htb]
    \centering
	\fbox{\includegraphics[width=.6\textwidth]{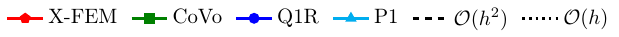}}\\
    \begin{subfigure}[b]{0.32\textwidth}
        \centering 
        \includegraphics[width=\textwidth]{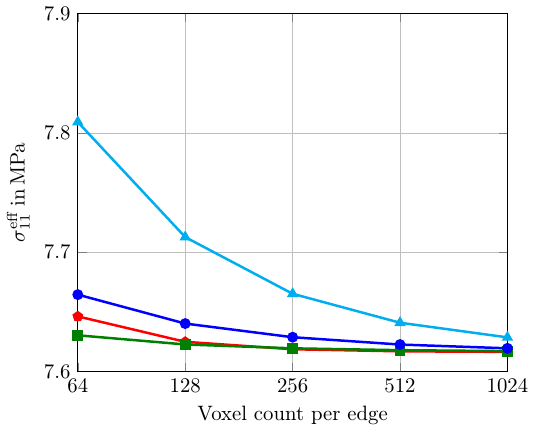}
         \subcaption{Effective stress}
         \label{fig:EffStressGrain}
    \end{subfigure}
    \begin{subfigure}[b]{0.32\textwidth}
        \centering 
        \includegraphics[width=\textwidth]{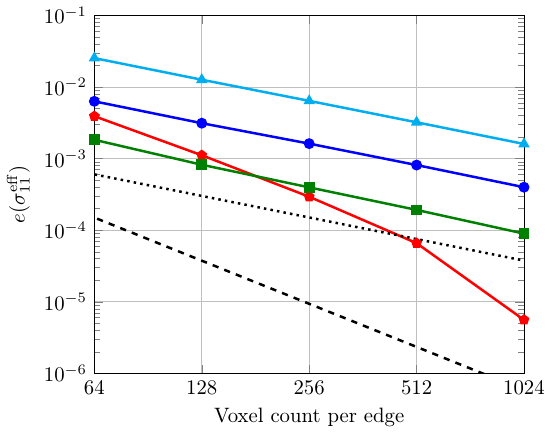}
         \subcaption{Error in the effective stress}
         \label{fig:ErrEffStressGrain}
    \end{subfigure}
        \begin{subfigure}[b]{0.32\textwidth}
        \centering 
        \includegraphics[width=0.95\textwidth]{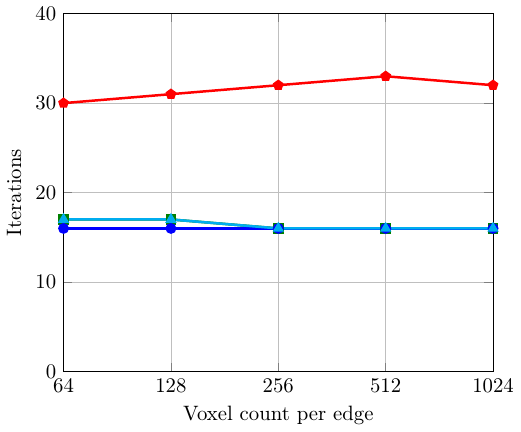}
         \subcaption{Iteration count}
         \label{fig:iterationsGrains}
    \end{subfigure}
    \caption{Effective stress and iteration count for the rock-cement microstructure.}
\end{figure}

\begin{figure}[htb]
    \centering  
    \begin{subfigure}[b]{0.3\textwidth}
        \centering 
          \includegraphics[width=\textwidth]{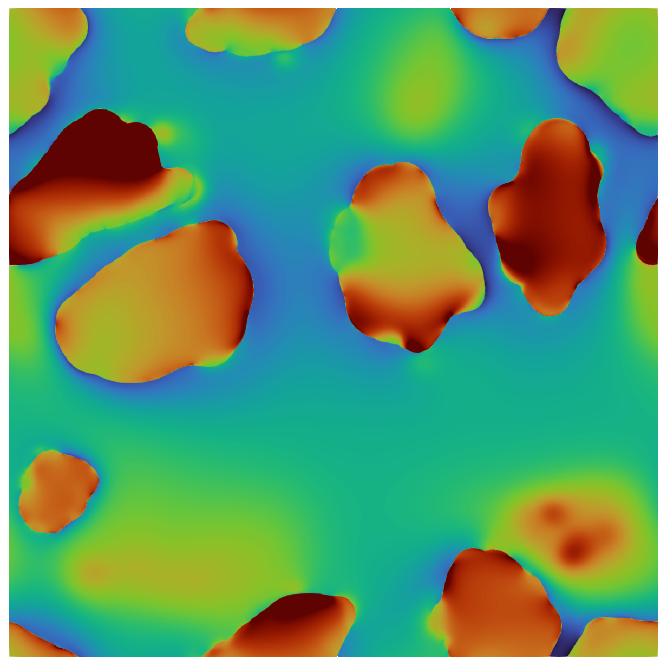}
         \subcaption{Reference}
    \end{subfigure}
        \begin{subfigure}[b]{0.3\textwidth}
        \centering 
        \includegraphics[width=\textwidth]{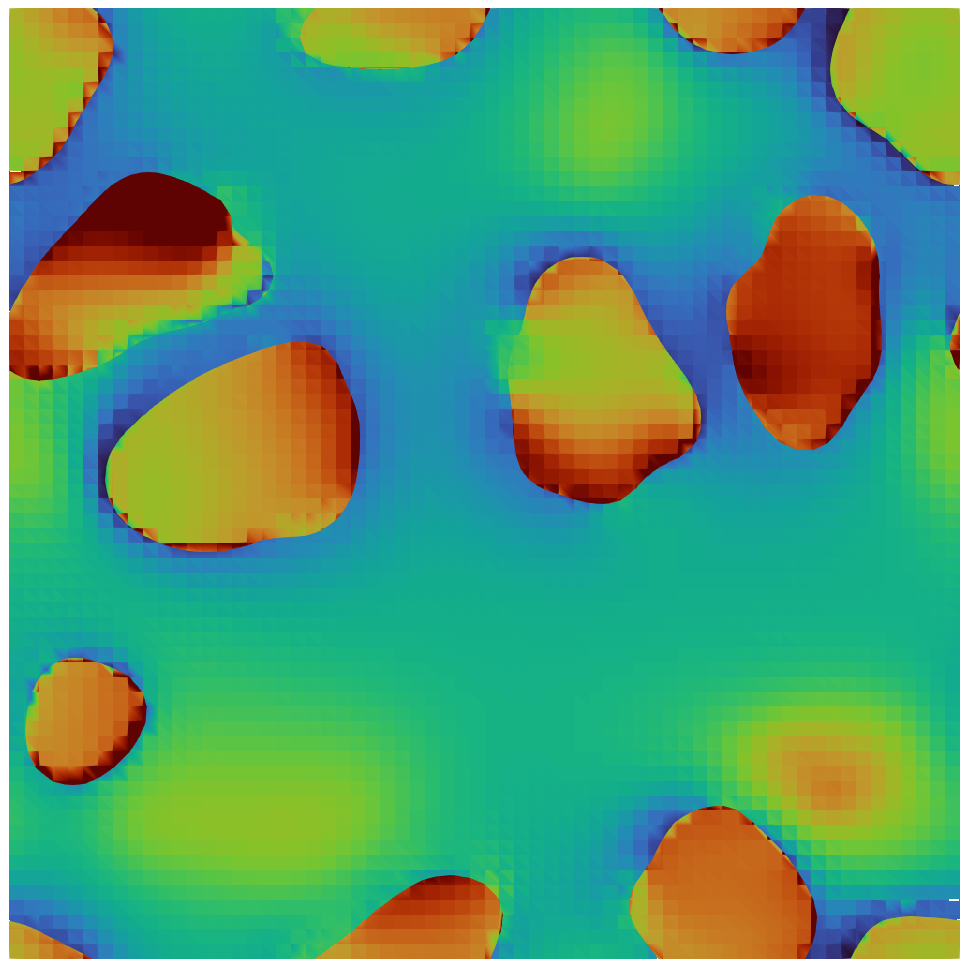}
         \subcaption{X-FEM}
    \end{subfigure}
    \begin{subfigure}[b]{0.30\textwidth}
        \centering 
        \includegraphics[width=\textwidth]{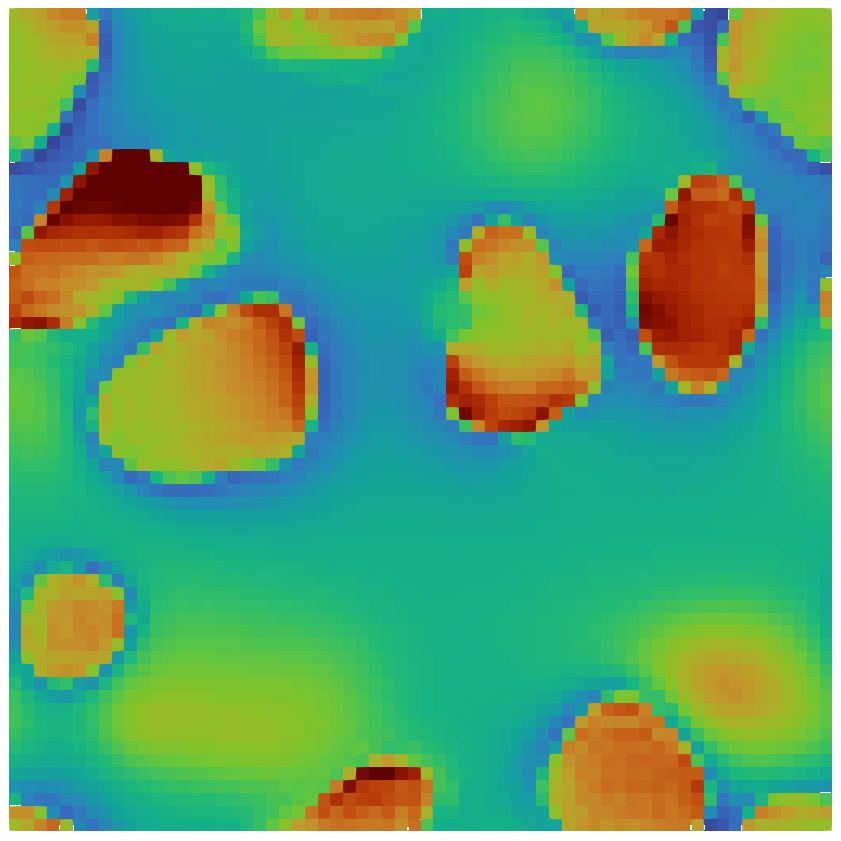}
         \subcaption{CoVo}
    \end{subfigure} 
    \begin{subfigure}[b]{0.30\textwidth}
        \centering 
        \includegraphics[width=\textwidth]{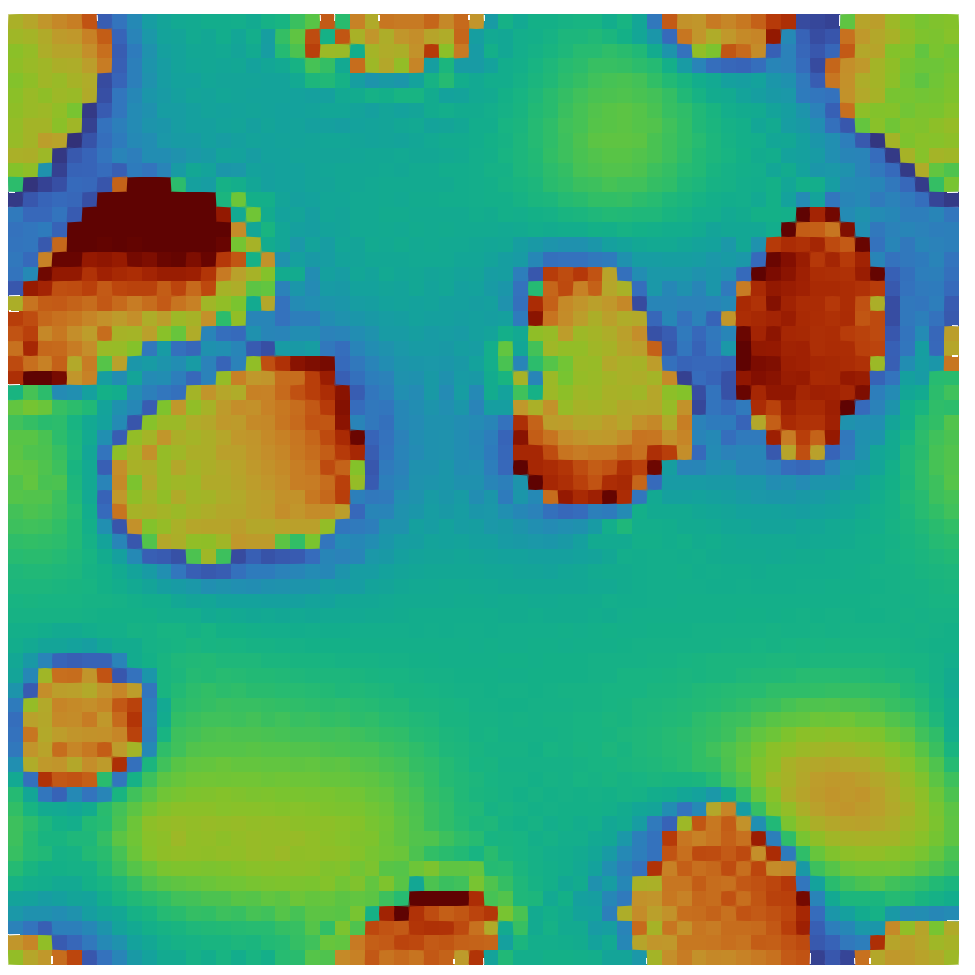}
         \subcaption{Q1R}
    \end{subfigure}    
    \begin{subfigure}[b]{0.30\textwidth}
        \centering 
        \includegraphics[width=\textwidth]{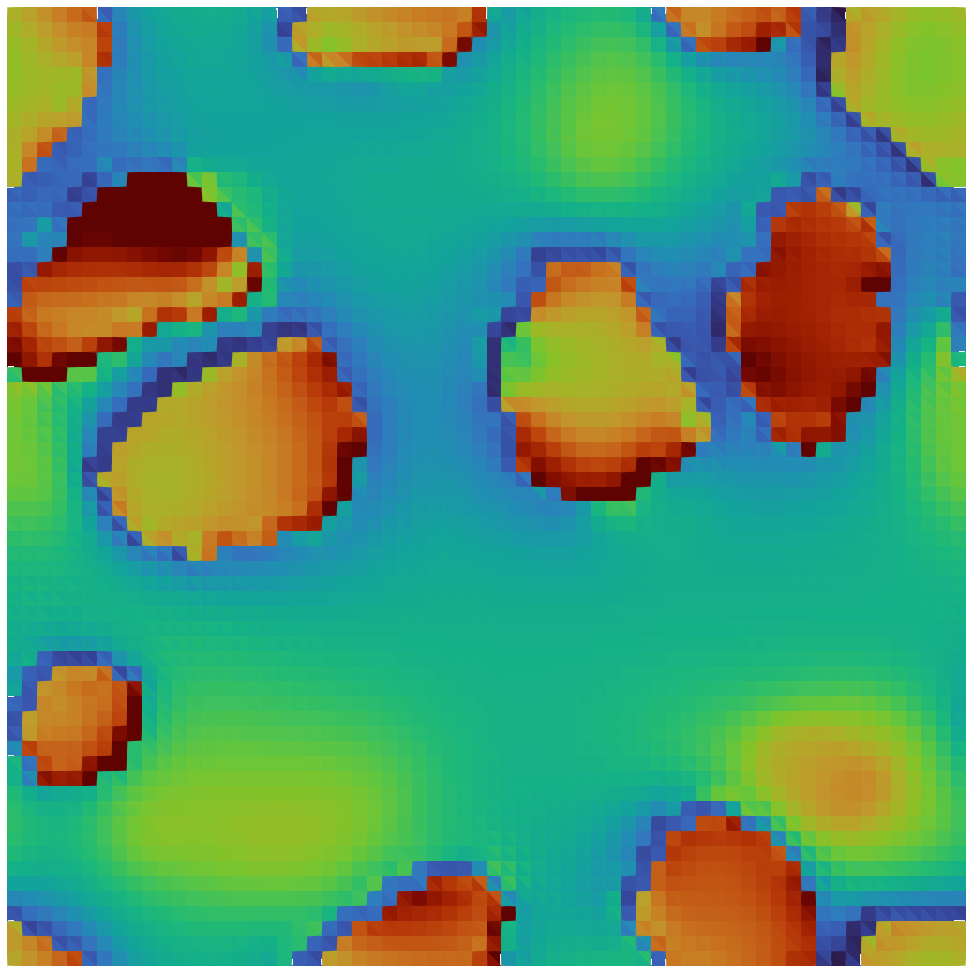}
         \subcaption{P1}
    \end{subfigure}
    \begin{subfigure}[b]{0.075\textwidth}
        \centering 
        \includegraphics[width=\textwidth]{color_bar_grains.pdf}
    \end{subfigure}
    \caption{Slice in the y-z plane of the local stress field at voxel count \mbox{$N=64$} for the rock-cement microstructure.}
    \label{fig:localGrains}
\end{figure}

Fig.~\ref{fig:EffStressGrain} shows the xx-component of the effective stress. We observe that the P1 discretization strongly overestimates the effective stress at low resolutions, followed by the Q1R discretization. The X-FEM discretization and the CoVo discretization approximate the xx-component of the effective stress the most accurately at the lowest resolution considered. As the voxel count per axis increases, the X-FEM discretization appears to converge faster in the effective stress than the CoVo discretization.
To investigate the accuracy of the effective stress approximations more closely, we consider the relative error in the xx-component of the effective stress with reference to the X-FEM result at voxel count \mbox{$N=1200$} in Fig~\ref{fig:ErrEffStressGrain}. Other reference solutions are possible, yet come with disadvantages. To retain the fluidity of reading, these considerations are outsourced to Apx.~\ref{apx:reference}.
We observe that at low resolutions, the CoVo discretization approximates the xx-component of the effective stress the most accurately. However, for voxel counts greater than $128$ voxels per axis, the X-FEM discretization approximates the xx-component of the effective stress the closest. Moreover, for X-FEM, we observe a quadratically decreasing error with increasing resolution, whereas for the CoVo, the P1 and the Q1R discretization the decrease in the error is linear. The convergence rates confirm those identified for Hashin's neutral inclusion in section~\ref{sec:hashin}.

We qualitatively analyze the local accuracy at the voxel count \mbox{$N=64$}. For each discretization, the local fields are resolved and interpolated using the finite elements discussed in Tab.~\ref{tab:Discretizations}. For X-FEM, we show the interpolation based on the additional shape functions in the subtetrahedra used for the numerical integration. In Fig.~\ref{fig:localGrains}, the xx-component of the local stress field is analyzed at the location that is indicated by the pink box in Fig.~\ref{fig:referenceGrain}. The reference solution is provided by an X-FEM discretization at voxel count \mbox{$N=1200$} and is indistinguishable across all other discretizations at the same resolution. In Fig.~\ref{fig:localGrains}, we observe that the X-FEM discretization at voxel count \mbox{$N=64$} approximates the reference solution the most closely. Due to the lower resolution, the X-FEM discretization at voxel count \mbox{$N=64$} does not capture all the details of the rock inclusion, but it detects the general shape and accurately measures the stress magnitude, except in some small areas. The CoVo, Q1R and P1 discretization do not detect the outlines as accurately as the X-FEM discretization. In addition, the Q1R discretization shows checkerboarding, i.e., solution artifacts known from the literature~\cite{willot2015fourier,leuschner2018fourier,ladecky_optimal_2023}. As the CoVo discretization at hand uses the Q1R discretization in voxels that do not contain an interface it is not surprising that for the CoVo discretization slight solution artifacts are observed as well.

For the conforming Galerkin discretizations, namely, the X-FEM and the P1 discretization, inequality~\eqref{eq:upperBoundOnEffStiffnessError} holds, such that the error in the local strain and stress fields is bounded from above by the square root of the error in the effective stress. As we observed in Fig~\ref{fig:ErrEffStressGrain} that for P1 the error in the effective stress decreases with the rate $\mathcal{O}(h)$ with decreasing mesh parameter, the local stress must decreases with the rate $\mathcal{O}(\sqrt{h})$ with decreasing mesh parameter. For X-FEM, the error in the effective stress decreases at the rate $\mathcal{O}(h^2)$, implying that the error in the local stress decreases at the rate $\mathcal{O}(h)$ as the mesh parameter decreases. This rate is identical to the rate of interface-conforming FEM with linear shape functions~\cites[Theorem 5.4.8]{brenner_mathematical_2008}[p.190]{hughes_finite_1987}. Therefore, our X-FFT solver achieves interface-conforming accuracy also for smooth rock inclusions in a cement matrix.

\subsubsection*{Numerical efficiency}
    \begin{figure}[htb]
    \centering
       \fbox{\includegraphics[width=0.7\textwidth]{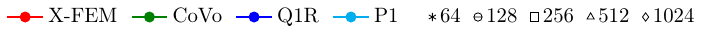}}
        \begin{subfigure}[b]{0.32\textwidth}
        \includegraphics[width=\textwidth]{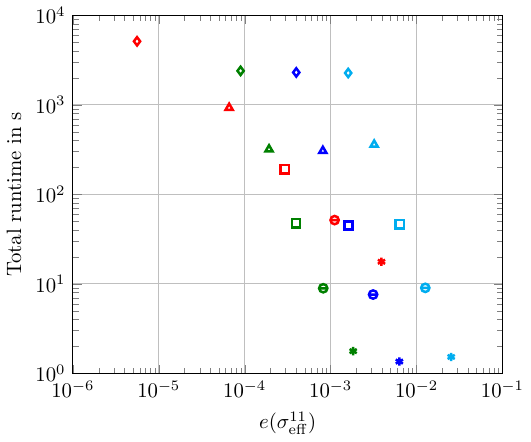}
        \caption{Total rutime}
        \label{fig: TimeGrains}
        \end{subfigure}
        \begin{subfigure}[b]{0.32\textwidth}
        \includegraphics[width=\textwidth]{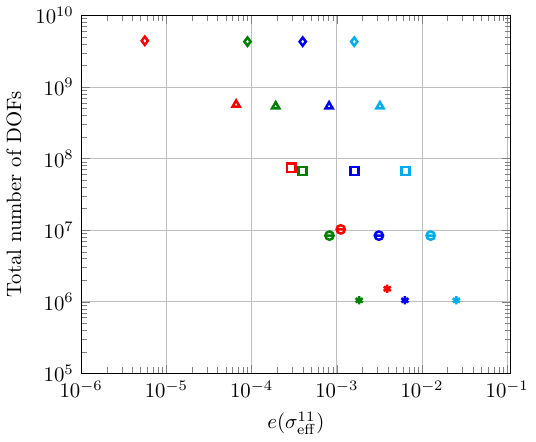}
        \caption{Degrees of freedom}
        \label{fig: DOFGrains}
        \end{subfigure}
        \caption{Numerical efficiency for the rock-cement microstructure.}
    \end{figure}

In Fig.~\ref{fig:iterationsGrains}, we observe that the CoVo, the P1 and the Q1R discretization require an iteration count of 16-17 iterations in linear CG until the residual~\eqref{eq:convergence} reaches the desired tolerance of $10^{-5}$. In contrast, the X-FEM discretization shows an iteration count between 30-33 iterations, i.e, approximately twice as many iterations are required to reach the desired tolerance compared to the other discretizations.

We show the total runtime of our implementation versus the accuracy of the effective axial stress $\sigma^{11}_\mathrm{eff}$ in Fig.~\ref{fig: TimeGrains}. At low resolutions, X-FEM comes with a runtime which is up to ten times higher compared to the other discretizations. At high resolutions, the time difference between X-FEM and the other discretizations reduces to a factor of two. The underlying reasons for the increased runtime of X-FEM are the same as the ones provided in the discussion following Fig.~\ref{fig: TimeHashin}. At low desired accuray, the CoVo discretization offers a cheaper solution in terms of runtime than X-FEM. Still, to reach an error in the effective stress that is below 0.01\%, X-FEM offers the fastest solution of all considered discretizations: 927s for $512^3$ voxels.

To assess the effort in terms of \emph{dofs}, we show in Fig.~\ref{fig: DOFGrains} the total number of \emph{dofs} per field versus the relative error in the xx-component of the effective stress. We observe that for a desired accuracy of 0.1\% the CoVo discretization at \mbox{$N=64$} offers the cheapest solution in terms of \emph{dofs}. To reach an accuracy below 0.01\%, the X-FEM discretization at \mbox{$N=512$} offers the cheapest solution in terms of \emph{dofs}. Thus, for the smooth rock inclusion microstructure the question of which discretization offers the best balance of accuracy and efficiency depends on the desired level of accuracy. However, the results of this section confirm that our X-FFT solver may achieve interface-conforming accuracy also for complex microstructures with smooth inclusions while maintaining numerical efficiency.

\subsection{Long fiber reinforced composite}

\subsubsection*{Microstructure}
\begin{figure}[htb]
    \centering
        \begin{subfigure}[b]{0.48\textwidth}
        \centering 
        \includegraphics[width=0.82\textwidth]{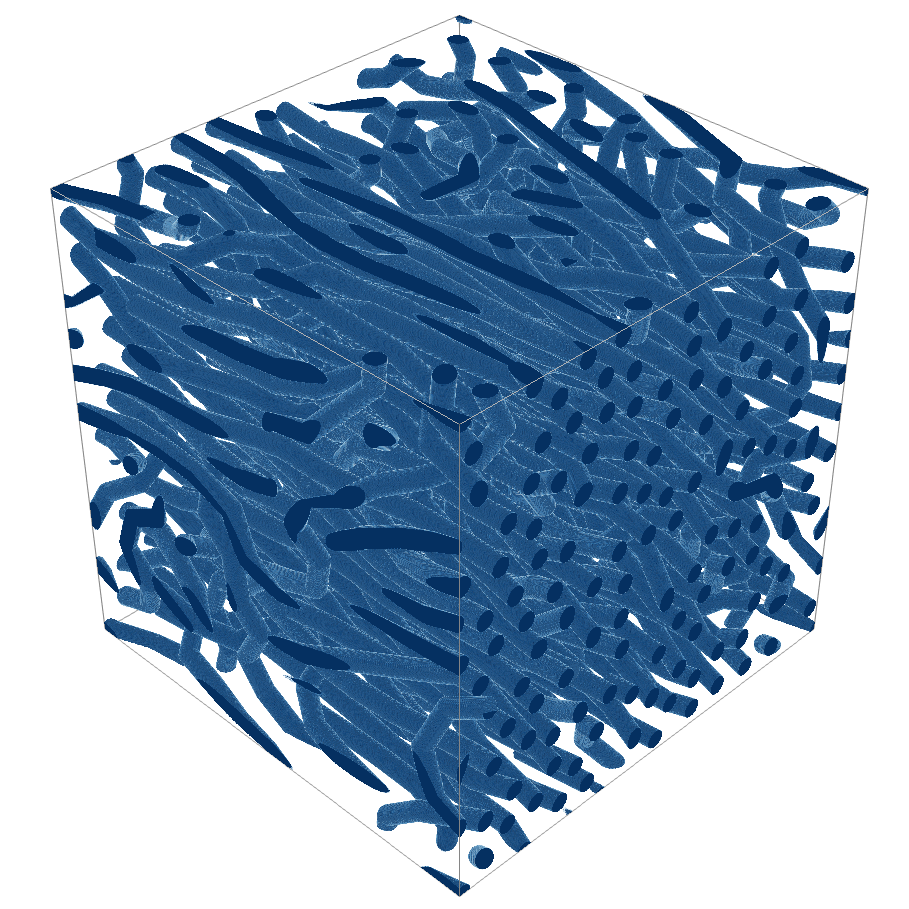}
        \includegraphics[width=0.16\textwidth]{axes.png}
         \subcaption{Geometry}\label{fig:geometryFiber}
    \end{subfigure}
        \begin{subfigure}[b]{0.48\textwidth}
        \centering 
          \includegraphics[width=0.82\textwidth]{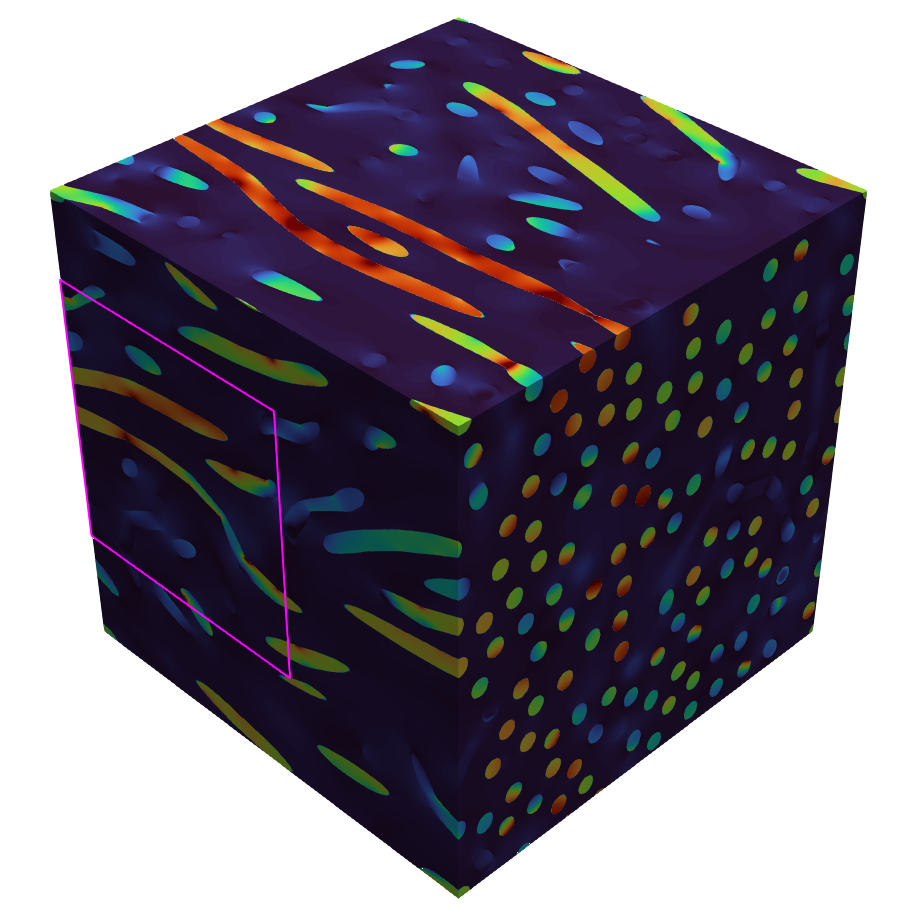}
          \includegraphics[width=0.16\textwidth]{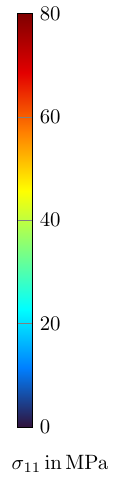}
         \subcaption{Reference}\label{fig:referenceFiber}
    \end{subfigure}
    \caption{Geometry and reference solution of the long fiber reinforced composite.}
\end{figure}

In this paragraph, we examine a microstructure with non-smooth interfaces, namely a long fiber reinforced composite microstructure. We use the fused sequential addition and migration method of Lauff et al.~\cite{lauff2024generating,lauff2025microstructure} to generate the microstructure shown in Fig.~\ref{fig:geometryFiber}. We assume a cell length of 64$\mu$m, a fiber diameter of 2.5$\mu$m, a fiber length of 252.96$\mu$m and use exact closure from the second-order fiber orientation tensor $\mathrm{diag}(0.77,0.17,0.6)$. The reached fiber volume fraction is 17.93 vol-$\%$, the average fiber curvature is 0.0718$\mu\text{m}^{-1}$ and the maximum bending angle is 45.80°. We consider a polypropylene matrix and E-glass fibers with the material parameters given in Tab.~\ref{tab:materialFiber} and prescribe unidirectional loading with the strain $\bar{\feps}_{11}=0.1\%$. All computations of this study were run on 16 threads using a solver tolerance of $10^{-5}$.

\begin{table}[htb]
        \centering
        \begin{tabular}{l|ll}
          & Polypropylene matrix&E-glass fibers\\
          \hline
        Young's modulus&1.25 GPa&72 GPa\\
        Poisson's ratio&0.35&0.22
        \end{tabular}
        \caption{Material parameters~\cite{dow2003c711,glass2013tufrov} for the long fiber reinforced composite.}
        \label{tab:materialFiber}
\end{table}

\subsubsection*{Solution scheme}

\begin{figure}[htb]
  \centering
    \begin{subfigure}[c]{0.40\textwidth}
        \centering 
        \includegraphics[width=\textwidth]{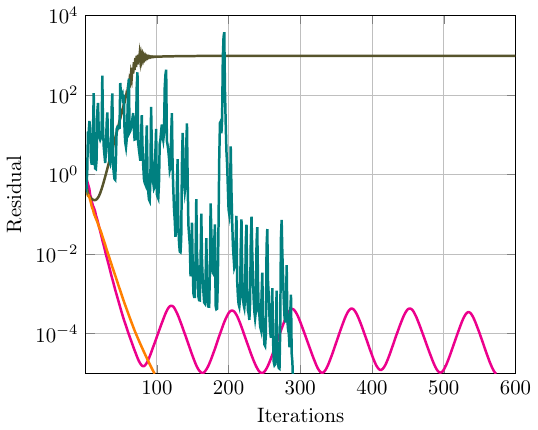}
    \end{subfigure}
    \begin{subfigure}[c]{0.18\textwidth}
        \fbox{\includegraphics[width=\textwidth]{legendSolvers.pdf}}\\
    \end{subfigure}
\caption{Solver convergence for X-FEM at voxel count \mbox{$N=128$} for the long fiber reinforced composite.}
\label{fig:solverLongFiber}
\end{figure}

To select the solution scheme for this section, we compare the convergence of linear CG~\cite{zeman2010accelerating}, the basic scheme~\cite{moulinec1994fast,moulinec1998numerical}, the Barzilai-Borwein scheme~\cite{schneider2019barzilai}, and the nonlinear CG scheme~\cite{schneider2020dynamical}, for X-FEM at voxel count \mbox{$N=128$}. Fig.~\ref{fig:solverLongFiber} shows the convergence of the residual~\eqref{eq:convergence} with the iteration count of each solver.
We observe that the residual of the basic scheme increases after reaching a minimum value and remains constant after approximately 100 iterations. Thus, the choice of the step size~\eqref{eq:stepsize} for the basic scheme is not recommended. The Barzilai-Borwein scheme shows its characteristic non-monotonic convergence behavior and reaches the desired tolerance of $10^{-5}$ at iteration count 291. The linear CG scheme shows some sinusoidal behavior and requires 574 iterations to reach the desired tolerance. In contrast, the nonlinear CG scheme converges monotonically and reaches the desired tolerance after 98 iterations. Although we still consider linear elasticity, the non-smooth fiber geometry appears to affect the conditioning of the X-FEM system so that nonlinear CG shows faster convergence than linear CG for the long fiber reinforced composite. Therefore, we select nonlinear CG as the solver for this section.

\subsubsection*{Accuracy}

\begin{figure}
    \centering 
	\fbox{\includegraphics[width=.6\textwidth]{legendGrain.pdf}}\\
        \begin{subfigure}[b]{0.32\textwidth}
        \centering 
        \includegraphics[width=\textwidth]{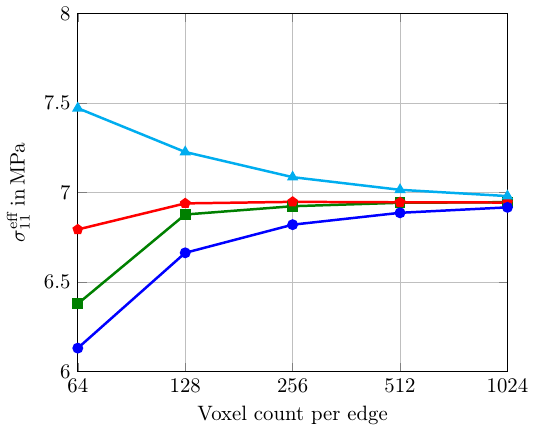}
         \subcaption{Effective stress}
         \label{fig:EffStressLongFiber}
    \end{subfigure}
    \begin{subfigure}[b]{0.32\textwidth}
        \centering 
        \includegraphics[width=\textwidth]{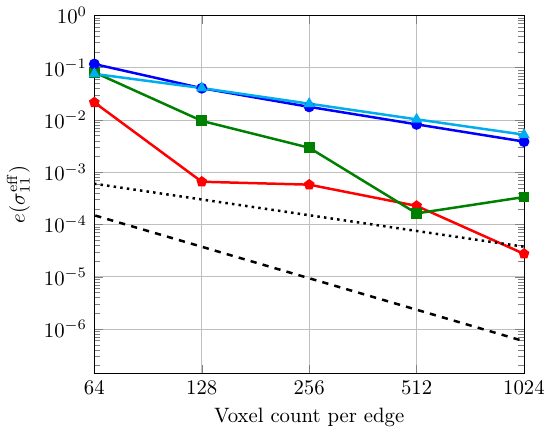}
         \subcaption{Error in the effective stress}
         \label{fig:ErrEffStressLongFiber}
    \end{subfigure}
    \begin{subfigure}[b]{0.32\textwidth}
        \centering 
        \includegraphics[width=0.95\textwidth]{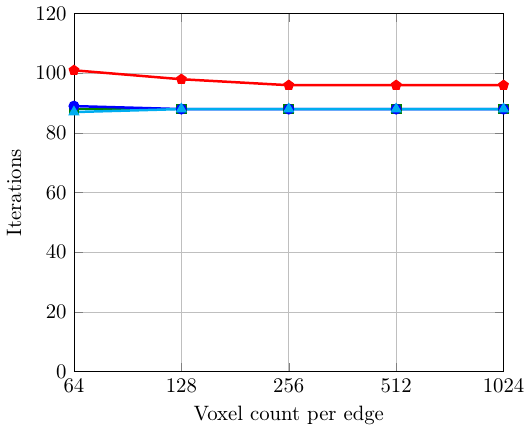}
         \subcaption{Iteration count}
         \label{fig:IterationsLongFiber}
    \end{subfigure}
    \caption{Effective stress and iteration count for the long fiber reinforced composite.}
\end{figure}

\begin{figure}
    \centering  
    \begin{subfigure}[b]{0.3\textwidth}
        \centering 
          \includegraphics[width=\textwidth]{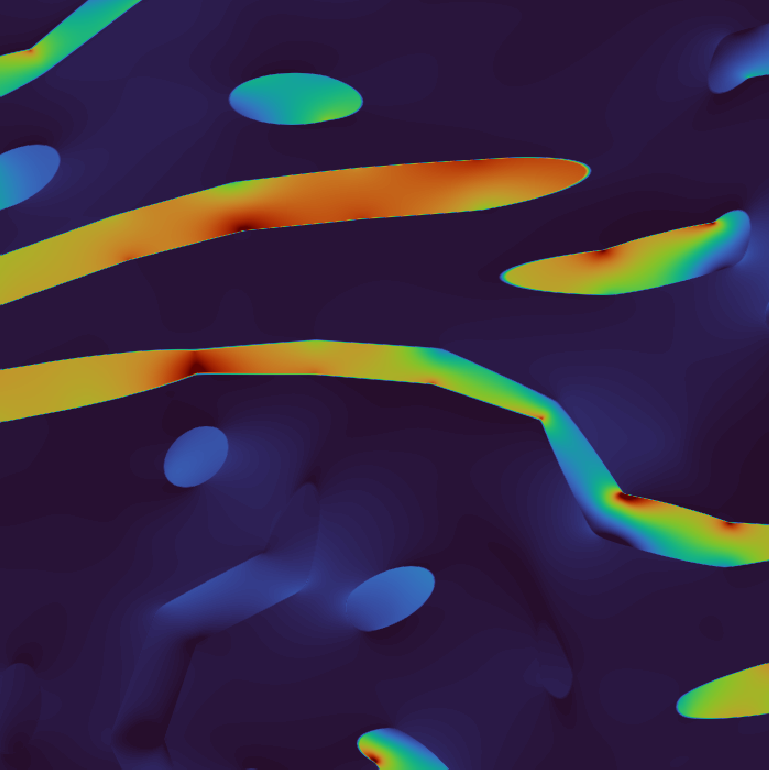}
         \subcaption{Reference}
    \end{subfigure}
        \begin{subfigure}[b]{0.3\textwidth}
        \centering 
        \includegraphics[width=\textwidth]{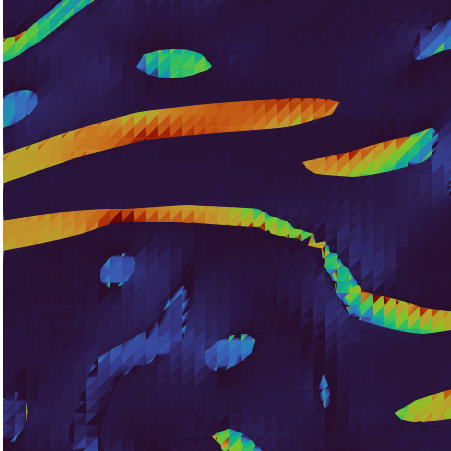}
         \subcaption{X-FEM}
    \end{subfigure}
    \begin{subfigure}[b]{0.30\textwidth}
        \centering 
        \includegraphics[width=\textwidth]{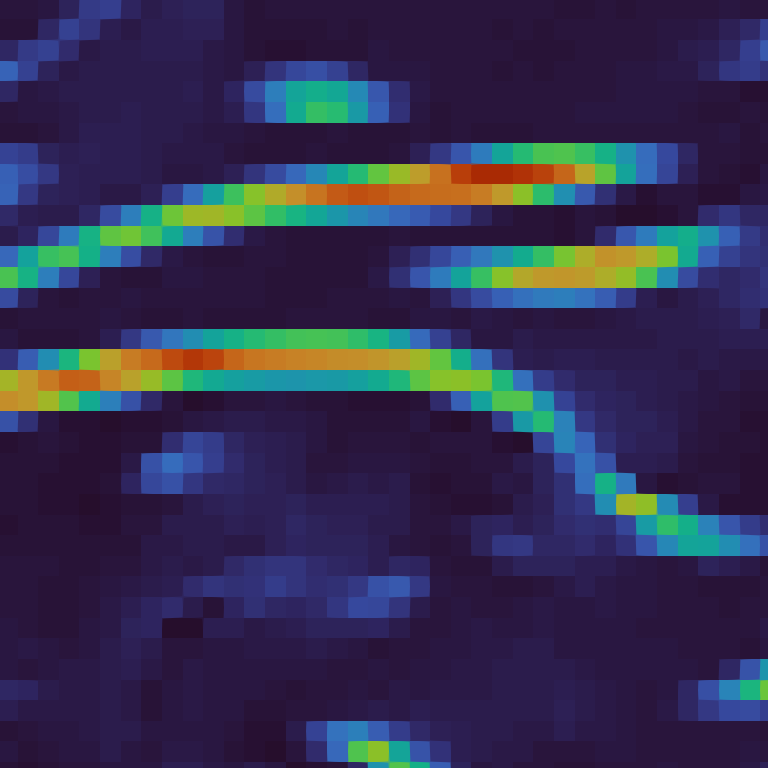}
         \subcaption{CoVo}
    \end{subfigure} 
    \begin{subfigure}[b]{0.30\textwidth}
        \centering 
        \includegraphics[width=\textwidth]{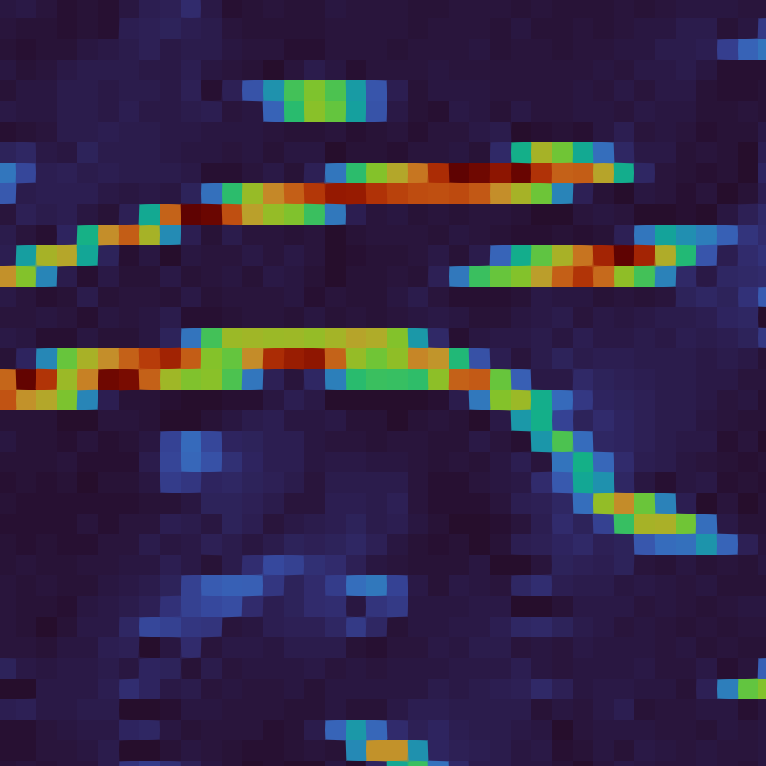}
         \subcaption{Q1R}
    \end{subfigure}    
    \begin{subfigure}[b]{0.30\textwidth}
        \centering 
        \includegraphics[width=\textwidth]{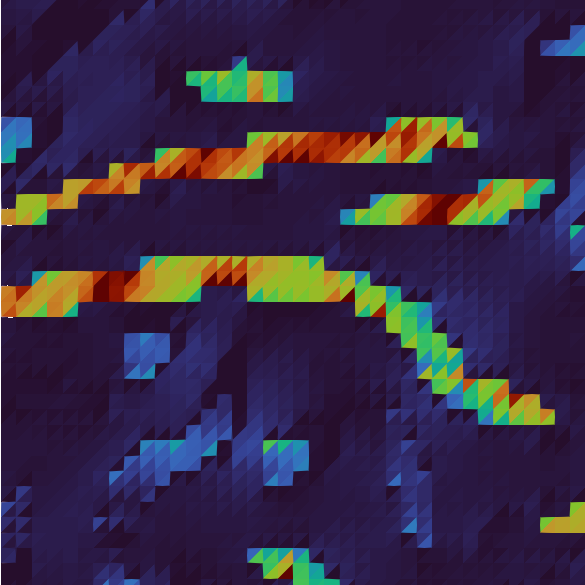}
         \subcaption{P1}
    \end{subfigure}
    \begin{subfigure}[b]{0.075\textwidth}
        \centering 
        \includegraphics[width=\textwidth]{color_bar_longFiber.pdf}
    \end{subfigure}
    \caption{Zoomed view of local stress field at voxel count \mbox{$N=64$} for the long fiber reinforced composite.}
     \label{fig:localLongFiber}
\end{figure}

The xx-component of the effective stress is shown in Fig.~\ref{fig:EffStressLongFiber}. X-FEM appears to most accurately approximate the xx-component of the effective stress overall. For a more detailed analysis we show in Fig.~\ref{fig:ErrEffStressLongFiber} the relative error in the xx-component of the effective stress, where we select X-FEM at voxel count \mbox{$N=1200$} as reference solution. For details on the choice of the reference solution, we refer to Appendix~\ref{apx:reference}.
We observe that X-FEM achieves the lowest error at most resolutions, followed by CoVo. P1 and Q1R yield the highest errors. As expected, the errors of P1, Q1R, and CoVo converge linearly. However, the X-FEM errors also converge less than quadratically for the long fiber reinforced composite. This subquadratic convergence rate in X-FEM is due to the non-smooth fiber geometry. Even when refining the mesh, the modified abs enrichment may not accurately approximate the non-smooth fiber ends. For X-FEM with the modified abs enrichment, we expect the error in the xx-component of the effective stress to increase with the number of non-smooth geometries in the microstructure. 

To evaluate the quality of the local fields, we assess the xx-component of the local stress field for each discretization and compare it to a reference solution provided by X-FEM with at voxel count \mbox{$N=1200$}. The pink box in the reference solution, shown in Fig.~\ref{fig:referenceFiber}, encloses the image section that we analyze below.
In Fig.~\ref{fig:localLongFiber}, we observe that X-FEM with a voxel count \mbox{$N=64$} approximates the local stress field of the reference solution rather closely. In contrast, CoVo significantly underestimates the xx-component of the local stress and offers only a crude approximation of the fibers. Furthermore, CoVo and Q1R display checkerboarding artifacts, as discussed in Fig.~\ref{fig:localGrains}. For Q1R, the stress values are in the correct order of magnitude, but the fiber geometry is not approximated correctly. The latter is also visible for P1 combined with an overestimation of the xx-component of the local stress in the fibers.

\subsubsection*{Numerical efficiency}
\begin{figure}[h]
    \centering
       \fbox{\includegraphics[width=0.7\textwidth]{legendDOFGrains.pdf}}
    \begin{subfigure}[b]{0.32\textwidth}
        \includegraphics[width=\textwidth]{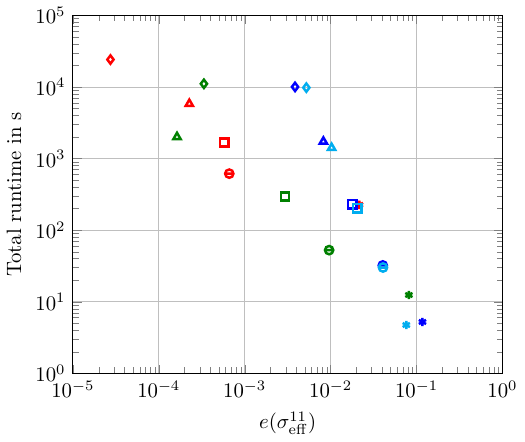}
        \caption{Total runtime}
        \label{fig: TimeLongFiber}
        \end{subfigure}
        \begin{subfigure}[b]{0.32\textwidth}
        \includegraphics[width=\textwidth]{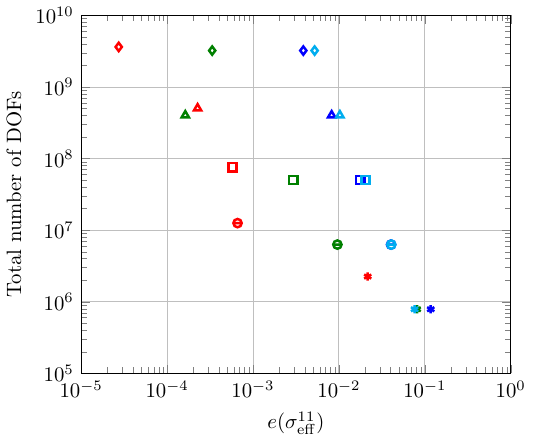}
        \caption{Degrees of freedom}
        \label{fig: DOFLongFiber}
        \end{subfigure}
        \caption{Numerical efficiency for the long fiber reinforced composite microstructure.}
    \end{figure}

The iteration count until convergence of nonlinear CG is shown in Fig.~\ref{fig:IterationsLongFiber}. We observe that P1, Q1R and CoVo require 87-89 iterations to converge. For X-FEM, an increased iteration count of 98-101 iterations is observed. We note that the iteration count of X-FEM could be reduced by improving the selection of the step size~\eqref{eq:stepsize}.

The total runtime versus the error in the effective stress $\sigma^{11}_\mathrm{eff}$ is shown in Fig.~\ref{fig: TimeLongFiber}. We observe an increased runtime of X-FEM compared to the other discretizations, similar to what was observed for the previous examples. However, for a desired error below 0.1\%, X-FEM offers the fastest solution with a runtime of 615s at resolution \mbox{$N=128$}.

In Fig.~\ref{fig: DOFLongFiber} the computational effort in terms of degrees of freedom (\emph{dofs}) per field as a measure of the memory footprint is analyzed. We observe that the X-FEM discretization at voxel count
\mbox{$N=128$} provides the most memory-efficient solution for achieving a desired accuracy of 0.1\%. Furthermore, to reach an accuracy below 0.01\%, the X-FEM discretization at voxel count
\mbox{$N=1024$} offers the most efficient solution in terms of \emph{dofs}.
Overall, the X-FEM method approximates the local and effective stress fields the most accurately for the long fiber-reinforced composite in question. However, we found that non-smooth interfaces can degrade the conditioning of the X-FEM system and reduce its accuracy.

\newpage
\subsection{Single cubic inclusion}

\subsubsection*{Microstructure}
\begin{figure}[htb]
    \centering
        \begin{subfigure}[b]{0.4\textwidth}
        \centering 
        \includegraphics[width=\textwidth]{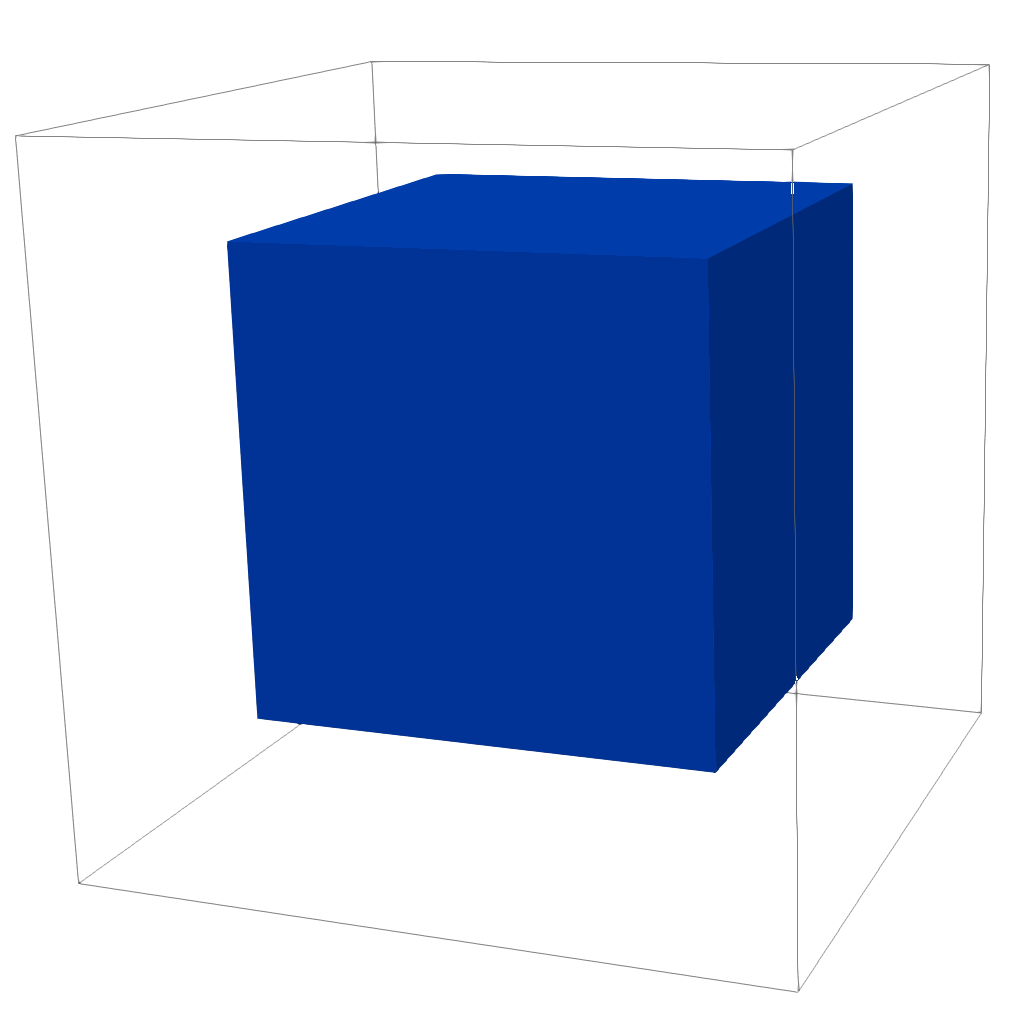}
         \subcaption{Inclusion parallel to axes}
    \end{subfigure} 
    \begin{subfigure}[b]{0.08\textwidth}
        \includegraphics[width=\textwidth]{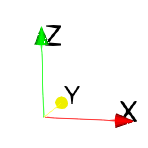}
    \end{subfigure}       
    \begin{subfigure}[b]{0.4\textwidth}
        \centering 
        \includegraphics[width=\textwidth]{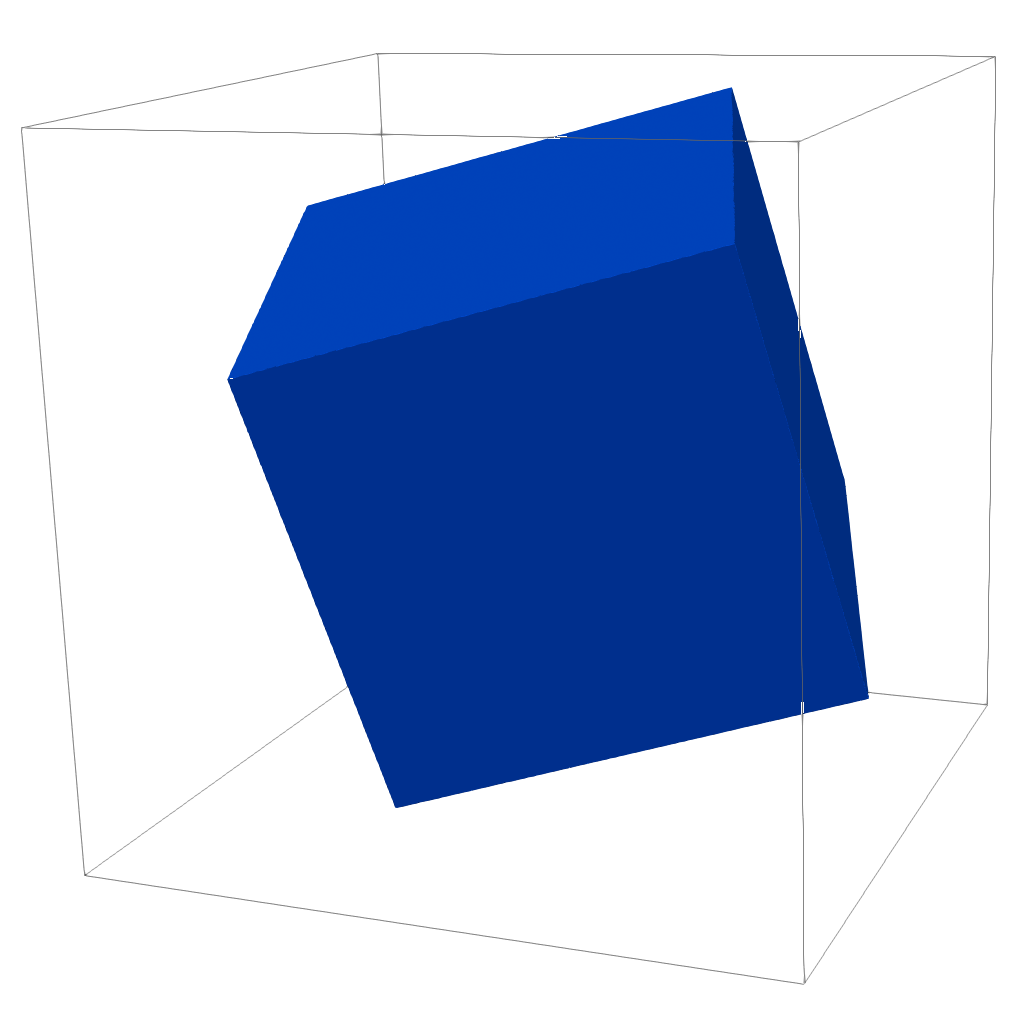}
         \subcaption{Inclusion rotated by 15° along each axis}
    \end{subfigure}
    \caption{Geometry of the cubic inclusion configurations.}
    \label{fig:geometryCuboid}
\end{figure}

Our long fiber reinforced composite study indicates that non-smooth interfaces affect both the conditioning of the X-FEM system and the accuracy of the X-FEM discretization negatively. For a closer analysis of this effect, we consider a single cubic inclusion that is non-smooth at the corner points in this section.
We assume a cubic cell with an edge length of $128\,\mu$m that contains a single cubic inclusion with an edge length of $82.7854\,\mu$m, which is centered at the point $67.4142(\fe_\mathrm{x}+\fe_\mathrm{y}+\fe_\mathrm{z})$. Two configurations are considered: 
\begin{enumerate}
    \item The cubic inclusion is parallel to the axes and thus parallel to the mesh.
    \item The cubic inclusion is rotated at its center by 15° along each axis.

\end{enumerate}
The geometries of both configurations are shown in Fig.~\ref{fig:geometryCuboid}. We furnish the matrix with the material parameters of polyamide, whereas the inclusions are assumed to be made of E-glass. We re-use the material parameters of E-glass shown in Tab.~\ref{tab:materialFiber} and use a Young's modulus of $2.1\,\mathrm{GPa}$ and a Poisson's ratio of $0.22$ for the polyamide~\cite{doghri2011second}. We prescribe unidirectional loading with the strain $\bar{\feps}_{11}=0.1\%$, use four threads per computation and set the solver tolerance to $10^{-5}$.

\subsubsection*{Solution scheme}

\begin{figure}[htb]
  \centering
    \begin{subfigure}[c]{0.32\textwidth}
        \centering 
        \includegraphics[width=\textwidth]{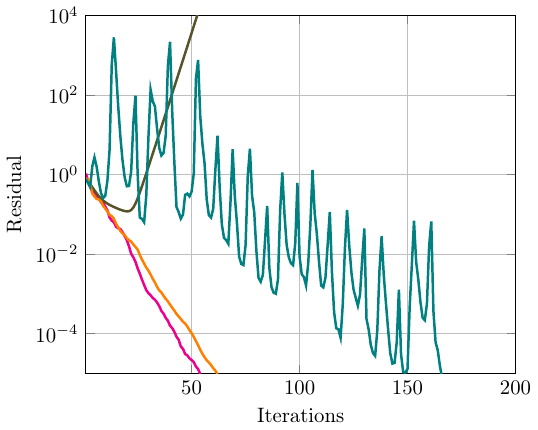}
         \subcaption{Parallel inclusion}
            \label{fig:solverCuboid}
    \end{subfigure}
        \begin{subfigure}[c]{0.32\textwidth}
        \centering 
        \includegraphics[width=\textwidth]{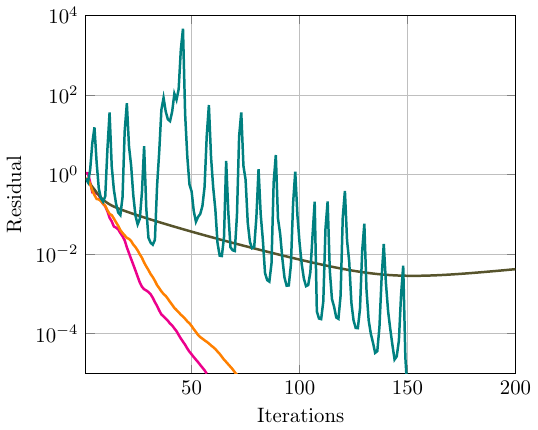}
         \subcaption{Rotated inclusion}
        \label{fig:solverRotatedCuboid}
    \end{subfigure}    
    \begin{subfigure}[c]{0.18\textwidth}
        \centering
        \fbox{\includegraphics[width=\textwidth]{legendSolvers.pdf}}
    \end{subfigure}
\caption{Solver convergence for X-FEM at voxel count \mbox{$N=128$} for the cubic inclusion.}
\end{figure}

We compare the convergence of linear CG~\cite{zeman2010accelerating}, the basic scheme~\cite{moulinec1994fast,moulinec1998numerical}, the Barzilai-Borwein scheme~\cite{schneider2019barzilai}, and the nonlinear CG scheme~\cite{schneider2020dynamical}, for X-FEM in both configurations of the single cubic inclusion at resolution \mbox{$N=128$}. 

In Fig.~\ref{fig:solverCuboid}, we show the convergence of the residual~\eqref{eq:convergence} versus the iteration count for the inclusion which is parallel to the axes. We observe that linear CG reaches the desired tolerance of $10^{-5}$ with the least number of iterations, closely followed by nonlinear CG. Barzilai-Borwein requires approximately a factor three more iterations and the residual of the basic scheme diverges indicating that the chosen step size is too large for the microstructure at hand. At other resolutions, similar trends to those discussed above are observed. Therefore, we use linear CG for the studies on the parallel cubic inclusion.

For the inclusion that is rotated at its center by 15° along each axis, we show the solver convergence in Fig.~\ref{fig:solverRotatedCuboid}. We observe that the residual of the basic scheme increases after reaching a minimum value, due to the non-optimal choice of the step size, which we already discussed for the other studies. The behavior of the Barzilai-Borwein scheme, the linear CG scheme, and the nonlinear CG scheme, is similar to that observed for the parallel inclusion study at all resolutions. The linear CG scheme requires the least number of iterations to reach the desired tolerance of $10^{-5}$. Therefore, for the rest of the manuscript, we use linear CG for the rotated cubic inclusion.

\subsubsection*{Accuracy}

\begin{figure}[htb]
    \centering 
	\fbox{\includegraphics[width=.6\textwidth]{legendGrain.pdf}}\\
        \begin{subfigure}[b]{0.32\textwidth}
        \centering 
        \includegraphics[width=\textwidth]{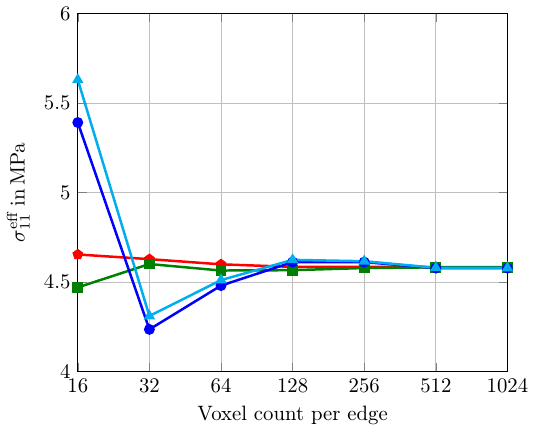}
         \subcaption{Effective stress}
         \label{fig:EffStressCuboid}
    \end{subfigure}
    \begin{subfigure}[b]{0.32\textwidth}
        \centering 
        \includegraphics[width=\textwidth]{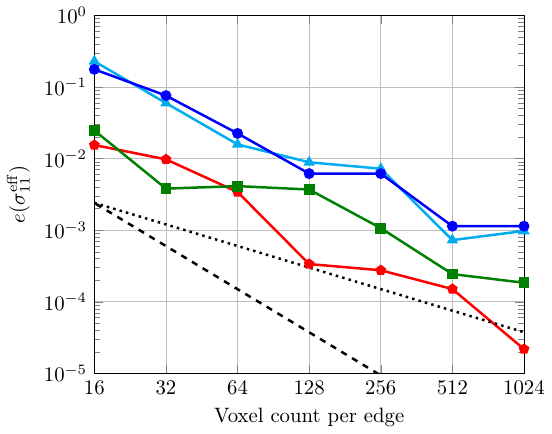}
         \subcaption{Error in the effective stress}
         \label{fig:ErrEffStressCuboid}
    \end{subfigure}
    \begin{subfigure}[b]{0.32\textwidth}
        \centering 
        \includegraphics[width=0.95\textwidth]{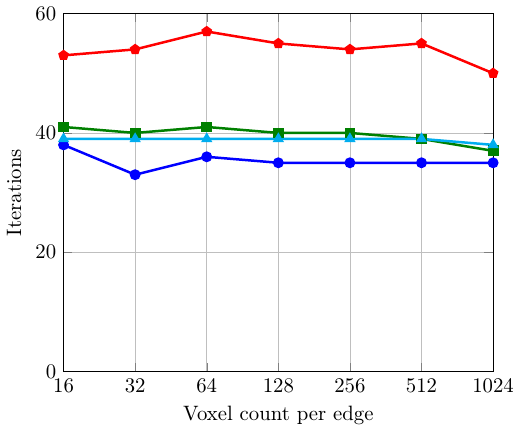}
         \subcaption{Iteration count}
         \label{fig:IterationsCuboid}
    \end{subfigure}
    \caption{Effective stress and iteration count for the parallel cubic inclusion.}
\end{figure}

\begin{figure}
    \centering  
    \begin{subfigure}[b]{0.3\textwidth}
        \centering 
          \includegraphics[width=\textwidth]{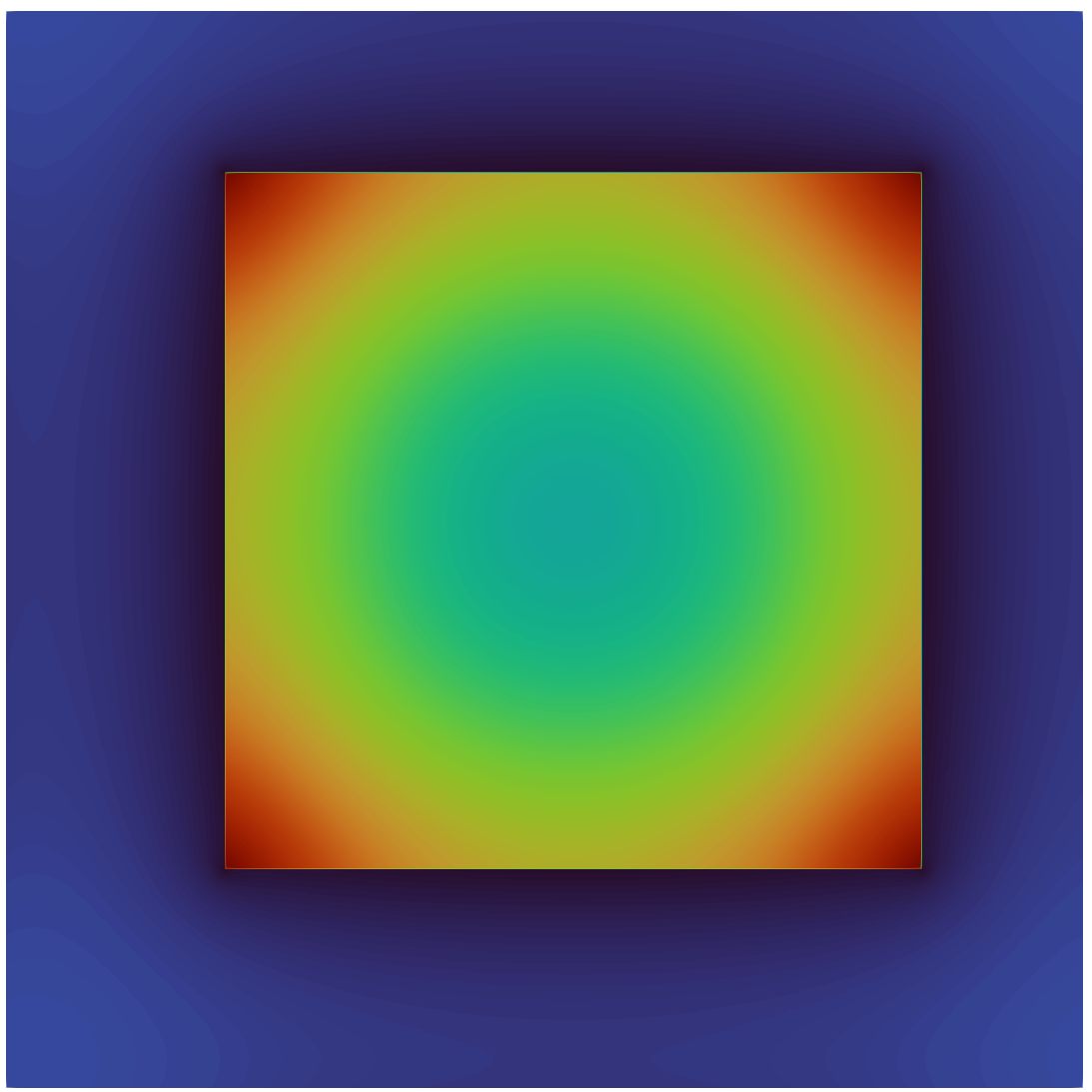}
         \subcaption{Reference}
    \end{subfigure}
        \begin{subfigure}[b]{0.3\textwidth}
        \centering 
        \includegraphics[width=\textwidth]{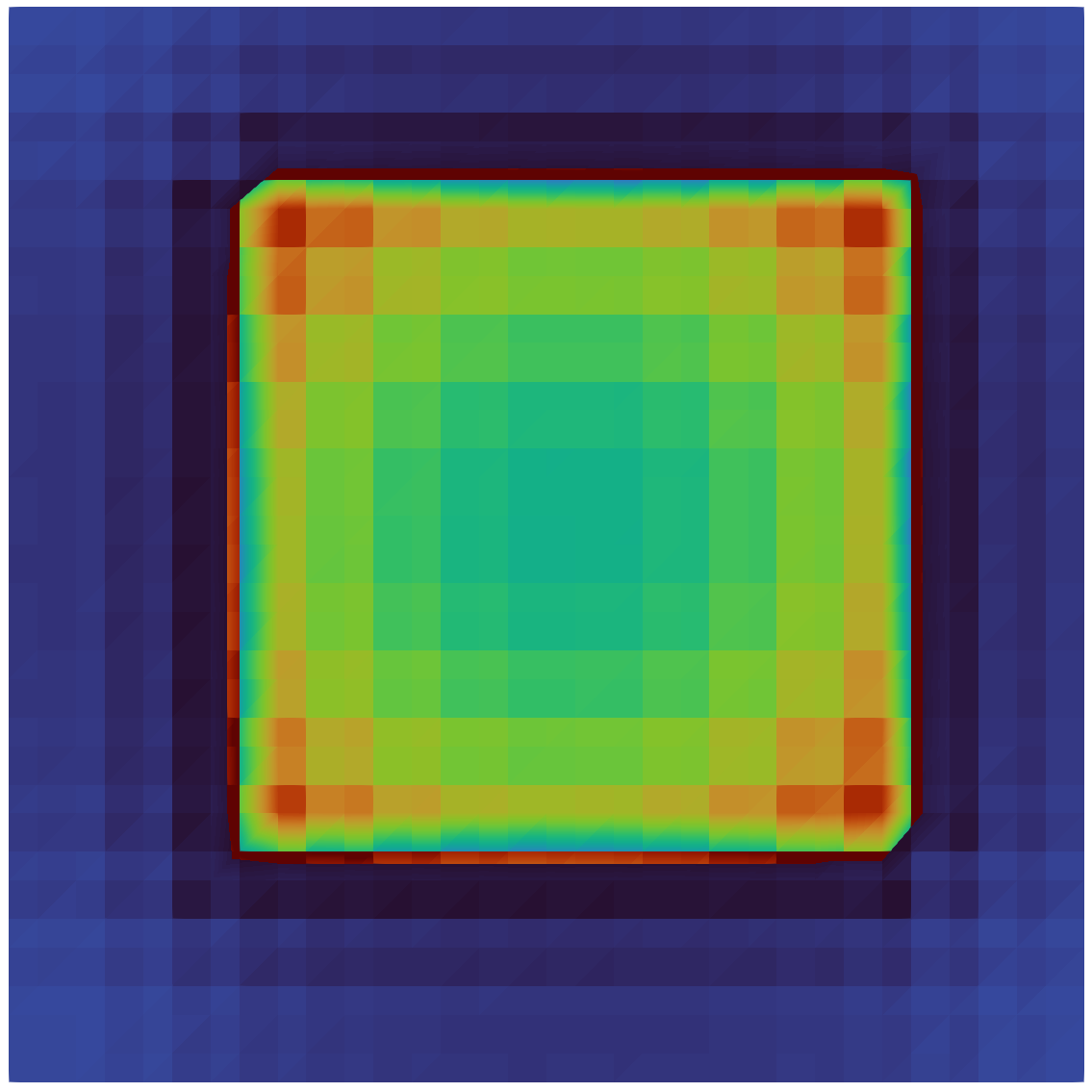}
         \subcaption{X-FEM}
    \end{subfigure}
    \begin{subfigure}[b]{0.30\textwidth}
        \centering 
        \includegraphics[width=\textwidth]{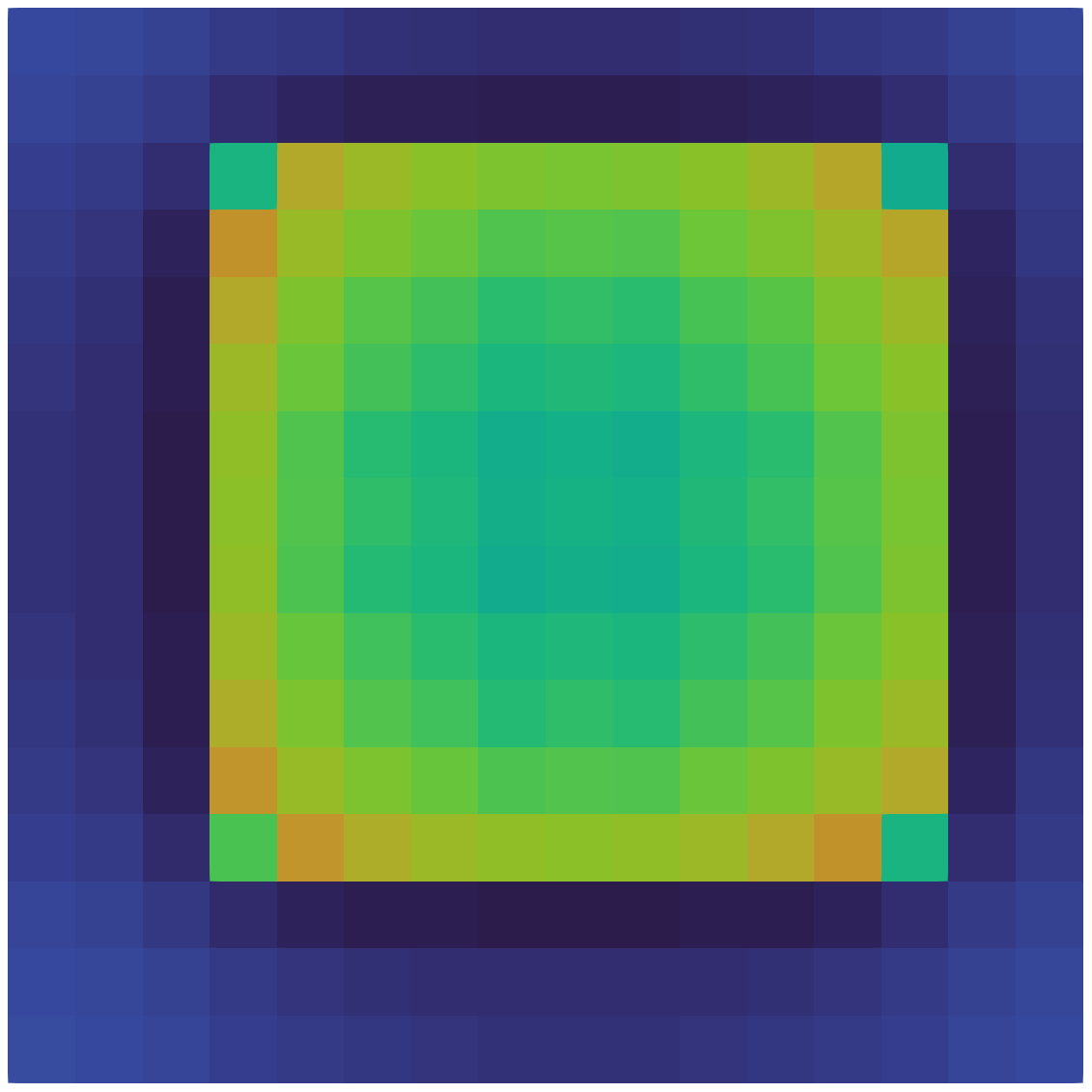}
         \subcaption{CoVo}
    \end{subfigure} 
    \begin{subfigure}[b]{0.30\textwidth}
        \centering 
        \includegraphics[width=\textwidth]{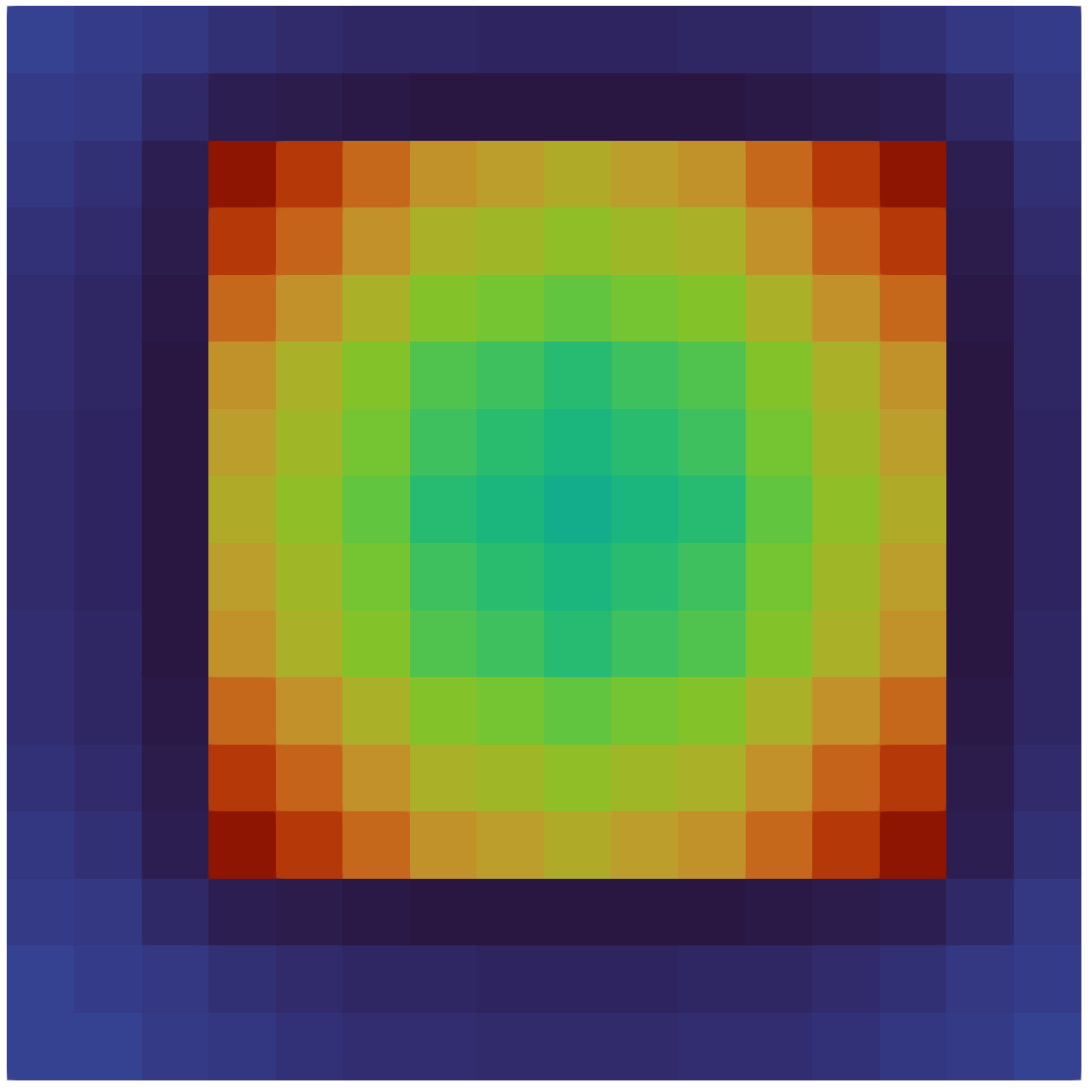}
         \subcaption{Q1R}
    \end{subfigure}    
    \begin{subfigure}[b]{0.30\textwidth}
        \centering 
        \includegraphics[width=\textwidth]{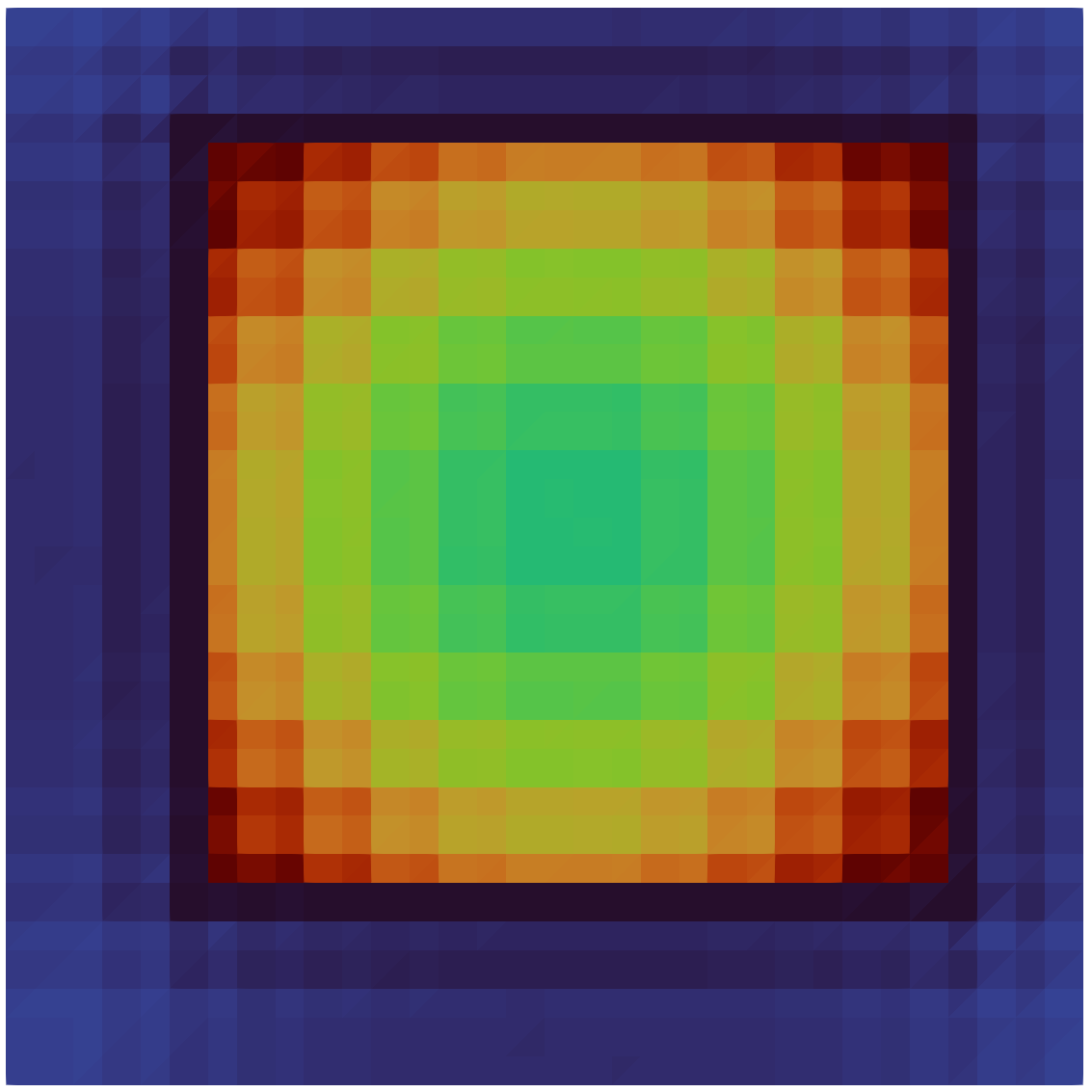}
         \subcaption{P1}
    \end{subfigure}
    \begin{subfigure}[b]{0.075\textwidth}
        \centering 
        \includegraphics[width=\textwidth]{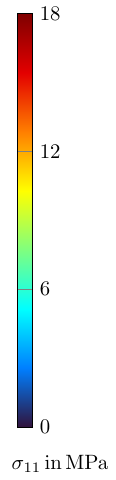}
    \end{subfigure}
    \caption{Local stress field at voxel count \mbox{$N=16$} for the parallel cubic inclusion.}
     \label{fig:localCuboid}
\end{figure}

\begin{figure}  
    \centering 
	\fbox{\includegraphics[width=.6\textwidth]{legendGrain.pdf}}\\
        \begin{subfigure}[b]{0.32\textwidth}
        \centering 
        \includegraphics[width=\textwidth]{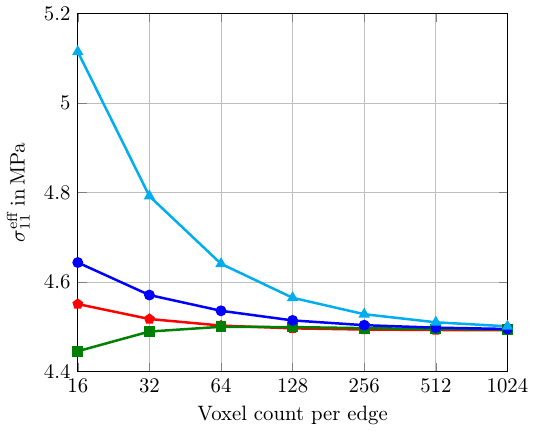}
         \subcaption{Effective stress}
         \label{fig:EffStressRotatedCuboid}
    \end{subfigure}
    \begin{subfigure}[b]{0.32\textwidth}
        \centering 
        \includegraphics[width=\textwidth]{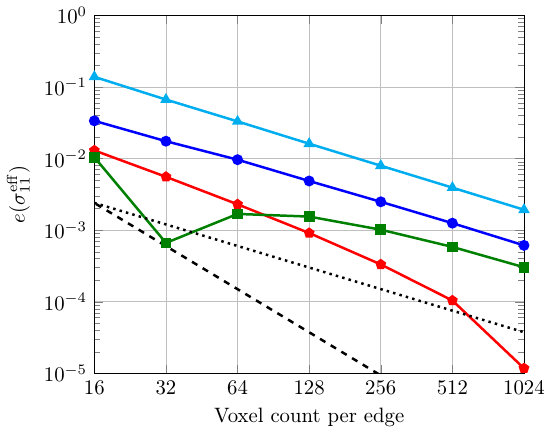}
         \subcaption{Error in the effective stress}
         \label{fig:ErrEffStressRotatedCuboid}
    \end{subfigure}
    \begin{subfigure}[b]{0.32\textwidth}
        \centering 
        \includegraphics[width=0.95\textwidth]{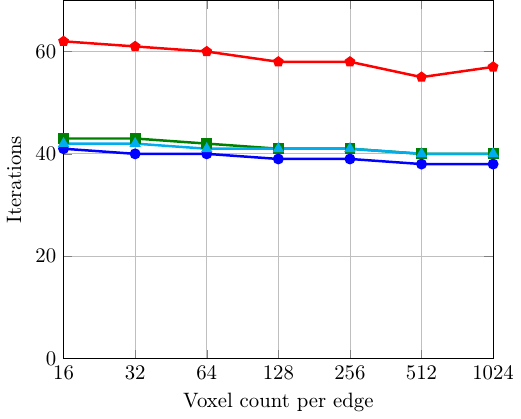}
         \subcaption{Iteration count}
         \label{fig:IterationsRotatedCuboid}
    \end{subfigure}
    \caption{Effective stress and iteration count for the rotated cubic inclusion.}
\end{figure}

\begin{figure}[htb] 
    \centering  
    \begin{subfigure}[b]{0.3\textwidth}
        \centering 
          \includegraphics[width=\textwidth]{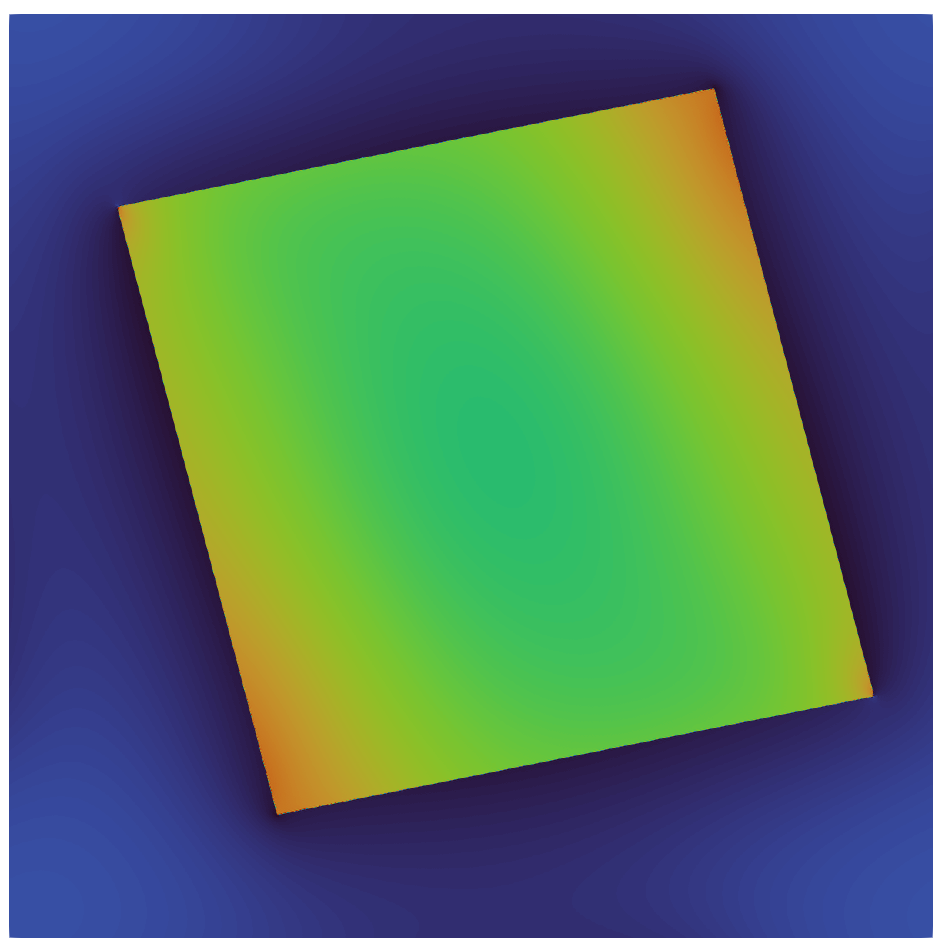}
         \subcaption{Reference}
    \end{subfigure}
        \begin{subfigure}[b]{0.3\textwidth}
        \centering 
        \includegraphics[width=\textwidth]{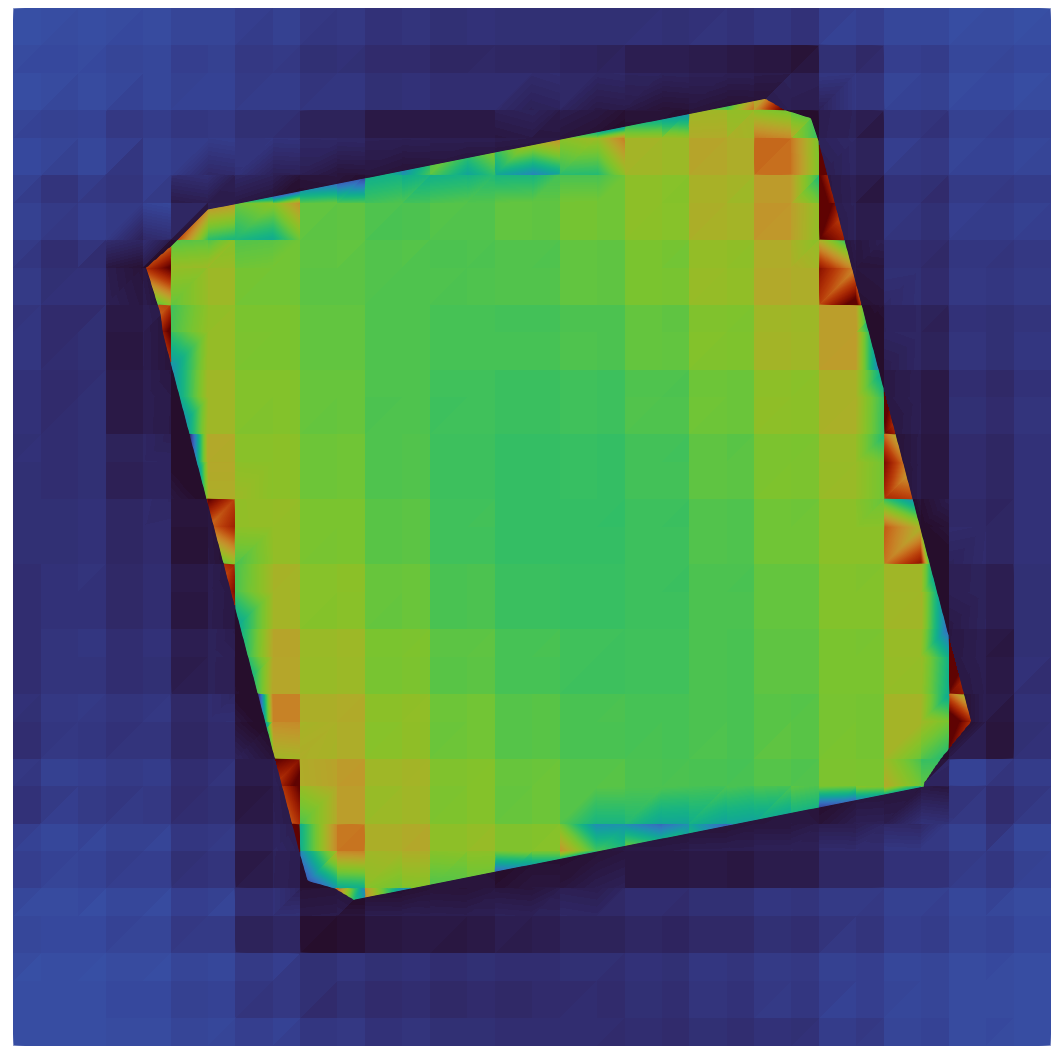}
         \subcaption{X-FEM}
    \end{subfigure}
    \begin{subfigure}[b]{0.30\textwidth}
        \centering 
        \includegraphics[width=\textwidth]{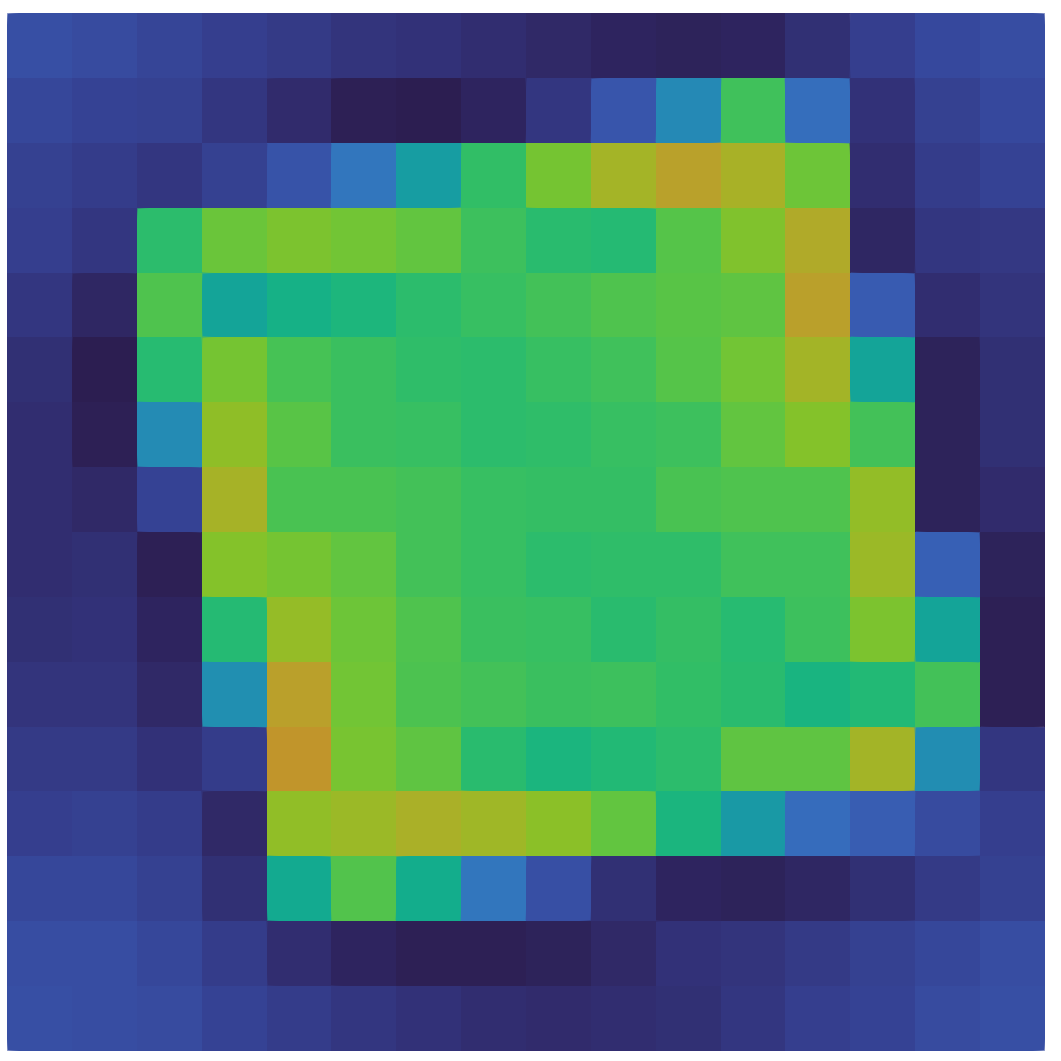}
         \subcaption{CoVo}
    \end{subfigure} 
    \begin{subfigure}[b]{0.30\textwidth}
        \centering 
        \includegraphics[width=\textwidth]{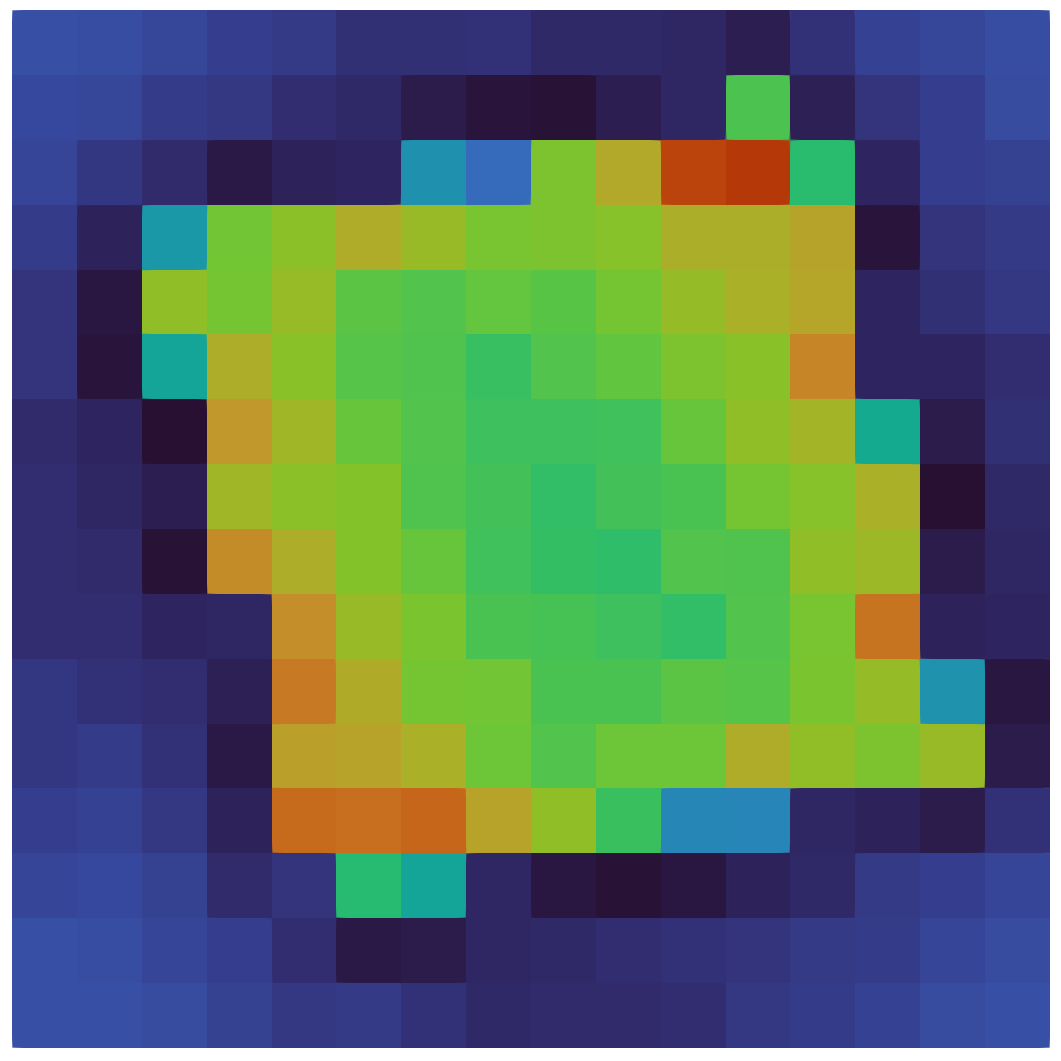}
         \subcaption{Q1R}
    \end{subfigure}    
    \begin{subfigure}[b]{0.30\textwidth}
        \centering 
        \includegraphics[width=\textwidth]{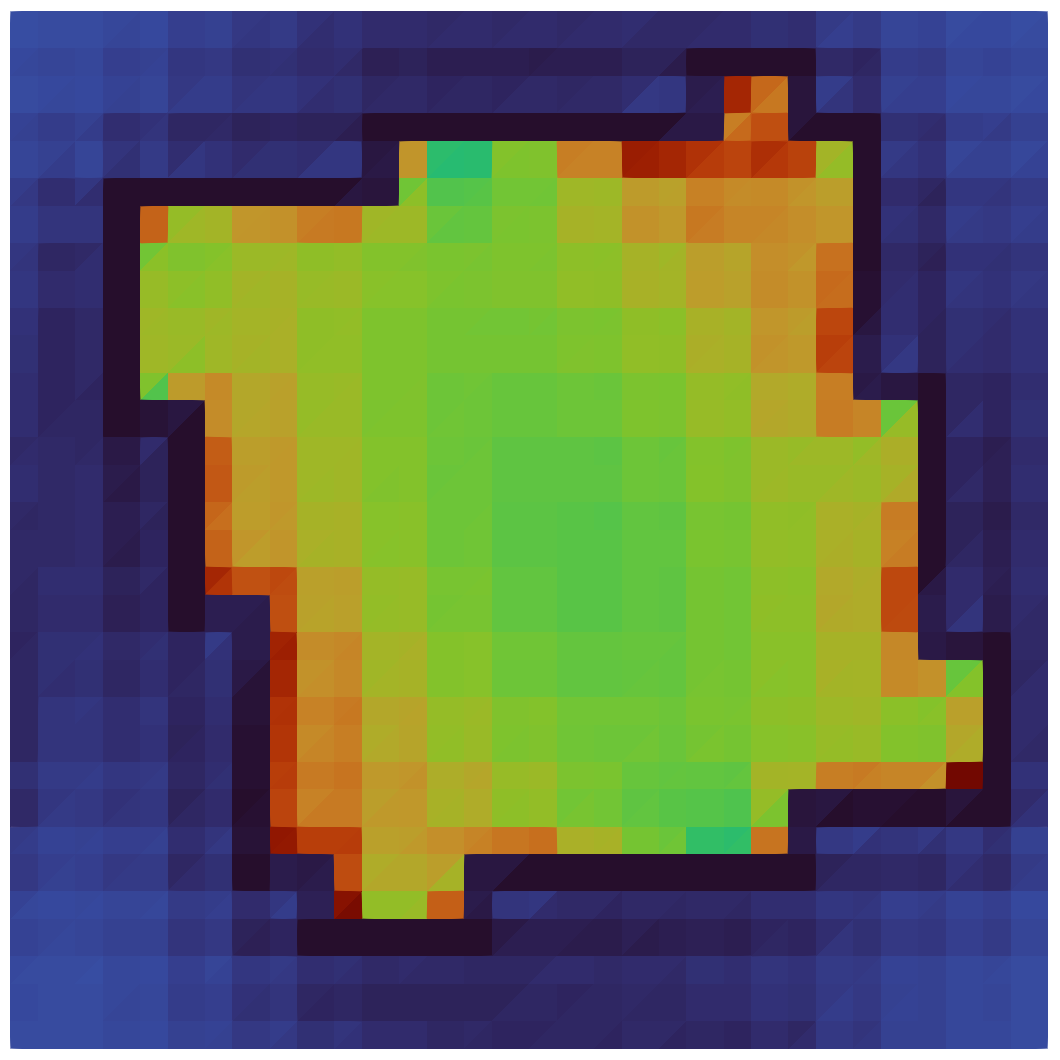}
         \subcaption{P1}
    \end{subfigure}
    \begin{subfigure}[b]{0.075\textwidth}
        \centering 
        \includegraphics[width=\textwidth]{color_bar_Cuboid.pdf}
    \end{subfigure}
    \caption{Local stress field at voxel count \mbox{$N=16$} for the rotated cubic inclusion.}
     \label{fig:localRotatedCuboid}
\end{figure}

For the cubic inclusion that is parallel to the mesh, we show the xx-component of the effective stress versus the resolution in Fig.~\ref{fig:EffStressCuboid}. At low resolutions, the Q1R and the P1 discretization feature a fluctuating behavior of over- and underestimating the stress depending on the resolution. This fluctuating behavior mirrors the over-/underestimation of the volume fraction at low resolutions. The X-FEM discretization shows the most accurate approximation at resolution \mbox{$N=16$} followed by the CoVo discretization. All discretizations converge to the same value around $4.58\,$MPa at the highest resolution \mbox{$N=1024$}.

In Fig.~\ref{fig:ErrEffStressCuboid}, we consider the error in the effective stress for the cubic inclusion that is parallel to the mesh, using X-FEM at resolution \mbox{$N=1200$} as the reference solution. For details on the choice of the reference solution, we refer to Appendix~\ref{apx:reference}. Compared to the other studies, the error rates appear more ragged. Still the decrease in the error with decreasing mesh parameter seems to follow a linear trend for the Q1R, P1 and CoVo discretization. For X-FEM, the trend appears to be superlinear, but not quadratic, with decreasing mesh parameter. The subquadratic trend for X-FEM is not surprising, because the quadratic convergence of X-FEM hinges on the $H^1$-regularity of the strain field, which holds for smooth inclusions but is violated for the example at hand. Still, X-FEM provides the most accurate solution for the parallel cubic inclusion at most resolutions.

In Fig.~\ref{fig:localCuboid}, we show a slice of the local stress field, where the cutting plane is parallel to the y-z-plane and intersects the center of the parallel cubic inclusion. We compare the individual discretization results at the resolution \mbox{$N=16$} to a reference solution computed via X-FEM at \mbox{$N=1200$}, which appears indistinguishable across all other discretizations at the same resolution. We observe that P1 leads to local over- and underestimations of the local stress. The Q1R discretization shows a good agreement in the order of the stress values but overestimates the volume of the inclusion leading to an error in the location of the stress minima and maxima. CoVo tackles this problem by applying a laminate mixing rule to the voxels at the interface, such that at the interface the stress maxima are underestimated, and the stress minima are overestimated. Overall, the X-FEM discretization approximates the outlines of the cubic inclusion the most accurately. However, we observe some overestimation of the local stress close to the interface.

For the cubic inclusion that is rotated around its center by 15° along each axis, we show the xx-component of the effective stress versus the resolution in Fig.~\ref{fig:EffStressRotatedCuboid}. We observe a strong overestimation of the effective stress of the P1 discretization followed by the Q1R discretization. CoVo and X-FEM approximate the effective stress the most accurately at low resolutions. All discretizations converge to the same value around $4.49\,$MPa at the highest resolution \mbox{$N=1024$}.

To get a more detailed impression, we analyze the error in the effective stress for the cubic inclusion that is rotated, using X-FEM at resolution \mbox{$N=1200$} as the reference solution. The results are shown in Fig.~\ref{fig:ErrEffStressRotatedCuboid}. For the P1, Q1R and CoVo discretization, we observe a roughly linear decrease in the error with increasing resolution. For X-FEM, the trend appears to be superlinear but below quadratic, which is a consequence of the non-smoothness of the corners of the cubic inclusion. Still, X-FEM provides the most accurate results for resolutions exceeding a voxel count of $64$ voxels per edge.

In Fig.~\ref{fig:localRotatedCuboid}, a slice of the local stress field is shown, where the cutting plane is parallel to the y-z-plane and intersects the center of the rotated cubic inclusion. The individual discretization results at the resolution \mbox{$N=16$} are compared to a reference solution provided by X-FEM with $1200$ voxels per edge. The P1 and Q1R discretizations show strong over- and underestimation in the local stresses and do not accurately approximate the inclusion shape. The approximation of the inclusions' outline is also an issue for CoVo and the stress values at the interface get smoothed by the laminate mixing rule. X-FEM approximates the outlines of the cubic inclusion the most accurately. Yet, we observe some over- and underestimation of the local stress fields in the inclusion. 

\subsubsection*{Numerical efficiency}
\begin{figure}[htb]
    \centering
       \fbox{\includegraphics[width=0.8\textwidth]{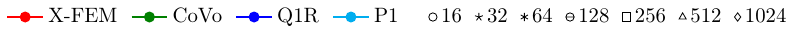}}
    \begin{subfigure}[b]{0.45\textwidth}
        \includegraphics[width=0.8\textwidth]{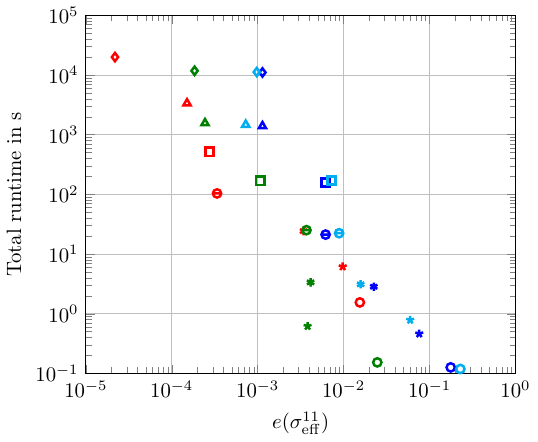}
        \caption{Total runtime for the parallel inclusion}
        \label{fig: TimeCuboid}
        \end{subfigure}
        \begin{subfigure}[b]{0.45\textwidth}
        \includegraphics[width=0.8\textwidth]{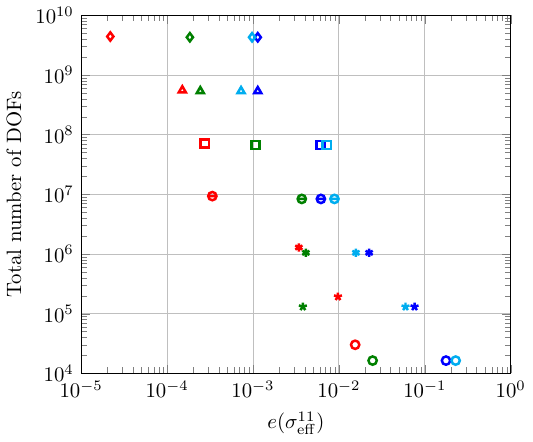}
        \caption{Degrees of freedom for the parallel inclusion}
        \label{fig: DOFCuboid}
        \end{subfigure}\\
        \begin{subfigure}[b]{0.45\textwidth}
        \includegraphics[width=0.8\textwidth]{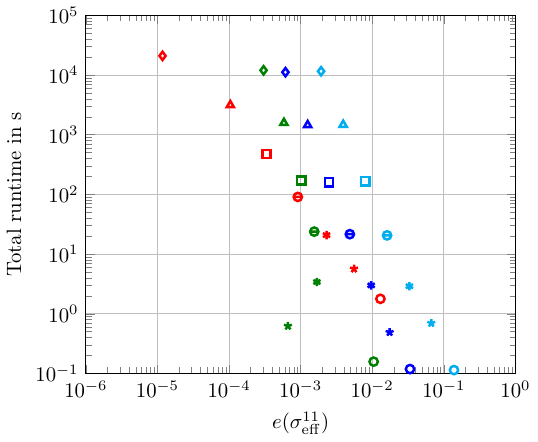}
        \caption{Total runtime for the rotated inclusion}
        \label{fig: TimeRotatedCuboid}
        \end{subfigure}
        \begin{subfigure}[b]{0.45\textwidth}
        \includegraphics[width=0.8\textwidth]{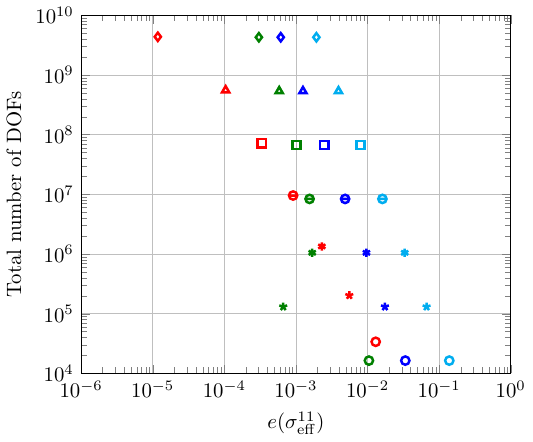}
        \caption{Degrees of freedom for the rotated inclusion}
        \label{fig: DOFRotatedCuboid}
        \end{subfigure}
        \caption{Numerical efficiency for the cubic inclusion studies.}
    \end{figure}

In Fig.~\ref{fig:IterationsCuboid}, we show the iteration count required for reaching the stopping criterion~\eqref{eq:stoppingCriterion} with a tolerance of $10^{-5}$ for linear CG in the parallel cubic inclusion study. The smallest iteration count is required by Q1R with approximately 35 iterations, followed by P1 and CoVo with around 40 iterations. At 50-57 iterations, X-FEM shows the highest iteration count. However, the iteration count of X-FEM appears to be stable with reference to the resolution.

We analyze the total runtime versus the accuracy for the parallel cubic inclusion in Fig.~\ref{fig: TimeCuboid}. We observe an increased total runtime of X-FEM compared to the other discretizations at the same resolution. The reasons for the increased runtime of X-FEM are identical to those discussed in Fig.~\ref{fig: TimeHashin}. Still, if an error in the effective stress below 0.1\% is desired, X-FEM offers the fastest solution. For example, at resolution \mbox{$N=128$}, X-FEM reaches an error of 0.03\% within 104s. 

In addition to the iteration count and the runtime, the memory footprint is relevant to analyze the numerical efficiency of each discretization in the parallel cubic inclusion study. Thus, in Fig.~\ref{fig: DOFCuboid}, we show the total number of \emph{dofs}, which influences the memory footprint as discussed in section~\ref{sec:hashin}. We observe a slight increase in \emph{dofs} for X-FEM compared to the other discretizations. Thus, X-FEM also offers the best benefit-cost ratio if an error below 0.1\% is desired.

For the rotated cubic inclusion study, we show the number of iterations until convergence of linear CG in Fig.~\ref{fig:IterationsRotatedCuboid}. We observe that the iteration count of Q1R, CoVo and P1 is around 40 iterations for all resolutions. For X-FEM, the iteration count is around 60 and appears to be stable with reference to the resolution. 

In Fig.~\ref{fig: TimeRotatedCuboid}, we show the total runtime versus the accuracy for the rotated cubic inclusion. The increased runtime of the X-FEM discretization is caused by the aspects discussed in Fig.~\ref{fig: TimeHashin}. Due to the high accuracy of CoVo, X-FEM only offers the fastest solution for a desired error below 0.05\%. 

To analyze the numerical efficiency for the rotated cubic inclusion study in terms of required \emph{dofs}, we show the total number of \emph{dofs} versus the accuracy in Fig.~\ref{fig: DOFRotatedCuboid}. There is only a small increase in the number of \emph{dofs} of X-FEM compared to the other discretizations. For a desired error in the effective stress that is below 0.05\%, X-FEM offers the cheapest solution in terms of \emph{dofs}.

The studies of the single cubic inclusion confirm that non-smooth interfaces may lead to suboptimal error convergence rates, as the quadratic convergence of the effective stress in X-FEM relies on the $H^1$-regularity of the strain field, a condition violated at the corner points of the cubic inclusion. In contrast to the investigation of the long fiber-reinforced composite, no conditioning issues were observed in the cubic inclusion studies. All in all, X-FEM provides the best benefit-cost ratio across most resolutions.

\section{Conclusion and perspectives}\label{sec:conclusion}
The work at hand introduced a three-dimensional X-FFT solver specifically designed for solving computational homogenization problems in linear elasticity with finite material contrast. This X-FFT solver achieves interface-conforming accuracy for matrix-inclusion problems with smooth interface geometry, while maintaining the numerical efficiency of FFT-based methods. Our computational studies demonstrated that interface-conforming convergence rates may be achieved not only for simplistic examples, such as Hashin's neutral inclusion, but also for more complex microstructures with smooth interfaces, including smooth rock inclusions in a cement matrix.
Despite featuring an increased iteration count, higher runtimes and an increased memory footprint compared to other discretizations in FFT-based methods, the X-FFT solver still offers the best balance between accuracy and numerical efficiency in most cases. 
The X-FFT solver is particularly useful for those interested in local fields because it produces remarkably accurate local fields, even at low resolutions.

To the best of our knowledge, applying the concept of strongly stable X-FEM~\cite{babuska_strongly_2017} to three-dimensional solid mechanics is novel, and using an FFT-based method to solve a mechanical homogenization problem discretized via X-FEM is an innovative strategy. We highlight that other X-FEM applications could benefit from using FFT-based solvers as well, e.g., modeling the propagation of cracks in heterogeneous microstructures~\cite{moes1999finite}. 
Furthermore, improving the current X-FFT framework to achieve interface-conforming accuracy in matrix-inclusion problems with non-smooth interfaces, such as fiber ends, would be beneficial. 
Based on the strategy discussed in Kästner et al.~\cite{kastner2011multiscale}, the X-FFT solver may be extended to allow for two interfaces per element. Extending it to nonlinear materials would also be feasible, although the large number of integration points could present a challenge, so that advanced strategies~\cite{gehrig2025element} for dealing with internal variables might become necessary. Moreover, improving the choice of the step size is recommended when extending it to nonlinear materials as linear CG is not suited for nonlinear systems.

\section*{Acknowledgements}
The authors would like to gratefully acknowledge the support of the European Research Council within the Horizon Europe program, project 101040238. The authors thank the anonymous reviewers for their valuable feedback.
\section*{Data availability statement}
The data that support the findings of this study are available from the corresponding author upon reasonable request.


\appendix
\section{Error bounds in conforming Galerkin methods}\label{apx:Bounds}
In this appendix, we derive the inequalities~\eqref{eq:upperBoundOnEffStiffnessError} and~\eqref{eq:upperBoundOnLocalStrainError}, which hold for conforming Galerkin methods. 

For linear elasticity, we consider a linear mapping between strain and stress fields based on the local stiffness matrix $\ffC\in\mathrm{Lin}\left(\Sym{3}\right)$, where the inequalities
\begin{equation}
    C_-\norm{\feps}^2\leq \feps:\ffC:\feps \leq C_+\norm{\feps}^2 \quad \text{for all } \feps\in\Sym{3}
\end{equation}
hold for all points $\fx\in Y$, with the positive constants $C_-,C_+$.
Under the assumption of exact integration over the cell $Y$, which is fulfilled by conforming Galerkin methods, the inequalities 
\begin{equation}\label{eq: inequalitiesOverCell}
    C_-\norm{\feps}_{L^2}^2\leq \left<\feps:\ffC:\feps\right>_Y \leq C_+\norm{\feps}_{L^2}^2
\end{equation}
result for all fields $\feps \in L^2(Y;\Sym{d})$.
We consider the difference between the simulated strain $\feps_h$ and the exact strain $\feps_*$. For this strain difference, the inequalities~\eqref{eq: inequalitiesOverCell} take the form
\begin{equation}\label{eq:inequalitiesBeforeSchneider1819}
    C_-\norm{\feps_* - \feps_h}_{L^2}^2\leq \left<\left(\feps_* - \feps_h\right):\ffC:\left(\feps_* - \feps_h\right)\right>_Y \leq C_+\norm{\feps_* - \feps_h}_{L^2}^2.
\end{equation}
Based on Schneider~\cite[eq.(18)]{schneider2022superaccurate}, we may rewrite the cell-averaged error of the elastic energy in the form
\begin{equation}
    \left<\left(\feps_* - \feps_h\right):\ffC:\left(\feps_* - \feps_h\right)\right>_Y =   \left<\feps_*:\ffC:\feps_* \right>_Y- \left<\feps_h:\ffC:\feps_* \right>_Y.
\end{equation}
For both terms on the right-hand side of the equation above, we follow the approach discussed in Schneider~\cite[eq.(19)]{schneider2022superaccurate}. More precisely, we use the definition of the exact strain
\begin{equation}
    \feps_* = \bar{\feps} + \nabla^s \fu_*,
\end{equation}
apply integration by parts and use the validity of the balance of linear momentum for the exact strain such that the relation
\begin{align*}
    \left<\feps_*:\ffC:\feps_* \right>_Y- \left<\feps_h:\ffC:\feps_* \right>_Y 
    &= \left<\feps_*:\ffC:\feps_* \right>_Y- \left<\feps_*:\ffC:\feps_h \right>_Y \\
    &= \left<\left(\bar{\feps} + \nabla^s \fu_*\right):\ffC:\feps_* \right>_Y- \left< \left(\bar{\feps} + \nabla^s \fu_*\right):\ffC:\feps_h \right>_Y \\
    &= \bar{\feps} \left<\ffC:\feps_* \right>_Y - \bar{\feps} \left< \ffC:\feps_h \right>_Y \\
    &= \bar{\feps}:\ffC^\mathrm{eff}_*:\bar{\feps} - \bar{\feps}:\ffC^\mathrm{eff}_h:\bar{\feps}
\end{align*} 
results with the simulated effective stiffness $\ffC^\mathrm{eff}_h$ and the exact effective stiffness $\ffC^\mathrm{eff}_*$.
In summary, the equation
\begin{equation}
    \left<\left(\feps_* - \feps_h\right):\ffC:\left(\feps_* - \feps_h\right)\right>_Y = \bar{\feps}:\left(\ffC^\mathrm{eff}_*-\ffC^\mathrm{eff}_h\right):\bar{\feps}
\end{equation}
holds for the cell-averaged error of the elastic energy, so that the inequalities~\eqref{eq:inequalitiesBeforeSchneider1819} take the form
\begin{equation}
    C_-\norm{\feps_* - \feps_h}_{L^2}^2\leq \bar{\feps}:\left(\ffC^\mathrm{eff}_*-\ffC^\mathrm{eff}_h\right):\bar{\feps} \leq C_+\norm{\feps_* - \feps_h}_{L^2}^2.
\end{equation}

The error in the effective elastic energy may thus be bounded from above by a value which is proportional to the quadratic error in the local strain fields via the relation
\begin{equation}
     \bar{\feps}:\left(\ffC^\mathrm{eff}_*-\ffC^\mathrm{eff}_h\right):\bar{\feps}\leq C_+\norm{\feps_* - \feps_h}_{L^2}^2.
\end{equation}
Moreover, the error in the local strain fields may be bounded from above by a value which is proportional to the square root of the error in the effective elastic energy via the inequality
\begin{equation}
    \norm{\feps_* - \feps_h}_{L^2} \leq \sqrt{\frac{\ \bar{\feps}:\left(\ffC^\mathrm{eff}_*-\ffC^\mathrm{eff}_h\right):\bar{\feps}}{C_-}}.
\end{equation}

\section{Reference values for computational studies}\label{apx:reference}

If no analytical solution is available, we use X-FEM at \mbox{$N=1200$} as a reference solution in our computational studies.
In the following, we show the errors that result if the individual discretizations at \mbox{$N=1200$} are chosen as the reference solution for their lower resolutions results.

\begin{figure}[H]    
    \centering 
	\fbox{\includegraphics[width=.6\textwidth]{legendGrain.pdf}}\\
    \begin{subfigure}[b]{0.32\textwidth}
        \centering 
        \includegraphics[width=\textwidth]{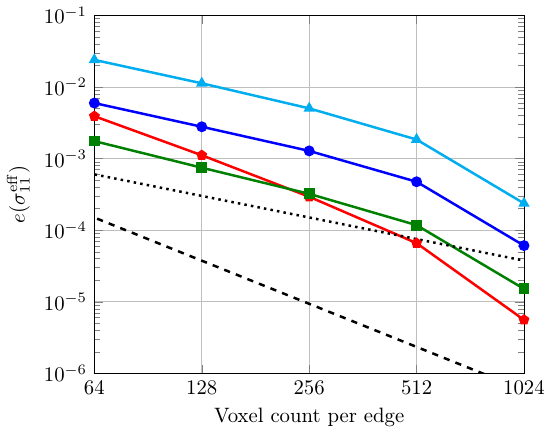}
         \subcaption{Rock-cement microstructure}\label{fig:comparisonRockIndi}
    \end{subfigure}
    \begin{subfigure}[b]{0.32\textwidth}
        \centering 
        \includegraphics[width=\textwidth]{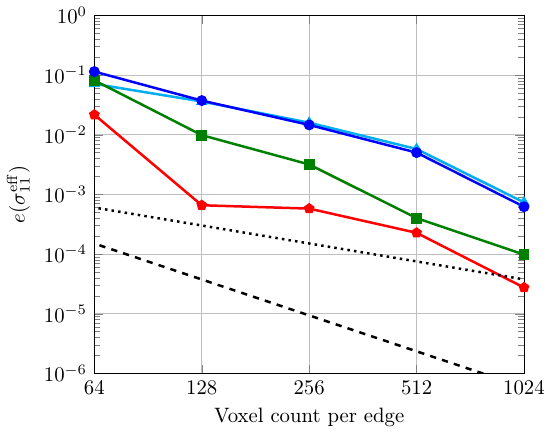}
         \subcaption{Long fiber reinforced composite}\label{fig:comparisonLongFiberIndi}
    \end{subfigure}\\
    \begin{subfigure}[b]{0.32\textwidth}
        \centering 
        \includegraphics[width=\textwidth]{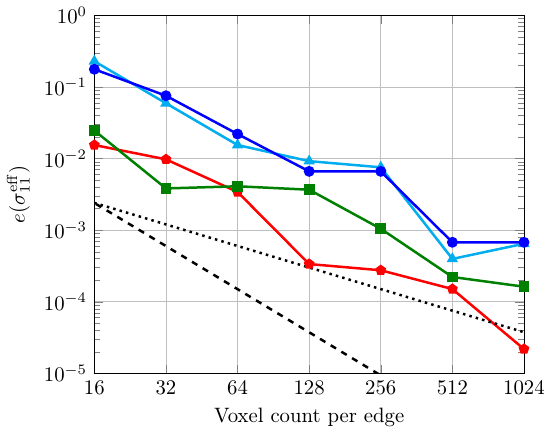}
         \subcaption{Parallel cubic inclusion}\label{fig:comparisonCuboidIndi}
    \end{subfigure}
        \begin{subfigure}[b]{0.32\textwidth}
        \centering 
         \includegraphics[width=\textwidth]{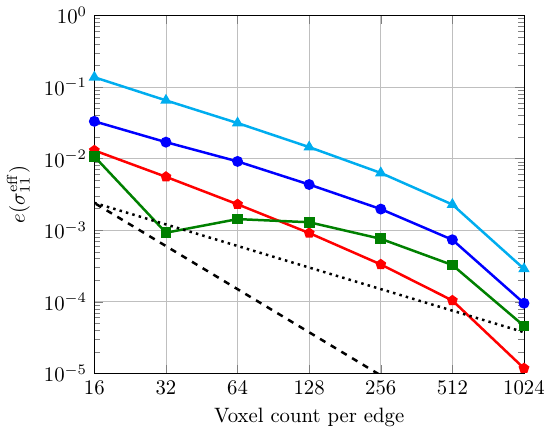}
         \subcaption{Rotated cubic inclusion}\label{fig:comparisonRotatedCuboidIndi}
    \end{subfigure}
    \caption{Error in the effective stress using the individual reference solution.}
\end{figure}

In Fig.~\ref{fig:comparisonRockIndi}, we show the results for the rock-cement microstructure if we use each individual discretization at resolution \mbox{$N=1200$} as reference. We observe that the errors are indistinguishable from those shown in Fig.~\ref{fig:ErrEffStressGrain} for all resolutions except at $1024$ voxels per edge. At this resolution, the errors in Fig.~\ref{fig:comparisonRockIndi} are below the general trend observed in Fig.~\ref{fig:ErrEffStressGrain}. This behavior indicates that the individual discretizations still experience a non-negligible error at \mbox{$N=1200$}. As the X-FEM discretization shows the most accurate results at \mbox{$N=1200$}, we chose to use it as a reference solution in the manuscript.

In Fig.~\ref{fig:comparisonLongFiberIndi}, the results of the long fiber reinforced composite are shown. For Q1R and P1, the errors are identical for all resolutions except at $1024$ voxels per edge due to the reasons discussed above. For CoVo, also the value at resolution \mbox{$N=512$} is slightly influenced by the change in reference.

Similar trends to those discussed above are observed when comparing the results of the parallel cubic inclusion shown in Fig.~\ref{fig:comparisonCuboidIndi} to those shown in Fig.~\ref{fig:ErrEffStressCuboid}, and when comparing the results of the rotated cubic inclusion shown in Fig.~\ref{fig:comparisonRotatedCuboidIndi} to those shown in Fig.~\ref{fig:ErrEffStressRotatedCuboid}.

\bibliographystyle{ieeetr}
{\footnotesize
\bibliography{references}
}
\end{document}